\documentclass[11pt,a4paper]{article}
\usepackage{jheppub,amsmath,  amssymb,slashed,url,bm,textgreek,upgreek}
\usepackage{graphicx}
\usepackage{epstopdf}
\usepackage{subcaption}
\usepackage[normalem]{ulem}
\usepackage{relsize}  
\usepackage{braket}
\usepackage{dsfont}
 
\def\g{\text{{\teneurm g}}}

\def\be{\begin{equation}}
\def\ee{\end{equation}}

\def\hat{\widehat}

\def\tilde{\widetilde}

\def\d{{\mathrm d}}

\def\R{{\mathbb R}}

\def\U{{\sf U}}

\def\[{\bigl [}

\def\]{\bigr ]}

\def\Tr{{\mathrm {tr}}}

\def\1{{\mathbf 1}}

\def\tilde{\widetilde}

\def\AdS{{\mathrm{AdS}}}

\def\SL{{{SL}}}
\def\i{{\mathrm i}}
\def\hSL{\widetilde{{SL}}}
\def\hSU{\widetilde{{SU}}}

\font\teneurm=eurm10 \font\seveneurm=eurm7  \font\fiveeurm=eurm5
\newfam\eurmfam
\textfont\eurmfam=\teneurm \scriptfont\eurmfam=\seveneurm
\scriptscriptfont\eurmfam=\fiveeurm

\font\teneusm=eusm10 \font\seveneusm=eusm7 \font\fiveeusm=eusm5
\newfam\eusmfam
\textfont\eusmfam=\teneusm \scriptfont\eusmfam=\seveneusm
\scriptscriptfont\eusmfam=\fiveeusm

\font\tencmmib=cmmib10 \skewchar\tencmmib='177
\font\sevencmmib=cmmib7 \skewchar\sevencmmib='177
\font\fivecmmib=cmmib5 \skewchar\fivecmmib='177
\newfam\cmmibfam
\textfont\cmmibfam=\tencmmib \scriptfont\cmmibfam=\sevencmmib
\scriptscriptfont\cmmibfam=\fivecmmib

\def\Tr{{\mathrm{Tr}}}

\title{Algebras and states in super-JT gravity}

\author{Geoff Penington$^{1}$ and Edward Witten$^2$}
\affiliation{$^1$Center for Theoretical Physics and Department of Physics, University of California,\\  Berkeley, CA 94720 USA}
\affiliation{$^2$School of Natural Sciences, Institute for Advanced Study,\\ 1 Einstein Drive, Princeton, NJ 08540 USA}

\abstract{In bosonic JT gravity, minimally coupled to bulk matter, there exists a single, delta-function-normalisable state in each $\hSL(2,\R)$ representation of the matter QFT for any pair of positive  energies $E_L, E_R$ at the left and right boundaries. In $\mathcal{N} = 2$ super-JT gravity coupled to matter, we show that there exists a single normalisable state in each $\hSU(1,1|1)$ matter representation (given appropriate R-charges) that has exactly zero energy at both boundaries. For non-BPS representations, these states have the peculiar property that they break all supersymmetry in the bulk, while preserving supersymmetry at both boundaries. Projecting the algebras of boundary observables onto these zero-energy states leads to  a Type II$_1$ von Neumann factor at each boundary that contains a single operator for each supersymmetric matter boundary primary with sufficiently small R-charge. For neutral boundary primaries, the Type II$_1$ factor has a natural action on the matter QFT Hilbert space (with no additional gravitational degrees of freedom) such that the QFT vacuum is the unique tracial state. Moreover, the product of neutral matter operators can be found very explicitly and has a remarkably simple form. When primaries with nonzero matter R-charge  are included, the trace can be written as a sum over matter vacuum expectation values associated to each allowed boundary R-charge $J_R$, with the terms in the sum weighted by $\mathrm{cos}(\pi J_R)$. In this way, the ground state algebras encode the ratios of the number of BPS microstates within each R-charge sector. In addition to the results on super-JT gravity described above, we provide a purely Lorentzian derivation of the algebraic structure of canonically quantised (bosonic) JT gravity plus matter, without appeal to the Euclidean gravitational path integrals used in previous work.}

\begin{document}\maketitle

\section{Introduction}
 Perhaps our greatest insight into nonperturbative quantum gravity is that the logarithm of the number of microstates of a black hole should be given by the Bekenstein-Hawking entropy \cite{Bekenstein:1973ur, Hawking:1975vcx}
\begin{align}\label{eq:bhentropy}
S_{BH} = \frac{A_\mathrm{hor}}{4G}.
\end{align}
In string theory, supersymmetric microstates can be counted directly at weak coupling where they are described by perturbative strings anchored on D-branes \cite{Strominger:1996sh}. Since the number of supersymmetric microstates, or at least the associated index, is independent of the string coupling, the same microstate count should also be found at strong coupling where the microstates are black holes. Indeed, in the semiclassical limit, the stringy calculation agrees with \eqref{eq:bhentropy}. 

Nevertheless an explicitly \emph{gravitational} description of the microstates of a black hole, whether in string theory or otherwise, has  proved elusive. Instead, the formula \eqref{eq:bhentropy} has  traditionally been justified within semiclassical gravity either  via the Hawking temperature \cite{Hawking:1975vcx} and the classical thermodynamic relation
\begin{align}\label{eq:clausius}
dE = TdS
\end{align}
 or by computing the classical gravitational action of a Euclidean black hole \cite{PhysRevD.15.2752}. While both approaches provide compelling indirect evidence that \eqref{eq:bhentropy} is correct, neither offers a direct statistical interpretation of \eqref{eq:bhentropy} as an entropy. 

In recent work \cite{Witten:2021unn, Chandrasekaran:2022eqq}, however, it was shown that the Bekenstein-Hawking entropy \eqref{eq:bhentropy} can be seen directly in the algebra $\mathcal{A}_{\rm bdy}$ of asymptotic boundary observables present in a particular $G_N \to 0$ limit of canonically quantised gravity.\footnote{Other recent progress on the algebraic structure of semiclassical gravity and its holographic duals includes \cite{Leutheusser:2021ab, Leutheusser:2021aa, Chandrasekaran:2022cip, Leutheusser:2022bgi, Witten:2023qsv, Jensen:2023yxy, Witten:2023xze, Kudler-Flam:2023qfl, Akers:2024bel, chen2024clock, Kudler-Flam:2024psh}.}  In this limit, the algebra $\mathcal{A}_{\rm bdy}$ becomes a type II$_\infty$ von Neumann factor, meaning that it has divergent entanglement, but has a ``trace'' that is unique up to an arbitrary multiplicative constant. Here, a trace is simply a suitably continuous linear map $\Tr: \mathcal{A}_{\rm bdy} \to \R$ satisfying 
\begin{align}
\Tr(a b) = \Tr(b a)
\end{align}
for all $a, b \in \mathcal{A}_{\rm bdy}$. Given any state $\ket{\Psi}$, such a trace allows us to define the reduced density matrix $\rho_{\rm bdy}$ by
\begin{align}
\Tr(\rho_{\rm bdy}\, a) = \braket{\Psi|a|\Psi},
\end{align}
 for all $a \in \mathcal{A}_{\rm bdy}$, and consequently the entropy 
\begin{align}\label{eq:typeIIentropy}
S(\rho_{\rm bdy}) = -\Tr(\rho_{\rm bdy }\log \rho_{\rm bdy}).
\end{align}
This entropy is unique up to a state-independent additive constant, which should be thought of as a renormalisation ambiguity that arises when subtracting the divergent entanglement present for all states in a Type II$_\infty$ algebra. In particular, differences in entropy between pairs of states are unambiguous. 

The entropy \eqref{eq:typeIIentropy} quantifies the statistical uncertainty of observables in the algebra  $\mathcal{A}_{\rm bdy}$ and is the algebraic generalization of the von Neumann entropy of a reduced density matrix on a tensor product subsystem of Hilbert space. It was shown in \cite{Witten:2021unn, Chandrasekaran:2022eqq} that the entropy associated to the algebra $\mathcal{A}_{\rm bdy}$ includes a contribution from the Bekenstein-Hawking entropy of the black hole, in addition to the entropy of quantum fields outside the black hole. The divergent entanglement present in the  algebra $\mathcal{A}_{\rm bdy}$ comes from the universal divergence in $S_{BH}$ as $G_N \to 0$ for black holes with $A_{\rm hor}$ fixed up to $O(G_N)$ corrections.

In two spacetime dimensions, there exist UV-complete toy models of quantum gravity where the full quantum algebraic structure can be studied without the need to take a semiclassical limit. In particular, the algebraic structure of canonically quantised JT gravity \cite{Jackiw:1984je, Teitelboim:1983ux}, coupled to arbitrary quantum matter, was studied in \cite{penington2023algebras, kolchmeyer2023neumann}. The Hilbert space of this theory is acted on by two commuting algebras $\mathcal{A}_L$ and $\mathcal{A}_R$ associated to the left and right boundaries. In pure JT gravity these algebras are commutative, but after adding matter they become Type II$_\infty$ factors, like the algebra $\mathcal{A}_{\rm bdy}$ of a semiclassical black hole. The two boundary algebras are commutants, meaning that any operator commuting with all right-boundary observables is contained in the left-boundary algebra and vice versa. And, just like $\mathcal{A}_{\rm bdy}$, entropies on the algebras $\mathcal{A}_L$ and $\mathcal{A}_R$ include a Bekenstein-Hawking contribution. The universal divergent entanglement comes from the topological Gauss-Bonnet term in the JT gravity action, which needs to be taken to infinity in a Euclidean path integral to suppress topology change and thereby match the canonically quantised theory. In sharp contrast to pure JT gravity, JT gravity plus matter is therefore a genuinely \emph{holographic} theory, in the specific sense that a) boundary operators form a complete basis for all observables in the theory  and b) semiclassical states have boundary entanglement entropies given by a holographic entropy prescription \cite{Ryu:2006ef, Hubeny:2007xt, Lewkowycz:2013nqa, Faulkner:2013ana, Engelhardt:2014gca}.

In this work, we carry out a similar analysis to \cite{penington2023algebras, kolchmeyer2023neumann} for $\mathcal{N}=2$ super-JT gravity \cite{Fu:2016vas, Lin:2022rzw, Lin:2022zxd} coupled to an arbitrary supersymmetric matter field theory. Our motivation for doing so is twofold. Firstly, we would like to better understand the connection between the algebraic derivations of the Bekenstein-Hawking entropy described in \cite{Witten:2021unn, Chandrasekaran:2022eqq, penington2023algebras, kolchmeyer2023neumann} and black hole microstate counts found using string theory \cite{Strominger:1996sh}. $\mathcal{N} = 2$ super-JT gravity is perhaps the simplest gravitational theory that shares important features with the stringy black holes studied in \cite{Strominger:1996sh}. Those features include, in particular, an exponential degeneracy of exact ground states in the Euclidean partition function \cite{fu2017supersymmetric, stanford2017fermionic}.

Secondly, our work is  inspired, in large part, by some very interesting observations made recently in \cite{Lin:2022rzw, Lin:2022zxd} about the physics of zero-energy wormholes in super-JT gravity plus matter. The authors of \cite{Lin:2022rzw, Lin:2022zxd} first noted that in super-JT gravity the two-point function between matter fields at the left and right boundaries, which can be computed using canonically quantised pure super-JT gravity, remains finite even as we take the inverse temperature $\beta \to \infty$. 

\begin{figure}
\begin{subfigure}{.48\textwidth}
  \centering
 \includegraphics[width = 0.75\linewidth]{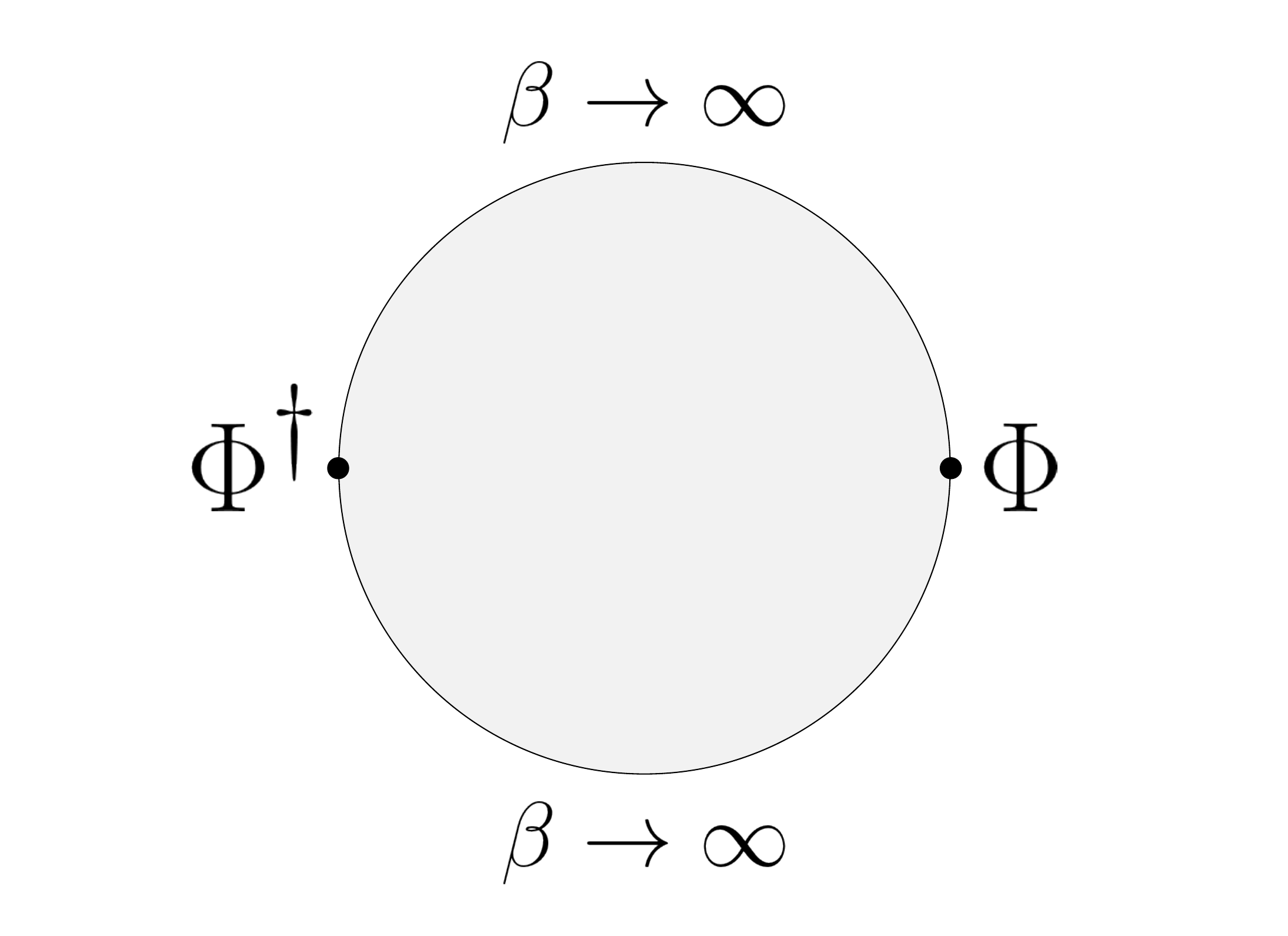}
\end{subfigure}
\begin{subfigure}{.48\textwidth}
  \centering
 \includegraphics[width = 0.75\linewidth]{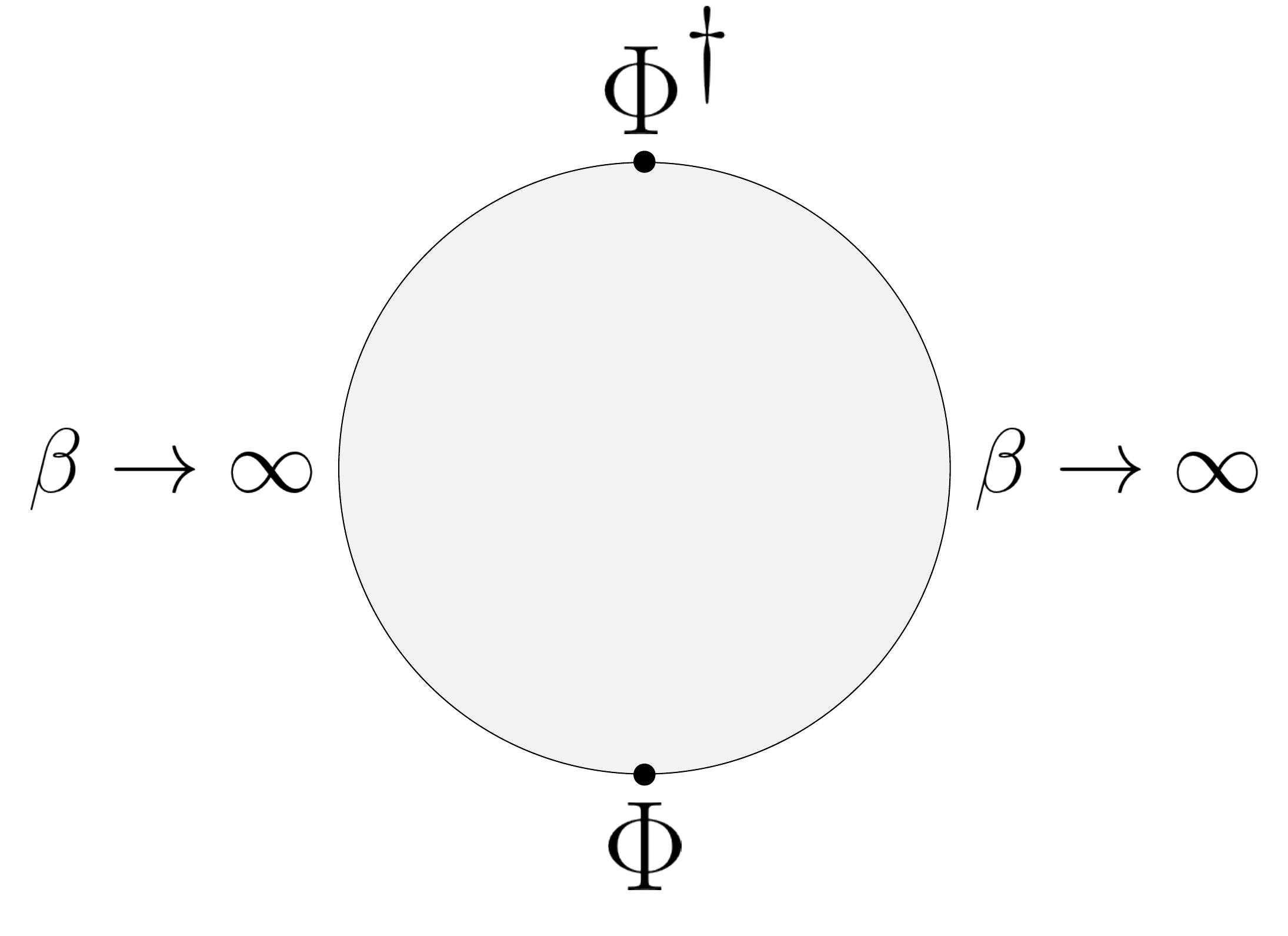}
\end{subfigure}
\caption{\footnotesize The same Euclidean path integral computes a left-right boundary two-point function (\emph{left}) that can be found using canonically quantised pure super-JT gravity and the inner product between two states with nontrivial matter excitations (\emph{right}) that has been projected to have zero energy at both the left and right boundaries. Interestingly, this two-point function turns out to have a finite nonzero limit, which implies the existence of states with nontrivial matter excitations in the bulk, but exactly zero energy at both boundaries.}
\label{fig:betainf}
\end{figure}

However, as shown in Figure \ref{fig:betainf}, the same gravitational path integral that computes this two-point function should also compute an inner product, in super-JT gravity plus matter, between two copies of a state where the matter fields are excited but both boundaries are projected into the zero energy subspace. Since matter one-point functions vanish, this state has zero overlap with an empty wormhole containing no matter excitations. But, because the two-point function is finite, it has finite norm. In other words, there exist normalisable states where bulk matter fields are excited, but both boundaries have exactly zero energy. Somewhat remarkably, this is true even if the matter operators involved are not BPS. In other words, there exist states that preserve all boundary supersymmetry despite breaking all bulk supersymmetry.

By explicitly constructing the canonically quantised theory of super-JT gravity coupled to matter, we are able to verify the conclusions of \cite{Lin:2022rzw, Lin:2022zxd}, without appealing to path integral methods. Moreover, we are able to precisely count the number of zero-energy states associated to each irreducible representation of the superisometry group in the matter Hilbert space. 

We first show that the Hilbert space of super-JT gravity can be described as the super tensor product\footnote{Here, a super-tensor product is the same as a tensor product except that fermionic operators acting on independent subsystems anticommute with each other (as usual).}
\begin{align}
\mathcal{H}_{\rm super-JT} \cong  \mathcal{H}_{\rm matt} \otimes L^2(\R) \otimes L^2([0,a_{\rm max}]) \otimes \mathcal{H}_{\rm fermion} \otimes \mathcal{H}_{\rm fermion}'. 
\end{align}
Here $ \mathcal{H}_{\rm matt}$ is the matter Hilbert space in a fixed $\AdS_2$ background, the position operator on $L^2(\R)$ describes the renormalised length $\ell = 2 \rho$ of a geodesic connecting the left and right boundaries of the spacetime, $L^2([0,a_{\rm max}])$ describes the phase $a$ of a Wilson line that stretches between the two boundaries for the R-charge gauge field  and $\mathcal{H}_{\rm fermion}$ and $\mathcal{H}_{\rm fermion}'$ are fermionic modes acted on by $\psi, \psi^\dagger$ and $\psi', \psi'^\dagger$ respectively. 
The matter Hilbert space is acted on by an $\hSU(1,1|1)$ Lie superalgebra whose bosonic part consists of the $\hSL(2,\R)$ isometry group of $\AdS_2$, with generators $j_1, j_2, j_3$, together with a $U(1)$ R-charge $j$. There are also four fermionic charges $g_\pm, g_\pm^\dagger$. In terms of these generators one can show that the left- and right-boundary Hamiltonians $H_L$ and $H_R$ can be written as
\begin{align}
H_L = \{Q_L,Q_L^\dagger\}=  \frac{1}{4}(p_{a}-j)^2 +& \frac{1}{4} (p_{\rho}-j_3)^2 + e^{-\rho} j_- + e^{-2\rho} + i e^{-\rho+ia} \psi' \psi^\dagger \\\nonumber&+ i e^{(-\rho+ia)/2} \psi' g_-^\dagger + ie^{-\rho-ia} \psi'^\dagger \psi+ i e^{(-\rho-ia)/2} \psi'^\dagger g_-\\\label{eq:introHR}
H_R =  \{Q_R, Q_R^\dagger\} =  \frac{1}{4} (p_a + j)^2 +& \frac{1}{4}(p_\rho + j_3)^2 + e^{-\rho} j_+ + e^{-2 \rho} - i e^{-\rho-ia} \psi \psi'^\dagger \\&- i e^{(-\rho-ia)/2} \psi g_+^\dagger - i e^{-\rho+ia} \psi^\dagger \psi' - i e^{(-\rho+ia)/2} \psi^\dagger g_+\nonumber,
\end{align}
where $p_\rho$ and $p_a$ are the conjugate momenta to $\rho$ and $a$. The boundary Hamiltonians are supersymmetric, with supercharges
\begin{align}
Q_L =& \frac{1}{2} \psi' \left(p_{\rho} - j_3 + i p_{a} - ij\right) +e^{-\rho-ia} \psi + e^{(-\rho-ia)/2}g_- \\
Q_R =& \frac{1}{2} \psi \left(p_\rho + j_3 - i p_a - ij\right)  - e^{-\rho+ia} \psi' - e^{(-\rho+ia)/2}g_+ \label{eq:introQR}
 \end{align}
Finally, the supercharges are charged under the boundary R-charges
\begin{align}
J_L =& -p_a+j + \frac{1}{2} [\psi',\psi'^\dagger]\\
J_R =&\, \,p_a+j + \frac{1}{2} [\psi,\psi^\dagger].
\end{align}

We are particularly interested in the states, discovered in  \cite{Lin:2022rzw, Lin:2022zxd}, where \emph{both} boundaries have exactly zero energy. However it is simplest to first consider the subspace of states with zero energy at the right boundary and arbitrary energy at the left boundary. After conjugation by an appropriate unitary, one can show that the right-boundary Hamiltonian and supercharges become equal to those of pure super-JT gravity (i.e. \eqref{eq:introHR} and \eqref{eq:introQR} with the matter $\hSU(1,1|1)$ charges set to zero). Pure super-JT gravity has a single normalisable ground state for each allowed value of the boundary R-charge $J_R$ with $|J_R| < 1/2$.\footnote{The precise set of allowed charges depends on a choice of quantisation for the $\hSU(1,1|1)$ R-charge.} In our case, however,  the full Hilbert space is a super-tensor product of the pure super-JT gravity Hilbert space with the matter Hilbert space. As a result, the subspace of states with $H_R = 0$, for any fixed  $-1/2 < J_R <1/2$, is naturally isomorphic to the matter Hilbert space $\mathcal{H}_{\rm matt}$.

The left-boundary Lie superalgebra commutes with the right Hamiltonian $H_R$ and so preserves this subspace. As a result, the restrictions $\tilde Q_L, \tilde Q_L^\dagger, \tilde H_L, \tilde J_L$ of the left-boundary Lie superalgebra to the $H_R = 0$ (and fixed $J_R$) subspace become operators acting on the matter Hilbert space. With some effort, one can show that
\begin{align}
\tilde Q_L = - (1 + g_+^\dagger g_+)^{-1/4} (1 + &g_+ g_+^\dagger)^{1/4} g_- (1+g_+^\dagger g_+)^{1/4} (1+g_+g_+^\dagger)^{-1/4} + i\frac{J_R}{2} (1+j_+)^{-1/2} g_+\nonumber\\ \label{eq:tildesuperalgebraintro}
&\tilde H_L = \{\tilde Q_L, \tilde Q_L^\dagger\} ~~~~~~~~
\tilde J_L = -J_R + 2 j.
\end{align}
The set of states in super-JT gravity with $H_R = H_L = 0$ are therefore in one-to-one correspondence the set of matter QFT states with $\tilde H_L = 0$ or, equivalently, with the cohomology of $\tilde Q_L$. Since $\tilde Q_L$ only involves matter $\hSU(1,1|1)$ charges, this cohomology splits as a direct sum over the matter $\hSU(1,1|1)$ representations that make up the QFT Hilbert space $\mathcal{H}_{\rm matt}$. One finds that in fact there exists exactly one $H_L = H_R = 0$ state with right R-charge $|J_R| < 1/2$ for each matter representation with primary R-charge $u$  such that 
\begin{align}
|\tilde J_L| = |J_R - 2 u| < 1/2.
\end{align}
It is worth emphasizing the somewhat peculiar nature of this conclusion. The set of $\hSU(1,1|1)$ representations present in $\mathcal{H}_{\rm matt}$ include every state that can be described by the matter theory, modulo isometries of $\AdS_2$, phase rotations proportional to the R-charge, and their superpartners. This includes e.g. states that  describe
complicated dynamical processes involving a large number of high-energy particles. And yet, so long as the total R-charge of a representation is sufficiently small, there will exist a state within that
representation in which both boundaries have exactly zero energy. Such states will in general have highly nontrivial bulk dynamics, but no boundary dynamics at all. The situation is somewhat reminiscent of quantum gravity in closed cosmologies, where there is no asymptotic boundary or physical time evolution operator and yet the Wheeler-de Witt constraint can apparently accommodate all the dynamics of general relativity. 

An advantage of ground states in super-JT gravity over closed cosmologies, however, is that, while boundary time evolution is trivial, there is still a rich algebra $\widetilde{\mathcal{A}}_R$ of gauge-invariant \emph{boundary observables} \cite{Lin:2022rzw, Lin:2022zxd} that preserve the subspace of ground states. This algebra is generated by right-boundary matter primaries $\Phi^i_{R}(t)$ projected onto states with $H_R = 0$. The explicit action $\tilde \Phi^i_{R}$ of these operators on the matter QFT Hilbert space is messy and mostly unenlightening. Even in the simplest case, where $J_R = 0$ and the boundary primary $\Phi^i_{R}(t)$ is R-charge neutral, we have the rather complicated expression
\begin{align} \label{eq:tildePhimessyform}
\tilde \Phi^i_{R} &= \frac{8}{\pi}\int_{-\infty}^\infty d \rho\, \exp(-(2\Delta_i +2)\rho) \,\times\\\nonumber & ~~~~\left[ \sqrt{1 + g_+ g_+^\dagger} K_{1/2}(2 \sqrt{1 + j_+} e^{-\rho}) \Phi_{R,0}^i(0) K_{1/2}(2 \sqrt{1 + j_+} e^{-\rho})  \sqrt{1 + g_+ g_+^\dagger}\right.\\ \nonumber&~~~~~~\left.  + g_+^\dagger K_{1/2}(2 \sqrt{1 + j_+} e^{-\rho}) \Phi_{R,0}^i(0) K_{1/2}(2 \sqrt{1 + j_+} e^{-\rho})  g_+\right.\\\nonumber&~~~~~~\left. + g_+ K_{1/2}(2 \sqrt{1 + j_+} e^{-\rho}) \Phi_{R,0}^i(0) K_{1/2}(2 \sqrt{1 + j_+} e^{-\rho})  g_+^\dagger\right.\\ &~~~~~~\left. + \sqrt{1 + g_+^\dagger g_+ } K_{1/2}(2 \sqrt{1 + j_+} e^{-\rho}) \Phi_{R,0}^i(0) K_{1/2}(2 \sqrt{1 + j_+} e^{-\rho})  \sqrt{1 + g_+^\dagger g_+}\right]\nonumber,\end{align}
where $K_{\nu}(z)$ is a modified Bessel function. Note here that unlike the original boundary primary $\Phi^i_{R}(t)$, which is a local operator that depends on the boundary time $t$ at which it acts, the operator $\tilde \Phi^i_{R}$ is a single, densely defined operator that is time independent. 

Somewhat miraculously, the algebra $\widetilde{\mathcal{A}}_{R,0}$ spanned by operators of the form \eqref{eq:tildePhimessyform} closes, such that one can always write
\begin{align}\label{eq:simpleproduct}
\tilde \Phi^i_{R} \tilde \Phi^j_{R} = \sum_{k} A^{ij}_k \tilde \Phi^k_{R},
\end{align}
with $A^{ij}_k$ $c$-numbers. (Here, the sum over matter primaries $k$ needs to include the identity operator.) Even more remarkably, the coefficients $A^{ij}_k$ can be found explicitly, and are given by
\begin{align}
A^{ij}_k  = C_{ijk} \frac{4^{\Delta_k - \Delta_i -\Delta_j} \Gamma(1 + 2\Delta_i) \Gamma(1+ 2\Delta_j)}{\Gamma(1+ \Delta_i + \Delta_j +\Delta_k)},
\end{align}
where $\Delta_i$ is the scaling dimension of $ \Phi^i_{R}$ and the matter three-point function $\braket{ \Phi^i_{R}  \Phi^j_{R}  \Phi^k_{R} }$ is proportional to $C_{ijk}$. By comparison, the operator product expansion of matter primaries in bosonic JT gravity leads to a complicated sum of terms that all involve the generalized hypergeometric function $_4F_3$.

Like the boundary algebras in bosonic JT gravity, the algebra $\widetilde{\mathcal{A}}_{R,0}$ has divergent entanglement. However, because of the projection onto zero-energy states,  its (renormalised) entanglement is bounded from above, making it a Type II$_1$ von Neumann factor. The trace on $\widetilde{\mathcal{A}}_{R,0}$ is simply
\begin{align}\label{eq:introtrace}
\Tr(\tilde a) = \braket{\Omega_{\rm matt}| \tilde a|\Omega_{\rm matt}}
\end{align}
where $\ket{\Omega_{\rm matt}}$ is the matter vacuum state. In particular, we have $\Tr(\mathds{1}) = 1$ and $\Tr(\tilde \Phi^i_{R}) = 0$ for all other primaries $\Phi^i_{R}$. The result \eqref{eq:introtrace} is consistent with intuition from AdS/CFT that suggests a zero-energy wormhole without matter excitations should be holographically dual to the zero-temperature limit of a thermofield double state and, consequently, should be maximally entangled on the subspace of one-sided zero-energy microstates. 

Independent of its gravitational significance, the existence of the algebra $\widetilde{\mathcal{A}}_{R,0}$ is an interesting property of supersymmetric quantum field theories in $\AdS_2$. To our knowledge, it is the first natural example of a Type II$_1$ von Neumann algebra that acts on a QFT Hilbert space.\footnote{An example of a Type II$_\infty$ algebra that acts on the Hilbert space of a scalar inflaton field in de Sitter space -- and that also has a natural gravitational interpretation -- was found recently in \cite{chen2024clock}. }
 Moreover it is a Type II$_1$ algebra for which the QFT vacuum is tracial. The product \eqref{eq:simpleproduct} is easy to write down abstractly. However, one would have no reason to suspect that it is associative without knowing about its explicit realisation via the operators \eqref{eq:tildePhimessyform}. Indeed, the associativity of \eqref{eq:simpleproduct} implies an interesting set of crossing relations for the three-point function coefficients $C_{ijk}$ that might be of potential interest in the conformal bootstrap program.\footnote{These crossing relations should follow from the full crossing relations found from the matter boundary four-point function after integration against zero-energy super-JT gravity propagators, although we have not demonstrated this explicitly.} 

It would be particularly interesting to know whether similar algebras to $\widetilde{\mathcal{A}}_{R,0}$ appear in higher-dimensional supersymmetric quantum field theories in anti-de Sitter space, or in their close cousins superconformal field theories (SCFTs). An example that is somewhat similar in spirit is the two-dimensional chiral algebra associated to a four-dimensional SCFT \cite{Beem:2013sza}. However, that algebra only depends on the structure of semi-short multiplets in the SCFT called Schur operators. In contrast, the algebra $\widetilde{\mathcal{A}}_{R,0}$ encodes information about neutral boundary primaries, which one would not typically expect to be highly constrained by supersymmetry.

We can extend the algebra $\widetilde{\mathcal{A}}_{R,0}$ to a larger algebra $\widetilde{\mathcal{A}}_R$ that includes primaries with nonzero R-charge $q$ (although only primaries $\Phi^i_{R}$ with $|q| < 1/2$ lead to nonzero operators  $\tilde\Phi^i_{R} \in \widetilde{ \mathcal{A}}_R$). Since charged operators $\tilde\Phi^i_{R}$ do not commute with $J_R$, they most naturally act on a direct sum over copies of $\mathcal{H}_{\rm matt}$ with one copy for each allowed value of $J_R$. The QFT vacuum for any fixed R-charge $J_R$ is no longer tracial once charged operators are included. However, one can construct a tracial state from a superposition over the vacuum states in each $J_R$ sector,  weighted by $\sqrt{\cos(\pi J_R)}$. 

Since the trace of the projector onto a particular R-charge $J_R$ is proportional to the number of microstates with that charge, the algebra $\widetilde {\mathcal{A}}_R$ (with charged operators included) thereby knows that the number of zero-energy microstates in each R-charge sector is proportional to $\cos(\pi J_R)$, as indeed one finds from a Euclidean partition function. If the lesson of \cite{Witten:2021unn, Chandrasekaran:2022eqq, penington2023algebras, kolchmeyer2023neumann} is that canonically quantised gravity can count black hole microstates (up to an overall factor), one lesson of the present paper is therefore that even ordinary quantum field theory can count black hole microstates (up to the same overall factor) so long as you know the correct algebra of observables to study.

The layout of this paper is as follows. In Section \ref{sec:JT} we review the structure of canonically quantised bosonic JT gravity plus matter. Much of the material in this section (and indeed in Section \ref{sec:JTalgebras}) is not new, and expert readers looking to  get immediately to the results on super-JT gravity can safely skip over them, referring back only as needed. Our principal novel contribution in Section \ref{sec:JT} is a direct proof in canonically quantised JT gravity that the joint eigenvalue problem for the two boundary Hamiltonians has a single delta-function normalisable state, within each $\hSL(2,\R)$ matter representation, for any pair of positive energies $E_L, E_R >0$. (This result was previously shown via Euclidean path integrals in \cite{kolchmeyer2023neumann}.)  Using this result, in Section \ref{sec:JTalgebras}, we provide a completely self-contained proof of the structure of the JT gravity boundary algebras within the Lorentzian canonically quantised theory, without any need for the Euclidean gravitational path integral arguments that were used in \cite{penington2023algebras, kolchmeyer2023neumann}.

Then, in Section \ref{sec:superJT}, we turn to super-JT gravity. We first construct the canonically quantised theory of super-JT gravity coupled to matter and impose the gauge constraints. We explain the isomorphism between states with right-boundary energy $E_R = 0$ and the matter Hilbert space and show that there exists a single state with both $E_L = E_R = 0$ within each $\hSU(1,1|1)$ representation of the matter theory. Finally, in Section \ref{sec:algebras} we study the boundary algebras in super-JT gravity, focusing in particular on the Type II$_1$ algebras found by projecting onto the zero-energy subspaces. 

Various technical calculations, along with a quantisation of JT gravity plus matter in a ``highest-weight gauge'' that connects to the chord Hilbert space for the double-scaled SYK model found in \cite{Lin:2022rbf, Lin:2023trc}, are given in appendices.

\section{JT gravity with matter} \label{sec:JT}

We start by reviewing the analysis of the algebraic structure of canonically quantised JT gravity plus matter given in \cite{penington2023algebras, kolchmeyer2023neumann} in a language that will naturally generalise to the supersymmetric case. JT gravity, both with and without matter, has been studied from numerous perspectives over the last decade; for a necessarily incomplete list of some of the most important results see \cite{almheiri2015models, MSY-1, Kitaev:2018wpr, Harlow:2018tqv, Yang:2018gdb, Saad:2019lba, LMZ, Jafferis:2019wkd, Harlow:2021dfp, Jafferis:2022wez}. 

\subsection{The boundary particles}\label{sec:group}
 
The action of JT gravity with negative cosmological constant on a spacetime $M$ can be written, in the notation of \cite{Harlow:2018tqv, Jafferis:2019wkd}, as
\be\label{sjt}I_{JT}=\int_M \d^2x\sqrt{-g} \upphi (R+2) +2  \int_{\partial M} \d t \sqrt{|\gamma|}\upphi(K-1)+\cdots,\ee
where $g$ is the bulk metric with curvature scalar $R$, $\gamma$ is the induced metric on the boundary,  $K$ is the extrinsic curvature of the boundary, and 
we have omitted a topological term that controls the classical entropy $S_0$. The scalar field $\upphi$ is known as the dilaton. The boundary conditions are
\be\label{bc}\gamma_{tt}=-\frac{1}{\epsilon^2},~~~\upphi|_{\partial M}=\frac{\upphi_b}{\epsilon}, \ee
with $\epsilon \to 0$. For convenience, we will henceforth work in units where the renormalised boundary dilaton value is $\upphi_b = 1/4$. 

Upon integrating first over $\upphi$ to impose the equation of motion $R+2=0$, the action reduces to the boundary term
\be\label{ibdry}I_{\partial M}=\frac{1}{2\varepsilon}  \int_{\partial M} \d t \sqrt{|\gamma|}(K-1). \ee
The condition $R+2=0$ implies  that $M$ is locally isomorphic to a portion of $\AdS_2$.
The metric for $\AdS_2$ can be written as
\be\label{adstwo} \d s^2=\d\sigma^2-\cosh^2\sigma \,\d T^2,~~~~-\infty<\sigma,T<\infty.\ee
There is  a ``right'' conformal boundary at $\sigma\to +\infty$ and a ``left'' conformal boundary at $\sigma\to -\infty$.    

In the limit $\epsilon \to 0$, the spacetime $M$ becomes ``almost all'' of $\AdS_2$ in the following sense. After embedding $M$ in $\AdS_2$, the right boundary of $M$ can be described by functions $\sigma(t)$, $T(t)$ of the physical right-boundary time $t$; similarly the left boundary can be described by functions $\sigma'(t')$, $T'(t')$ of the left-boundary time $t'$. As $\epsilon \to 0$, we take $\sigma, -\sigma' \to +\infty$ so that $T(t), T'(t')$ remain finite. For small $\epsilon$, the condition $\gamma_{tt}=-1/\epsilon^2$ reduces to
\be\label{relno} e^{\sigma}=\frac{2}{\epsilon}\frac{1}{\dot T}, ~~e^{-\sigma'}=\frac{2}{\epsilon}\frac{1}{\dot T'},\ee where dots 
represent derivatives with respect to $t$ and $t'$ for the right and left boundaries respectively.  It follows that the renormalised coordinates
\be\label{usefuldef} \chi = -\sigma + \log\left(\frac{1}{2\epsilon}\right),~~~\chi'=\sigma' + \log\left(\frac{ 1}{2\epsilon}\right)\ee
must remain finite as $\epsilon \to 0$. The renormalised geodesic length between the left and right boundaries is
\be\label{eq:length}\ell =-\chi-\chi'+\log\left(\frac{1 + \cos (T - T')}{2}\right).\ee
With some work (see \cite{penington2023algebras} and references therein for details), one finds that time evolutions of the left and right boundaries are generated respectively by the boundary Hamiltonians
\begin{align}\label{eq:hamL} \mathbf{H}_L &= \left(  p_{\chi'}^2+ 2 p_{T'} e^{\chi'}+ e^{2\chi'}\right)\\
\mathbf{H}_R&=\left(p_{\chi}^2+ 2 p_{T} e^{\chi} + e^{2\chi}\right). \label{eq:hamR} \end{align}

A convenient way to understand \eqref{eq:hamL} and \eqref{eq:hamR}, which will later prove helpful when we come to construct the supersymmetric version of the theory, is as follows. $\AdS_2$ is a homogeneous space for the Lie group $\hSL(2,\R)$ that is the universal cover of
$\SL(2,\R)$. The action of $\hSL(2,\R)$ on $\AdS_2$ is generated classically by the vector fields
\begin{align}
\mathcal{J}_1 &= i \partial_T \\
\mathcal{J}_2 &= i \cos T \tanh \sigma \partial_T + i \sin T \partial_\sigma  \\ 
\mathcal{J}_3 &= -i \sin T \tanh \sigma \partial_T + i \cos T \partial_\sigma
\end{align}
which satisfy $[\mathcal{J}_a, \mathcal{J}_b] = i {\epsilon_{ab}}^c \mathcal{J}_c$.\footnote{The signs of $\mathcal{J}_1$ and $\mathcal{J}_2$ are switched here relative to the conventions used in \cite{penington2023algebras}. This will mean that the action of $\mathcal{J}_1$ on the matter QFT is equal to the global Hamiltonian (rather than its negative).} Here $\epsilon_{abc}$ is totally antisymmetric with $\epsilon_{123} = 1$ and indices are raised and lowered with $\eta_{ab} = \mathrm{diag}(-1,1,1)$.

The quantum mechanics of a particle in a homogeneous space $M$ for a group action $G$ can be understood as a nonlinear sigma model with target space $G$, gauged by the right action of the stabiliser group $H_p$ of any fixed point $p$ in $M$.
If the point $p$ is at the right conformal boundary  of $\AdS_2$, the stabiliser group of $p$ is generated by a single null generator of $\hSL(2,\R)$. In particular, if we choose $p$ to be at $T = 0$, the relevant generator is $\mathcal{J}_- = \mathcal{J}_1 - \mathcal{J}_2$.

An element $g$ of $\hSL(2,\R)$ can be uniquely parameterised by its Iwasawa decomposition 
\begin{align}\label{eq:param1}
g = e^{i T \mathcal{J}_1} e^{-i \chi \mathcal{J}_3} e^{i \eta \mathcal{J}_-}.
\end{align} 
The left action $\mathcal{L}(h)$ of $\hSL(2,\R)$ on $\hSL(2,\R)$ wavefunctions $\mathbf{\Psi}(g)$  is given by $\left[\mathcal{L}[h]\mathbf{\Psi}\right](g) = \mathbf{\Psi}(h^{-1}g)$. In terms of the parameterisation $(T, \chi, \eta)$,  we have
\begin{align} \label{eq:leftaction}
\mathcal{L}[\mathcal{J}_1] &=  - p_{T} \\
\mathcal{L}[\mathcal{J}_2] &= \cos T \left[- p_{T} + e^{\chi} p_{\eta}\right] +  \sin T p_{\chi}  \\ 
\mathcal{L}[\mathcal{J}_3] &= -\sin T\left[-p_{T} + e^{\chi} p_{\eta}\right] + \cos T p_{\chi}.
\end{align}
Similarly, the right action $\mathcal{R}(h)$, defined by $\left[\mathcal{R}[h]\mathbf{\Psi}\right](g) = \mathbf{\Psi}(g h)$, can be written as
\begin{align}
\mathcal{R}[\mathcal{J}_1] &=  e^{\chi} p_T + \frac{1}{2} p_\eta + \eta p_\chi + \frac{1}{2}\eta^2 p_\eta\\
\mathcal{R}[\mathcal{J}_2] &= e^{\chi} p_T - \frac{1}{2} p_\eta  + \eta p_\chi + \frac{1}{2}\eta^2 p_\eta  \\ 
\mathcal{R}[\mathcal{J}_3] &= -p_\chi - \eta p_\eta .
\end{align}
It is easy to check that these are Hermitian with respect to the natural inner product
\begin{align}
\left(\mathbf{\Phi}, \mathbf{\Psi}\right) = \int dg\,\, \mathbf{\Phi}^*(g) \mathbf{\Psi}(g)
\end{align}
induced by the bi-invariant Haar measure
\begin{align}\label{eq:Haarbosonic}
dg = f \,dT d\chi d\eta = e^{-\chi} d T d \chi d \eta.
\end{align}
To see that the right-hand side of \eqref{eq:Haarbosonic} is indeed the Haar measure on $\hSL(2,\R)$ note that invariance of $dg$ under the right action $\mathcal{R}[\mathcal{J}_-] = p_\eta$ requires $f$ to be independent of $\eta$, while the invariance under the left action $\mathcal{L}[\mathcal{J}_1]$ requires $f$ to be independent of $T$. Finally, the right action $\mathcal{R}[e^{i \chi_0 \mathcal{J}_3}]$ maps $\chi \to \chi - \chi_0$ and $\kappa \to e^{-\chi_0} \kappa$. Invariance of the measure therefore requires $f \sim e^{-\chi_L}$.

To describe a particle at the right-boundary of $\AdS_2$, we need to gauge the right action of the stabiliser $\mathcal{J}_-$. However, to reproduce the dynamics of JT gravity, the correct gauge constraint turns out not to be $p_{\eta} = 0$ but instead $p_{\eta} + 1 =0$. This is sometimes described as the boundary particle having ``nonzero charge'', because an analogous procedure applied to the group $SO(3)$ leads to the quantum mechanics of a charged particle on a two-sphere in the presence of a constant background magnetic flux.

Because the gauge group $\R$ is noncompact, the constraint $p_{\eta} + 1 = 0$ needs to be imposed using the method of coinvariants or ``group-averaging method'' \cite{Giulini:1998kf, Marolf:2000iq, Chandrasekaran:2022cip}. Given an action $\mathcal{K}$ of a gauge group $H$, one first imposes an equivalence relation on the space of wavefunctions $\Psi$ so that $\Psi \sim \mathcal{K}(h) \Psi$ for all $h \in H$. One then defines a ``coinvariant inner product'' by
\begin{align}
\braket{\Phi, \Psi} = \int_H dh \,(\Phi, \mathcal{K}(h) \Psi),
\end{align}
where $(\Phi, \Psi)$ is an inner product on the space of unconstrained wavefunctions and $dh$ is the Haar measure on the group $H$. So long as this inner product is gauge-invariant (as the inner product induced by the Haar measure is in our case), the coinvariant inner product $\braket{\Phi, \Psi}$ will respect the equivalence relation introduced above.

In our case, the action $\mathcal{K}$ of the gauge group $H \cong \R$ consists of translations of $\eta$ combined with multiplication by a phase. Any wavefunction $\mathbf{\Psi}_R(g) = \mathbf{\Psi}_R(T, \chi,\eta)$ is therefore gauge equivalent to a wavefunction of the form
\begin{align} \label{eq:waveform}
\mathbf{\Psi}_R(T, \chi,\eta) =  e^{\chi/2} \delta(\eta) \,\Psi_R(T,\chi),
\end{align}
where we have included an explicit factor of $e^{\chi/2}$ for later convenience. Wavefunctions of the form \eqref{eq:waveform} are not normalisable with respect to the unconstrained inner product
\begin{align}
\left(\mathbf{\Phi}_R,\mathbf{\Psi}_R\right) = \int_{\hSL(2,\R)} dg \,\mathbf{\Phi}_R^*\mathbf{\Psi}_R = \int dT d\chi d\eta e^{-\chi} \mathbf{\Phi}_R^* \mathbf{\Psi}_R.
\end{align}
However they are normalisable with respect to the coinvariant inner product
\begin{align}
\braket{\mathbf{\Phi}_R,\mathbf{\Psi}_R} = \int dT d\chi d \eta d \tilde \eta\, e^{-\chi} e^{i \tilde\eta} \mathbf{\Phi}_R(T,\chi,\eta)^*\mathbf{\Psi}_R(T,\chi,\eta+\tilde\eta) = \int dT d\chi\, \Phi_R^*(T,\chi) \Psi_R(T,\chi).
\end{align}
The two factors of $e^{\chi/2}$ from \eqref{eq:waveform} cancel the factor of $e^{-\chi}$ in the Haar measure \eqref{eq:Haarbosonic} to give the standard inner product on $L^2(\R^2)$.

The left action of $\hSL(2,\R)$ commutes with the gauge constraint and so defines an action of $\hSL(2,\R)$ on the gauged Hilbert space. Using the gauge equivalence $(p_\eta + 1) \mathbf{\Psi}_R \sim 0$, we obtain
\begin{align}
\mathcal{L}[\mathcal{J}_i] \mathbf{\Psi}_R \sim e^{\chi/2} \delta(\eta) (J_i^R \Psi_R)
\end{align}
where
\begin{align}
J_1^R &=  - p_{T} \\
J_2^R &= - \cos T p_{T}  + \sin T p_{\chi} - e^{\chi} \cos T - \frac{i}{2} \sin T. \\ 
J_3^R&= \sin T p_{T}  + \cos T p_{\chi} + e^{\chi} \sin T - \frac{i}{2} \cos T.
\end{align} 
The superscript here indicates that this is the $\hSL(2,\R)$ action for the right boundary particle, in contrast to the action for the left boundary particle which we will discuss momentarily. Importantly, it has nothing to do with the right action of $\hSL(2,\R)$ on itself, which no longer exists after imposing the gauge constraint.

The Hamiltonian of the gauged sigma model is the $\hSL(2,\R)$ Casimir
\begin{align} \label{eq:HRbos}
\mathbf{H}_R =\eta^{ab} J_a^R J_b^R - \frac{1}{4}  =p_{\chi}^2+ 2 p_{T} e^{\chi}+ e^{2\chi},
\end{align}
which indeed matches \eqref{eq:hamR}.

There is an alternative choice of parameterisation of (part of) $\hSL(2,\R)$ that will prove convenient when we come to study the supersymmetric JT gravity. We write
\begin{align}\label{eq:param2}
g = e^{i \tau \mathcal{J}_+} e^{i \rho \mathcal{J}_3} e^{i \kappa \mathcal{J}_-},
\end{align} 
where $\mathcal{J}_+  = \mathcal{J}_1 + \mathcal{J}_2$. Note that $\rho$ appears in \eqref{eq:param2} with the opposite sign that $\chi$ appeared in \eqref{eq:param1}; the signs here were chosen for consistency respectively with \cite{penington2023algebras} and with \cite{Lin:2022rzw, Lin:2022zxd}. Unlike the Iwasawa decomposition \eqref{eq:param1}, \eqref{eq:param2} does not cover all of $\hSL(2,\R)$. However, when defining the final gauge-invariant Hilbert space of JT gravity, it will turn out to be sufficient to consider wavefunctions with support localised near $T = \eta = 0$ or equivalently near $\tau = \kappa = 0$. And \eqref{eq:param2} provides a set of smooth coordinates for an open neighbourhood of that one-dimensional submanifold.

In the parameterisation \eqref{eq:param2}, the left action  of $\hSL(2,\R)$ is given by\footnote{Because the parameterisation \eqref{eq:param2} does not cover the whole of $\hSL(2,\R)$, the generators of the action $\mathcal{L}$ are symmetric but not essentially self-adjoint on the subspace of $\hSL(2,\R)$ covered by the coordinates $\tau, \rho, \kappa$. Again, this subtlety will end up being irrelevant after we have made all our gauge choices.}
\begin{align}
\mathcal{L}(\mathcal{J}_+)  &= - p_{\tau}\\
\mathcal{L}(\mathcal{J}_-) &= -\tau^2 p_{\tau} + 2\tau p_{\rho} - e^{-\rho} p_{\kappa}\\
\mathcal{L}(\mathcal{J}_3) &= + \tau p_{\tau} - p_{\rho},
\end{align}
while the Haar measure is 
\begin{align}
dg = e^{\rho} d\tau d\rho d\kappa. 
\end{align}
After imposing the gauge constraint
\begin{align}
\mathcal{R}[\mathcal{J}_-] + 1 = p_\kappa + 1 = 0,
\end{align}
we can restrict to wavefunctions of the form $\mathbf{\Psi}_R(\tau,\rho,\kappa) = e^{-\rho/2} \Psi_R(\tau,\rho)\delta(\kappa)$. Within the region where the parameterisation \eqref{eq:param2} is valid, the coinvariant inner product reduces to the usual inner product on the space of wavefunctions $\psi(\tau,\rho)$. The left action of $\hSL(2,\R)$ acts on $\Psi_R$ as
\begin{align}
J_+^R &= - p_{\tau}\\
J_-^R &= - \tau^2 p_{\tau} + 2 \tau p_{\rho} + i \tau + e^{-\rho}\\
J_3^R &= +\tau p_{\tau} - p_{\rho} - \frac{i}{2}
\end{align}
with the Hamiltonian
\begin{align} \label{eq:altHRbos}
\mathbf{H}_R = p_{\rho_R}^2 + e^{-\rho_R} p_{\tau_R}.
\end{align}
The coordinates $\tau,\rho$ can be physically interpreted as the location of the right boundary particle in Poincar\'{e}  coordinates for $\AdS_2$ (with $\rho$ suitably renormalised).\footnote{The Poincar\'{e} coordinates $(\tau,\rho)$ only cover part of the right boundary of $\AdS_2$, just like the parameterisation \eqref{eq:param2} only covered part of $\hSL(2,\R)$. But, again, this part will be all that we care about given later gauge choices.}

To understand the dynamics of the left boundary particle, the natural parameterisation of $\hSL(2,\R)$ is $(T', \chi', \eta')$ with
\begin{align}
g = e^{i T' \mathcal{J}_1} e^{i \chi' \mathcal{J}_3} e^{i \eta' \mathcal{J}_+}.
\end{align} 
In these coordinates, the left action $\mathcal{L}$ of $\hSL(2,\R)$ is
\begin{align} \label{eq:leftactionL}
\mathcal{L}[\mathcal{J}_1] &=  - p_{T'} \\
\mathcal{L}[\mathcal{J}_2] &= \cos T' \left[p_{T'} - e^{\chi'} p_{\eta'}\right] - \sin T' p_{\chi'}  \\ 
\mathcal{L}[\mathcal{J}_3] &= -\sin T'\left[p_{T'} - e^{\chi'} p_{\eta'}\right] - \cos T' p_{\chi'}.
\end{align}
The Haar measure is again $dg = e^{-\chi'} d T' d \chi' d \eta'$.

The stabiliser $H_p$ of a point $p$ at $T = 0$ on the left boundary is $\mathcal{J}_+ = \mathcal{J}_1 + \mathcal{J}_2$. So we should impose the constraint 
\begin{align}
\mathcal{R}[\mathcal{J}_+] + 1 = p_{\eta'} + 1 = 0. 
\end{align}
After gauging and restricting to wavefunctions of the form 
\begin{align}
\mathbf{\Psi}_L(T',\chi'\eta') = e^{\chi'/2} \Psi_L(T',\chi') \delta(\eta'),
\end{align}
the left action $\mathcal{L}$ acts on the wavefunction $\psi'$ as
\begin{align}
J_1^L&=  - p_{T'} \\
J_2^L &= \cos T' p_{T'}  - \sin T' p_{\chi'}  + e^{\chi'} \cos T' + \frac{i}{2} \sin T' \\ 
J_3^L &= -\sin T' p_{T'}  - \cos T' p_{\chi'} - e^{\chi'} \sin T' + \frac{i}{2} \cos T'.
\end{align}
The boundary Hamiltonian is again the Casimir
\begin{align} \label{eq:HLbos}
\mathbf{H}_L = \eta^{ab} J_a^L J_b^L - \frac{1}{4} = p_{\chi'}^2+ 2 p_{T'} e^{\chi'}+ e^{2\chi'},
\end{align}
which indeed matches \eqref{eq:hamL}.

Alternatively, we can use the parameterisation
\begin{align}\label{eq:param2l}
g = e^{i \tau' \mathcal{J}_-} e^{-i \rho' \mathcal{J}_3} e^{i \kappa' \mathcal{J}_+}.
\end{align} 
The left action is then
\begin{align}
\mathcal{L}(\mathcal{J}_+)  &= -\tau'^2 p_{\tau'} + 2\tau' p_{\rho'} - e^{-\rho'} p_{\kappa'}\\
\mathcal{L}(\mathcal{J}_-) &= - p_{\tau'}\\
\mathcal{L}(\mathcal{J}_3) &= - \tau' p_{\tau'} + p_{\rho'}
\end{align}
The Haar measure is 
\begin{align}
dg = e^{\rho'} d\tau' d\chi' d \kappa'.
\end{align}
After imposing the constraint 
\begin{align}
\mathcal{R}[\mathcal{J}_+] + 1 = p_{\kappa'} + 1 = 0
\end{align}
and restricting to wavefunctions of the form $\mathbf{\Psi}_L(\tau',\rho',\kappa') = e^{-\rho'/2} \Psi_L(\tau',\rho')\delta(\kappa')$, the left action becomes
\begin{align}
J_+^L &= - \tau'^2 p_{\tau'} + 2 \tau' p_{\rho'} + i \tau' + e^{-\rho'}\\
J_-^L &= - p_{\tau'}\\
J_3^L &= - \tau' p_{\tau'} + p_{\rho'} + \frac{i}{2}
\end{align}
and the Casimir is
\begin{align} \label{eq:altHLbos}
 \mathbf{H}_L = p_{\rho'}^2 +  e^{-\rho'} p_{\tau'}.
\end{align}

\subsection{The physical Hilbert space and Hamiltonians}
So far we have described only the gravitational sector of the theory. In addition to the two boundary particles, we also include a matter sector that does not couple directly to the dilaton. In the limit $\epsilon \to 0$,the spacetime $M$ covers almost all of $\AdS_2$ and finite energy matter excitations are unaffected by the presence of the boundary particles. As a result, the matter sector dynamics are described by quantum field theory on a fixed $\AdS_2$ background.

Let $\mathcal{H}_{\rm matt}$ be the Hilbert space of the matter theory quantised on $\AdS_2$. The Hilbert space $\mathcal{H}_{\rm matt}$ forms a representation of $\hSL(2,\R)$ with generators $j_i$. In particular, the operator $j_1$ is the global Hamiltonian of the matter theory and is therefore positive semi-definite. The nontrivial irreducible representations of $\hSL(2,\R)$ with $j_1 \geq 0$ are known as discrete series representations and are labelled by the smallest eigenvalue $\lambda > 0$ of $j_1$. The Hilbert space $\mathcal{H}_{\rm matt}$ decomposes as a direct sum of these representations.

States in JT gravity plus matter are described by wavefunctions 
\begin{align}
\mathbf{\Psi}_{JT}(T, \chi, T', \chi')\in \mathcal{H}_{\rm matt}
\end{align} 
that are valued in the $\AdS_2$ matter Hilbert space. This is not quite the full story, however, because there are two additional constraints on those wavefunctions that we have not yet taken into account. The first is that the two boundaries should be spacelike separated, and hence $\mathbf{\Psi}_{JT}$ should have support only where $|T - T'| < \pi$. The second is that the diagonal action  $\mathcal{D}$ of $\hSL(2,\R)$ on the boundary particles and $\mathcal{H}_{\rm matt}$, defined by 
\begin{align}
\mathcal{D}(\mathcal{J}_i) = J_i^L + j_i + J_i^R, 
\end{align}
describes a change in coordinates for the background $\AdS_2$ and hence forms a gauge symmetry of the quantum gravity theory.

We again impose this gauge constraint using the method of coinvariants. Whenever the two boundary particles are spacelike separated, we can use a gauge transformation to fix $T = T' = 0$ and $\chi = \chi'$. As a result, any wavefunction with support purely at spacelike separation is gauge equivalent to a wavefunction of the form
\begin{align}\label{eq:fullwaveformboson}
\mathbf{\Psi}_{JT}(T, T',\chi,\chi') = \delta(T) \delta(T') \delta(\chi - \chi') \Psi_{JT}(\chi)
\end{align}
where again $\mathbf{\Psi}_{JT}(\chi) \in \mathcal{H}_{\rm matt}$. The coinvariant inner product is
\begin{align}
\braket{\mathbf{\Phi}_{JT}|\mathbf{\Psi}_{JT}} = \int \,dg \int \,dT d\chi dT' d\chi' ( \mathbf{\Phi}_{JT}, \mathcal{D}(g) \mathbf{\Psi}_{JT})_{\rm matt} = \int \,d\chi \braket{\Phi_{JT}(\chi), \Psi_{JT}(\chi)}_{\rm matt},
\end{align}
where $\braket{\cdot,\cdot}_{\rm matt}$ is the inner product on $\mathcal{H}_{\rm matt}$. The physical Hilbert space of JT gravity plus matter is therefore 
\begin{align}\label{eq:bosHilbphys}
\mathcal{H}_{\rm JT} \cong L^2(\R) \otimes \mathcal{H}_{\rm matt},
\end{align}
where $L^2(\R)$ is the space of $L^2$ functions of $\chi$.

To identify the physical left and right Hamiltonians, we note that, on wavefunctions of the form \eqref{eq:fullwaveformboson}, we have
\begin{align}
\mathcal{D}(\mathcal{J}_1) \mathbf{\Psi}_{JT} &= \left(-p_T - p_{T'} + j_1\right) \mathbf{\Psi}_{JT} \sim 0\\
\mathcal{D}(\mathcal{J}_2) \mathbf{\Psi}_{JT} &= \left(-p_T + p_{T'} + j_2\right) \mathbf{\Psi}_{JT} \sim 0\\
\mathcal{D}(\mathcal{J}_3) \mathbf{\Psi}_{JT} &= \left(p_\chi - p_{\chi'} + j_3\right) \mathbf{\Psi}_{JT} \sim 0
\end{align}
Furthermore 
\begin{align}
(p_\chi + p_{\chi'}) \mathbf{\Psi}_{JT} = \delta(T) \delta(T') \delta(\chi - \chi') (p_\chi \Psi_{JT}(\chi)).
\end{align}
It follows that 
\begin{align}
\left(2 p_T -  j_+\right) \mathbf{\Psi}_{JT} \sim \left(2 p_{T'} - j_-\right) \mathbf{\Psi}_{JT} \sim 0,
\end{align} 
while 
\begin{align}
\left(2 p_\chi + j_3\right) \mathbf{\Psi}_{JT} \sim \left(2 p_{\chi'} - j_3\right) \mathbf{\Psi}_{JT} \sim \delta(T) \delta(T') \delta(\chi - \chi') (p_\chi \Psi_{JT}(\chi)).
\end{align}
With these substitutions, one finds that the Hamiltonians \eqref{eq:HRbos} and \eqref{eq:HLbos} act on the wavefunction $\Psi_{JT}$ as
\begin{align}\label{eq:Hamphyschi}
 H_R &= \frac{1}{4}(p_\chi - j_3)^2 + j_+ e^\chi + e^{2 \chi}\\
H_L &= \frac{1}{4}(p_\chi + j_3)^2 + j_- e^\chi + e^{2 \chi}.\nonumber
\end{align}
It is easy to check that $H_R$ and $H_L$ commute and are both positive semi-definite.

Suppose we instead used the alternative parameterisations \eqref{eq:param2} and \eqref{eq:param2l} of $\hSL(2,\R)$ for the boundary particles to write the full wavefunction (prior to imposing the gauge constraint) as 
\begin{align}
\mathbf{\Psi}_{JT}(\tau,\rho,\tau',\rho') \in \mathcal{H}_{\rm matt}.
\end{align}
In this parameterisation, wavefunctions of the form \eqref{eq:fullwaveformboson} become wavefunctions of the form
\begin{align}\label{eq:wavefnparam2}
\mathbf{\Psi}_{JT}(\tau,\rho,\tau',\rho') = \delta(\tau) \delta(\tau') \delta(\rho-\rho') \Psi_{JT}(\rho).
\end{align}
Note that \eqref{eq:wavefnparam2} only has support within the region where the parameterisations \eqref{eq:param2} and \eqref{eq:param2l} are valid. This justifies our previous claim that the failure of the coordinates  \eqref{eq:param2} to cover all of $\hSL(2,\R)$ or $\AdS_2$ does not result in any issues when constructing the final physical Hilbert space. The coinvariant inner product is
\begin{align}
\braket{\mathbf{\Phi}_{JT},\mathbf{\Psi}_{JT}} = \int \,d\rho\,  (\Phi_{JT}(\rho),\Psi_{JT}(\rho)).
\end{align}
We have
\begin{align}
\mathcal{D}(\mathcal{J}_+) \mathbf{\Psi}_{JT} &= \left(-p_\tau + e^{-\rho} + j_+\right) \mathbf{\Psi}_{JT} \sim 0\\
\mathcal{D}(\mathcal{J}_-) \mathbf{\Psi}_{JT}&= \left(e^{-\rho} - p_{\tau'} + j_-\right) \mathbf{\Psi}_{JT}\sim 0\\
\mathcal{D}(\mathcal{J}_3) \mathbf{\Psi}_{JT} &= \left(-p_\rho + p_{\rho'} + j_3\right) \mathbf{\Psi}_{JT} \sim 0.
\end{align}
Using the fact that 
\begin{align}
(p_\rho + p_{\rho'})\mathbf{\Psi}_{JT}= \delta(\tau) \delta(\tau') \delta(\rho - \rho') (p_\rho \Psi_{JT}(\rho)),
\end{align} we therefore find that the Hamiltonians \eqref{eq:altHRbos} and \eqref{eq:altHLbos} act on $\Psi_{JT}(\rho)$ as
\begin{align}\label{eq:Hamphysrho}
H_R &= \frac{1}{4}(p_\rho + j_3)^2 + j_+ e^{-\rho} + e^{-2 \rho}\\
 H_L &= \frac{1}{4}(p_\rho - j_3)^2 +  j_- e^{-\rho} +  e^{-2 \rho}.\nonumber
\end{align}
We immediately see that \eqref{eq:Hamphysrho} is the same as \eqref{eq:Hamphyschi} with $\chi \leftrightarrow -\rho$. This is exactly what we should expect: when $T= T' =\tau = \tau' = 0$, the parameterisation \eqref{eq:param2} is the same as \eqref{eq:param1} with $\eta, \eta' \to \kappa, \kappa'$ and $\chi, \chi' \to -\rho,-\rho'$. As we saw in \eqref{eq:length}, with these conditions $\chi$ and $\rho$ are also linearly related to the renormalised length 
\begin{align}
\ell = 2 \rho = - 2 \chi
\end{align}
between the two boundary particles.

\subsection{The spectra of the boundary Hamiltonians}\label{sec:spectra}

We can simplify the form of $H_R$, at the cost of breaking the symmetry between $H_R$ and $H_L$,  if we conjugate by $U_1 = \exp(-i \chi j_3)$. This leads to
\begin{align}
 U_1 H_R U_1^\dagger &= \frac{1}{4} p_\chi^2 + (1 +j_+) e^{2\chi} \\
U_1 H_L U_1^\dagger &= \frac{1}{4}(p_\chi + 2 j_3)^2 +  j_-  +  e^{2 \chi}.\nonumber
\end{align}
The same formulas (up to a rescaling of $\chi$ by a factor of two) can also be obtained by replacing \eqref{eq:fullwaveformboson} by a wavefunction of the form 
\begin{align}
\mathbf{\Psi}_{JT}(T, T',\chi,\chi') = \delta(T) \delta(T') \delta( \chi') \Psi_{JT}(\chi). 
\end{align}
We can then further simplify $U_1 H_R U_1^\dagger$ by conjugating by $U_2 = (1+j_+)^{-i p_\chi/2}$ to obtain
\begin{align} \label{eq:conjHR}
 U H_R U^\dagger &= \frac{1}{4} p_\chi^2 + e^{2\chi}
\end{align}
where $U = U_2 U_1$. Notably,  \eqref{eq:conjHR} is the Hamiltonian for both the left and right boundaries in pure JT gravity (without matter), as can be easily verified by setting all matter charges to zero in \eqref{eq:Hamphyschi} \cite{Harlow:2018tqv}. 

The eigenvalue equation 
\begin{align}\label{eq:eigenvalueeqnHRbos}
 \left(\frac{1}{4} p_\chi^2 + e^{2\chi}\right) \psi(\chi) = E_R \psi(\chi)
\end{align}
 is a second-order ordinary differential equation and so has two smooth solutions. As $\chi \to +\infty$, 
the two solutions scale as
\begin{align}\label{eq:chi+infscale}
\psi(\chi) \sim \exp(\pm 2e^{\chi})
\end{align}
and so one decays superexponentially while the other blows up at the same rate. As $\chi \to - \infty$, the two solutions look like
\begin{align}\label{eq:chi-infscale}
\psi(\chi) \sim \exp(\pm 2i \sqrt{E_R} \chi)
\end{align}
 and so both solutions are oscillatory for any $E_R >0$.  It follows that \eqref{eq:eigenvalueeqnHRbos} has a single delta-function normalisable solution for any $E_R > 0$ that looks like the decaying solution in \eqref{eq:chi+infscale} as $\chi \to +\infty$ and looks like a linear combination of the two solutions \eqref{eq:chi-infscale} as $\chi \to -\infty$. Indeed, in this case, the precise form of the delta-function normalised (generalised) eigenfunctions are well known. They can be written in terms of the modified Bessel function $K_\nu(z)$ as 
\begin{align}\label{eq:kets}
\ket{s} = \sqrt{\rho(s)} K_{2is}(2 e^{\chi})
\end{align}
 where $E_R = s^2$ and $\rho(s) =(8/\pi^2) s \sinh(2 \pi s)$. Nonetheless, the simple scaling arguments above were sufficient to prove the existence and uniqueness of the state \eqref{eq:kets}, even if we did not know its explicit form. We shall make heavy use of similar arguments throughout this paper. 

For $E_R < 0$, we know that no normalisable or delta-function normalisable solution can exist  because $H_R$ is positive semi-definite on $L^2(\R)$. We therefore conclude that, as an operator on $L^2(\R)$, \eqref{eq:conjHR}  has a nondegenerate, purely continuous spectrum with support on the entire half-line $\R^+ = (0,\infty)$. 

In pure JT gravity, the boundary Hamiltonians have the nondegenerate spectrum that we have just described. However, the operator $U H_R U^\dagger $ in JT gravity with matter does not act on $L^2(\R)$ but instead on the larger Hilbert space $\mathcal{H}_{JT} \cong L^2(\R) \otimes \mathcal{H}_{\rm matt}$. As a result, the spectrum of $U H_R U^\dagger$ (and hence also of $H_R$) is no longer nondegenerate. Instead, the space of delta-function normalisable eigenstates for any fixed $E_R > 0$ is isomorphic to the matter Hilbert space $\mathcal{H}_{\rm matt}$. An intuitive interpretation of this result if that we can produce any desired matter state, while keeping the right-boundary energy fixed, by acting with bulk operators that are gravitationally dressed to the left boundary (and so commute with $H_R$).

Unlike in pure JT gravity, the left and right Hamiltonians in JT gravity with matter are not equal, as we saw in \eqref{eq:Hamphyschi}.  It is therefore interesting to understand the \emph{joint eigenvalue problem}
\begin{align}
H_L \mathbf{\Psi}_{JT}=E_L \mathbf{\Psi}_{JT}, ~~~~~~ H_R \mathbf{\Psi}_{JT} =E_R \mathbf{\Psi}_{JT},
\end{align}
for  $E_L, E_R  > 0$. Since $[H_L,H_R] = 0$ such joint eigenstates form a basis for the JT gravity Hilbert space. But it is not obvious whether, and how many, states exist for any given pair $E_L, E_R$. 

A first observation is that $H_L, H_R$ act on $\mathcal{H}_{\rm matt}$ only via $\hSL(2,\R)$ generators. Hence they preserve irreducible representations of $\hSL(2,\R)$ within $\mathcal{H}_{\rm matt}$. If we assume that the matter QFT does not have nontrivial superselection sectors, the Hilbert space $\mathcal{H}_{\rm matt}$ will always contain exactly one $\hSL(2,\R)$ invariant state, the QFT vacuum $\ket{\Omega_{\rm matt}}$. If the operator
\begin{align}
\Pi_{\Omega} = \ket{\Omega_{\rm matt}}\bra{\Omega_{\rm matt}}
\end{align}
projects the matter fields into the vacuum state, we find that
\begin{align}
H_L \Pi_\Omega = \Pi_\Omega H_L = H_R \Pi_\Omega = \Pi_\Omega H_R = \left(\frac{1}{4} p_\chi^2 + e^{2\chi}\right) \Pi_\Omega
\end{align}
So the vacuum sector is isomorphic to the pure JT gravity Hilbert space and has a single, delta-function normalisable state for any $E_L = E_R > 0$. 

The remaining states in $\mathcal{H}_{\rm matt}$ decompose into discrete series representations of $\hSL(2,\R)$. To analyse these representations, it is helpful to realise them in $L^2(\R)$ by defining
\begin{align} \label{eq:discreterealisationboson}
j_3 = p_\sigma ~~~~~ j_+ =  e^{-\sigma} ~~~~~ j_-  &= (p_\sigma + i \lambda)e^\sigma (p_\sigma - i \lambda) \\&= p_\sigma e^\sigma p_\sigma + \lambda (\lambda -1) e^\sigma.\nonumber
\end{align}
for real $\lambda > 0$. Here, $\sigma$ is the position operator on $L^2(\R)$ and $p_\sigma = - i \partial_\sigma$ as usual. The Casimir
\begin{align} 
j_1^2 - j_2^2 - j_3^2 = - j_3^2 + \frac{1}{2} \{ j_+,j_-\} = \lambda (\lambda - 1)
\end{align}
as expected for a discrete series representation. Furthermore, the operator $j_1 = (j_+ + j_-)/2$ has an orthogonal basis of eigenstates $\psi_n(\sigma)$ with eigenvalues $\lambda + n$ for integer $n \geq 0$ given by
\begin{align}\label{eq:actualeigenfunctions}
\psi_n(\sigma) = \frac{2^\lambda}{\sqrt{\Gamma(2\lambda)}}L_n^{(2\lambda-1)}(2e^{-\sigma}) \exp( -\lambda \sigma) \exp(-e^{-\sigma})
\end{align}
where $L_n^{(\alpha)}(x)$ is a generalised Laguerre polynomial. It follows that \eqref{eq:discreterealisationboson} is indeed a realisation of a single copy of the discrete-series representation labelled by $\lambda > 0$, as claimed.

There is a minor subtlety here that arises when $0 < \lambda < 1$ and that is worth mentioning because it will play a role in our later discussion of the super-JT gravity.\footnote{Representations with $0 < \lambda < 1$ describe  stable tachyonic particles within the unitarity bound.} As a differential operator, $j_-$ is invariant under $\lambda \leftrightarrow (1-\lambda)$ as (trivially) are $j_+$ and $j_3$. It follows that the states
\begin{align} \label{eq:fakeeigenfunctions}
\tilde \psi_n(\sigma) = L_n^{(1 - 2 \lambda)}(2e^{-\sigma}) \exp( -(1 -\lambda) \sigma) \exp(-e^{-\sigma})
\end{align}
are also eigenfunctions of $j_1$ with eigenvalues $1 - \lambda + n$. They are normalisable for $\lambda < 1$. Importantly, these eigenfunctions are not orthogonal to our original eigenfunctions, in apparent violation of the spectral theorem for self-adjoint operators.

The origin of these peculiar excess eigenfunctions is that, when applying the spectral theorem to an unbounded Hilbert space operator $j_1$, the operator in question must be self-adjoint, meaning not only that the adjoint operator $j_1^\dagger$ has the same action as $j_1$ on states in the domain of $j_1$, but also that the domains of $j_1$ and $j_1^\dagger$ are identical. If only the former criterion is satisfied, and the domain of $j_1^\dagger$ is instead a strict superset of the domain of $j_1$, the operator $j_1$ is said to be symmetric but not self-adjoint. Often, there is nevertheless a unique extension of a symmetric operator $j_1$ to a larger domain on which it is self-adjoint; in that case the original operator is said to be essentially self-adjoint. But sometimes there is not.

A classic example of an operator where such subtleties are important is the Laplacian $\Delta = -d^2/dx^2$ on the Hilbert space $L^2([0,1])$. In order to integrate by parts and write
\begin{align}\label{eq:laplacianselfadjoint}
-\int_0^1 dx f(x)^* \frac{d^2 g}{d x^2} = \int_0^1 dx \frac{d f}{dx}^* \frac{d g}{d x} = - \int_0^1 dx \frac{d^2f}{dx^2}^* g(x)
\end{align}
we need to impose appropriate boundary conditions on $f$ and $g$ at $x=0,1$. In particular, if we want the operator $\Delta$ to be essentially self-adjoint, we must impose boundary conditions on wavefunctions $g$ in the domain of $\Delta$ such that the same boundary conditions on $f$ are both necessary and sufficient for \eqref{eq:laplacianselfadjoint} to hold for all allowed $f,g$. Examples of such boundary conditions include e.g. the Dirichlet boundary conditions $g(0) = g(1) = 0$ and the Neumann boundary conditions $dg/dx = 0$ for $x =0,1$. In both cases, the potential boundary terms 
\begin{align}
\left[f^*\, \frac{dg}{dx}\right]^1_0~~~~ \text{and} ~~~~ \left[\, \frac{df^*}{dx} \,g\right]^1_0
\end{align}
vanish if and only if the same boundary condition is applied to $f$. The resulting Dirichlet and Neumann Laplacians are self-adjoint and obey the spectral theorem, but their spectra are different and their eigenfunctions are not orthogonal. A Laplacian operator defined on $[0,1]$ without any boundary conditions would include both sets of eigenfunctions. This does not violate the spectral theorem because that Laplacian is not self-adjoint. 

In our case, for $j_-$ to be essentially self-adjoint we need to impose conditions on the asymptotic behaviour of wavefunctions $\psi$ in the domain of $j_-$ so that the boundary terms
\begin{align}\label{eq:j-boundaryterms}
\left[- i \phi^*(\sigma) e^\sigma (p_\sigma - i \lambda) \psi(\sigma) \right]^\infty_{-\infty} ~~~~~ \text{and} ~~~~~ \left[ - i [(p_\sigma - i \lambda)\phi(\sigma)]^* e^\sigma  \psi(\sigma) \right]^\infty_{-\infty}
\end{align}
vanish if and only if the same conditions are imposed on $\phi(\sigma)$. As $\sigma \to - \infty$, $e^\sigma \to 0$ and there is no ambiguity; the boundary terms will vanish for any normalisable wavefunctions $\phi$, $\psi$ where $j_- \phi$ and $j_-\psi$ are also normalisable. 

This is not true for $\sigma \to + \infty$ when $\lambda < 1$. For example, any attempt to use the proof of the spectral theorem to argue that the wavefunctions
\begin{align}
\phi(\sigma) = \exp( -\lambda \sigma) \exp(-e^{-\sigma})~~~~ \text{and} ~~~~\psi(\sigma) = \exp( -(1 -\lambda) \sigma) \exp(-e^{-\sigma})
\end{align}
should be orthogonal (since they have different eigenvalues for $j_1$) fails precisely because the boundary terms at $\sigma = +\infty$ in \eqref{eq:j-boundaryterms} does not vanish for this pair of wavefunctions. This is true even though $\phi(\sigma)$ and $\psi(\sigma)$ are both normalisable and are eigenfunctions of the differential operator $j_-$. To make $j_-$ essentially self-adjoint and obtain the desired discrete series representation of $\hSL(2,\R)$, we need to impose the Neumann-like boundary condition that $(p_\sigma - i \lambda) \psi(\sigma)$ must vanish faster than $e^{-\sigma}$ as $\sigma \to + \infty$. This condition makes $j_1$ essentially self-adjoint and is satisfied by the eigenstates \eqref{eq:actualeigenfunctions} but not by the functions \eqref{eq:fakeeigenfunctions}.

If we restrict $H_R$ and $H_L$ to a single discrete-series representation, described by the realisation \eqref{eq:discreterealisationboson},  they become
\begin{align} \label{eq:xlxRstuff}
H_R=& p_R^2+ e^{x_R}\left(1+e^{x_L}\right)\\
H_L=& p_L^2+\left(p_L-p_R+i \lambda\right) e^{x_L}\left(p_L-p_R-i \lambda\right)+ e^{x_L+x_R},
\end{align}
where $x_R = \chi-\sigma$, $x_L = \chi +\sigma$, $p_R = (p_\chi - p_\sigma)/2$ and $p_L = (p_\chi +p_\sigma)/2$.

Since $j_3 =p_\sigma$ generates translations of the matter fields on the $T=0$ slice, we can somewhat heuristically think of $x_L$ and $x_R$ as the renormalised distances to each boundary particle in a gauge where the location of the matter excitations is fixed. However the asymmetry between $H_R$ and $H_L$ makes this picture somewhat unsatisfactory. In Appendix \ref{app:fixhighestweight}, we describe an alternative description of JT gravity with matter, where the $\hSL(2,\R)$ gauge is explicitly fixed such that matter fields are in a highest-weight state. Unlike \eqref{eq:xlxRstuff}, this preserves the symmetry between the left and right boundaries -- at the cost of introducing a nonlocal inner product. However, for the purpose of analysing the joint eigenvalue problem, the realisation used in \eqref{eq:xlxRstuff} is considerably more practical. 

The fact that $H_R$ commutes with $x_L$ lets one deduce a factorized form for the joint eigenfunction $\Psi\left(x_L, x_R\right)$ where
\begin{align}\label{eq:factorisedform}
\Psi\left(x_L, x_R\right)=F\left(x_L\right) G\left(x_R+\log \left(1+e^{x_L}\right)\right),
\end{align}
with $G$ satisfying the equation
\begin{align}\label{eq:Gequation}
-G^{\prime \prime}(x)+e^x G(x)=2 E_R G(x) .
\end{align}
The form \eqref{eq:factorisedform} is really just a restatement of our earlier result that after conjugating by $U$ the right Hamiltonian depends only on $\chi$ and not on the matter fields. Indeed, \eqref{eq:Gequation} is just \eqref{eq:eigenvalueeqnHRbos} in somewhat different notation. By our previous arguments, there exists a unique, delta-function normalisable wavefunction $G(x)$ satisfying \eqref{eq:Gequation} for any $E_R > 0$. 

We want to use similar arguments to show that the eigenvalue equation $H_L \Psi = E_L \Psi$ uniquely determines $F$, and hence that there exists a unique delta-function normalisable solution to the joint eigenvalue problem for any  pair $E_L, E_R > 0$ within any discrete-series representation of the matter Hilbert space. After canceling terms proportional to $G^{\prime}$ and using \eqref{eq:Gequation}, the eigenvalue equation $H_L \Psi=E_L \Psi$ reduces to 
\begin{align}\label{eq:Fequation}
-\left(1+e^{x_L}\right) F^{\prime \prime}-e^{x_L} F^{\prime}+e^{x_L}\left(\lambda^2-\lambda\right) F+2 E_R \frac{e^{x_L}}{1+e^{x_L}} F=2 E_L F .
\end{align}
Again, this is a second-order differential equation and so has two smooth solutions. As $x_L \rightarrow-\infty$, \eqref{eq:Fequation} reduces to $-F^{\prime \prime}=2 E_L F$, and both solutions are oscillatory. As $x_L \rightarrow+\infty$,  \eqref{eq:Fequation} reduces to
\begin{align}
-F^{\prime \prime}-F^{\prime}+\left(\lambda^2-\lambda\right) F=0 .
\end{align}
The two linearly independent solutions of this equation are $e^{(\lambda-1) x_L}$ and $e^{-\lambda x_L}$. For $\lambda > 1$, only the latter is normalisable, while for $\lambda < 1$, only the latter satisfies the required boundary conditions that $(p_s - i \lambda) \psi(s)$ vanishes faster than $e^{-s}$. It follows that there is a unique delta-function normalisable solution to \eqref{eq:Fequation}, and hence a unique delta-function normalisable JT gravity eigenfunction within each nontrivial $\hSL(2,\R)$ matter representation, for any pair $E_L, E_R > 0$.

As a final comment, we note that, when $x_L, x_R \ll 0$, $H_L$ and $H_R$ become the Hamiltonians for free particles on $x_L$ and $x_R$ respectively. As a result, to build a normalisable state out of the joint eigenfunctions for $H_L$ and $H_R$ described above, we need to take independent superpositions over both $E_L$ and $E_R$. In particular, this means that the difference $(H_L - H_R)$ between the two Hamiltonians, like the Hamiltonians themselves, has purely continuous spectrum. This is in contrast to the vacuum sector of the theory, where $H_L$ and $H_R$ have purely continuous spectrum but $(H_L - H_R)$ vanishes.

\section{Boundary algebras in JT gravity} \label{sec:JTalgebras}
In quantum field theory, every causal diamond has an associated Type III von Neumann algebra that describes observables localised within that diamond. In a theory of quantum gravity, spacetime is fluctuating and so it is difficult to specify bulk spacetime regions in a nonperturbative, gauge-invariant manner. Instead, it is easier to define von Neumann algebras built out of asymptotic boundary observables. In the AdS/CFT correspondence, for example, local subalgebras of the conformal field theory are holographically dual to algebras of quantum gravity observables that are localised at the conformal boundary of the bulk spacetime. In the case of JT gravity, there are natural algebras associated to the left boundary particle, the right boundary particle, and to their union.  

It was shown in \cite{penington2023algebras, kolchmeyer2023neumann}  that in JT gravity with matter the algebras associated to the left and right boundaries are commutants and Type II$_\infty$ von Neumann factors. However, the derivations of those results in  \cite{penington2023algebras, kolchmeyer2023neumann} relied heavily on the use of Euclidean gravitational path integral calculations. Our primary contribution here is to give a self-contained derivation of the same results using only canonically quantised JT gravity in Lorentzian signature.

\subsection{Boundary operators with and without matter}

In pure JT gravity, the only (bounded) gauge-invariant right-boundary operators are (bounded) functions $f(H_R)$ of the Hamiltonian $H_R$. Since $H_R$ has completely nondegenerate spectrum, the algebra $\mathcal{A}_R$ of operators of this form is a maximal commuting subalgebra. It follows that $\mathcal{A}_R$ is equal to its own commutant $\mathcal{A}_R'$. Since $H_L = H_R$, the same operators also make up the left-boundary algebra 
\begin{align}
\mathcal{A}_L = \mathcal{A}_R = \mathcal{A}_R'. 
\end{align} 
Although the two boundary algebras are commutants, they are not factors (indeed they are both equal to their common centre!). As a result, they do not collectively generate the entire Hilbert space algebra $\mathcal{B}(\mathcal{H}_{\rm pure-JT})$. Instead, the algebra associated to the union of both boundaries is
\begin{align}
(\mathcal{A}_L \cup \mathcal{A}_R)'' = \mathcal{A}_R.
\end{align}
Gauge-invariant operators that do not commute with $H_R$, such as the renormalised wormhole length $\ell = 2\rho = - 2\chi$, cannot be written using any combination of operators at the two boundaries.  In this sense, canonically quantised pure JT gravity is not a holographic theory.

In JT gravity with matter there are additional gauge-invariant boundary operators associated to matter operators localised at the location of the boundary particles. Before describing these operators in detail, it is helpful to first describe the set of asymptotic boundary observables that act on $\mathcal{H}_{\rm matt}$ before including gravity. To do so, it will be helpful to appeal to matter Euclidean path integral constructions. However, we believe that the conclusions we reach should be taken as axiomatic for an algebraic quantum field theory in Lorentzian anti-de Sitter space. 

\begin{figure}
 \begin{center}
   \includegraphics[width=0.9\textwidth]{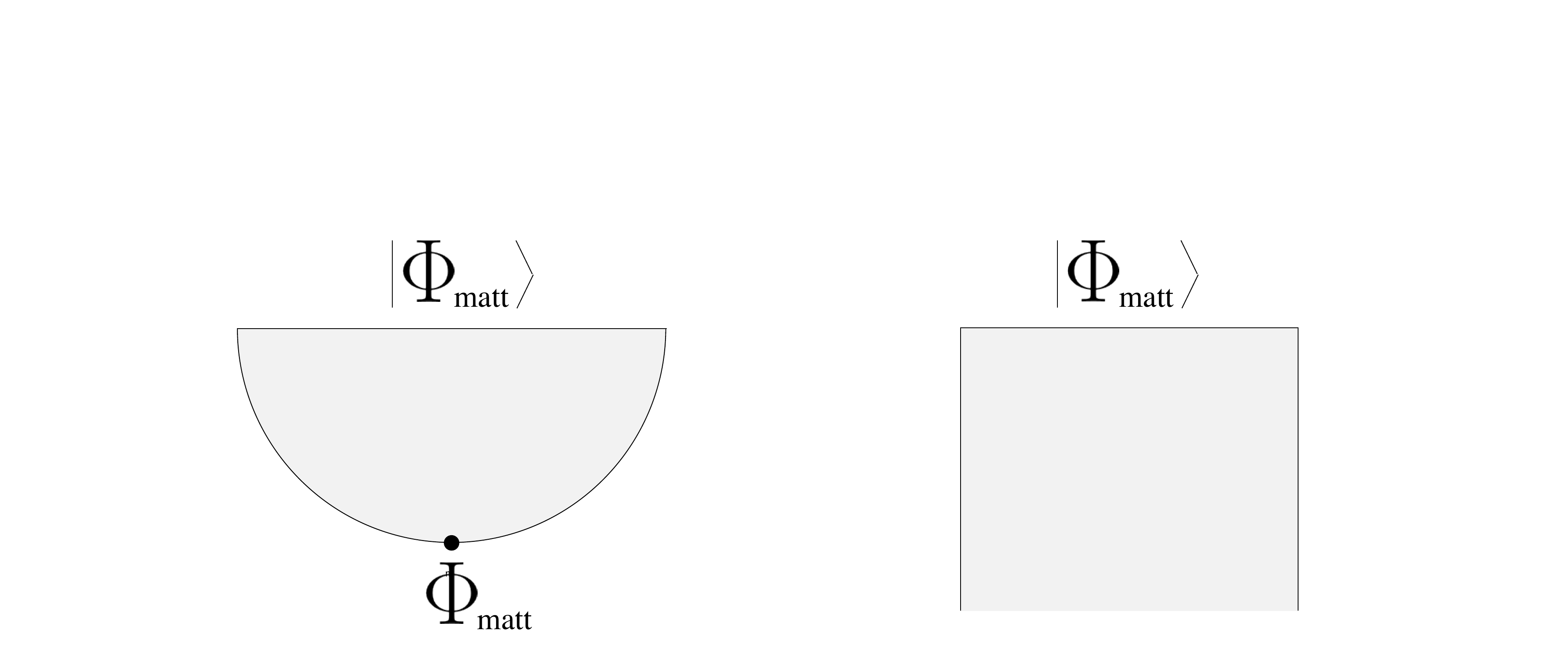}
 \end{center}
\caption{\footnotesize  Two views of half of Euclidean AdS$_2$.   Viewing AdS$_2$ as a hyperbolic disc, half of AdS$_2$ is the half-disc (\emph{left}); on the other hand, the AdS$_2$ metric can be put in the static form $\d\sigma^2+\cosh^2\sigma \d\tau^2$, and in this form, half of AdS$_2$ looks like a semi-infinite 
strip (\emph{right}).  This identification allows us to go back and forth between a state $\ket{\Phi_{\rm matt}}$ inserted at the bottom of the semi-infinite strip (and rescaled by $\exp[E\tau]$) and the corresponding local boundary operator $\Phi_{\rm matt}$ applied at the bottom of the hyperbolic disc.
 \label{fig:stateop}}
\end{figure}

Given a normalised state $\ket{\Phi_{\rm matt}} \in \mathcal{H}_{\rm matt}$ with global energy $E$ in a quantum field theory in anti-de Sitter space, we can define a corresponding asymptotic boundary operator $\Phi_{\rm matt}$, and vice versa, by equating two interpretation of the same Euclidean path integral, both illustrated in Figure \ref{fig:stateop}. In one interpretation, the state $\ket{\Phi_{\rm matt}}$ is inserted at the bottom of a Euclidean strip of divergent length $\tau \to \infty$ and is renormalised by a factor of $\exp(E \tau)$. In the other interpretation, the semi-infinite strip becomes the bottom half of the hyperbolic disc and the state $\ket{\Phi_{\rm matt}}$ becomes a local operator $\Phi_{\rm matt}$ at the boundary. This gives a map from states to local boundary operators. Conversely, given a boundary operator $\Phi_{\rm matt}$, we can define $\ket{\Phi_{\rm matt}}$ to be the state preprared by acting with $\Phi_{\rm matt}$ at the bottom of the hyperbolic disc. As a result, quantum field theories in anti-de Sitter space satisfy a ``state--boundary operator correspondence'' analogous to the state--operator correspondence in conformal field theories. 

Indeed, since isometries of anti-de Sitter space act as conformal transformations of the asymptotic boundary, Euclidean boundary correlation functions for a quantum field theory in anti-de Sitter space satisfy all the usual axioms for a conformal field theory except for the existence of a local (boundary) stress-energy tensor. Of course, the latter axiom is necessary for a ``true'' conformal field theory: quantum field theories in anti-de Sitter space are not holographic like AdS/CFT. However, a stress-energy tensor is not necessary to derive a state--operator correspondence. In the case of a quantum field theory in anti-de Sitter space, that state-operator correspondence is exactly the state--boundary operator correspondence that we have just described.

(As an aside, we note that, while a local quantum field theory in anti-de Sitter space does have a local \emph{bulk} stress-energy tensor, this stress-energy tensor is also not actually necessary for any of our discussion. All the principal conclusions of this paper will remain valid if we couple the JT gravity boundary particle to any theory satisfying the axioms of a CFT$_1$  (or the super-JT boundary superparticle to a theory satisfying the axioms of an SCFT$_1$) without any need for a stress-energy tensor in either the bulk or the boundary.)

The boundary operators $\Phi_{\rm matt}$ introduced above are not themselves $\hSL(2,\R)$ invariant. Instead they transform under conjugation by elements of $\hSL(2,\R)$ in the same discrete series representations as the states $\ket{\Phi_{\rm matt}}$. In particular, there exists a basis of local boundary operators $\Phi^i_{R,0}$ acting at Lorentzian time $T = 0$ on the right boundary that satisfy
\begin{align} \label{eq:scaling}
[ j_3, \Phi^i_{R,0}] =  i \Delta_i \Phi^i_{R,0}.
\end{align}
for some $\Delta_i > 0$ that is called the scaling dimension of $\Phi^i_{R,0}$. The corresponding state
\begin{align}
 \ket{\Phi^i_{\rm matt}} = e^{-\pi j_2/2} \Phi^i_{R,0}\ket{\Omega_{\rm matt}}
\end{align}
satisfies
\begin{align}
j_1 \ket{\Phi^i_{\rm matt}} = j_1 e^{-\pi j_2/2} \Phi^i_{R,0}\ket{\Omega_{\rm matt}} = -ie^{-\pi j_2/2} j_3 \Phi^i_{R,0}\ket{\Omega_{\rm matt}}  = \Delta_i \ket{\Phi_{\rm matt}}.
\end{align}
So the scaling dimension of a boundary operator is equal to the energy of the corresponding state.

A particular important class of boundary operators are boundary primaries, which are dual to the minimal energy state within an $\hSL(2,\R)$-representation and hence satisfy $\Delta = \lambda$. In addition to \eqref{eq:scaling}, boundary primaries satisfy
\begin{align}\label{eq:primary}
[j_-, \Phi_{R,0}] = 0,
\end{align}
since otherwise the operator $[j_-, \Phi_{R,0}]$ would satisfy
\begin{align}
[j_3 ,[j_-,\Phi_{R,0}]] = -[j_-,[\Phi_{R,0},j_3]] - [\Phi_{R,0},[j_3,j_-]] = +i (\Delta -1) [j_-,\Phi_{R,0}] .
\end{align}
No such operator exists within the discrete series representation. Finally, since the discrete series representations of $\hSL(2,\R)$ are real, we can always choose a basis of primary operators $\Phi_{R,0}^i$ that are all self-adjoint. Note that the identity operator satisfies all of these conditions (with $\Delta = 0$ but it will be more convenient for various reasons not to count it as a boundary primary. (We will change this convention when we come to study super-JT gravity.)

Now let
\begin{align}
\Phi_{R,0}(T_0) = e^{i j_1 T_0} \Phi_{R,0} e^{-i j_1 T_0}
\end{align}
act at the right boundary at time $T_0$. We can then define an operator
\begin{align}
\mathbf{\Phi}_R(t) = e^{\Delta \chi(t)} \Phi_{R,0}(T(t))
\end{align} 
that acts at the location $(T(t),\chi(t))$ of the right boundary particle at boundary time $t$. We claim that the operator $\mathbf{\Phi}_R(t)$ is invariant under the diagonal action $\mathcal{D}$ of $\hSL(2,\R)$. In particular, we have
\begin{align}
[j_1, \mathbf{\Phi}_R(t)] &= - i e^{\Delta \chi(t)}  \partial_T  \Phi_{R,0}(T(t)) =- [J^R_1,\mathbf{\Phi}_R(t)] \\
[j_2, \mathbf{\Phi}_R(t)] &=e^{i j_1 T(t)} [\cos T(t) j_2 + \sin T(t) j_3,\Phi_{R,0}]e^{-ij_1 T(t)} \\\nonumber &=  e^{\Delta \chi(t)}\left( -i\cos T(t) \partial_T  \Phi_{R,0}(T(t)) +i \Delta \sin T\, \Phi_{R,0}(T(t)) \right) = -[J_2^R,\mathbf{\Phi}_R(t)]
\\
[j_3, \mathbf{\Phi}_R(t)] &=e^{i j_1 T(t)} [\cos T(t) j_3 - \sin T(t) j_2,\Phi_{R,0}]e^{-ij_1 T(t)} \\\nonumber &=  e^{\Delta \chi(t)}\left( i\sin T(t) \partial_T  \Phi_{R,0}(T(t)) +i \Delta \cos T \,\Phi_{R,0}(T(t)) \right) = -[J_3^R,\mathbf{\Phi}_R(t)].
\end{align}
Since $\mathcal{D}(\mathcal{J}_i) =  J_i^R + j_i + J_i^L$ and trivially $[J_i^L, \mathbf{\Phi}_R(t)] = 0$, it follows immediately that
\begin{align}
[\mathcal{D}(\mathcal{J}_i),\mathbf{\Phi}_R(t)] = 0
\end{align}
as claimed. For wavefunctions of the form \eqref{eq:fullwaveformboson}, we have 
\begin{align}
\mathbf{\Phi}_R(0) \delta(T)\delta(T') \delta(\chi - \chi') \Psi_{JT}(\chi) = \delta(T)\delta(T') \delta(\chi - \chi') e^{\Delta \chi} \Phi_{R,0} \Psi_{JT}(\chi).
\end{align}
It follows that $\mathbf{\Phi}_R(t)$ acts on the physical Hilbert space \eqref{eq:bosHilbphys} as
\begin{align}
\Phi_R(t) = e^{iH_R t} e^{\Delta \chi} \Phi_{R,0} e^{-iH_R t}.
\end{align}
 It is easy to check that this commutes with the left-boundary Hamiltonian $H_L$. Note that one obtains the same operator $\Phi_R(t)$, up to the identification $\chi \leftrightarrow - \rho$, by starting from the gauge-invariant operator
\begin{align}
\mathbf{\Phi}_R(t) = e^{-\Delta \rho(t)} \,e^{i j_+ \tau(t)} \Phi_{R,0} e^{-ij_+ \tau(t)}
\end{align} 
in the parameterisation \eqref{eq:param2}. 

Similarly, a left-boundary primary operator $\Phi_{L,0}$ at time $T = 0$ satisfies
\begin{align}
[ j_3, \Phi_{L,0}] = - i \Delta \Phi_{L,0} , ~~~~ [j_+, \Phi_{L,0}] = 0 ~~~~\text{and}~~~~ e^{\pi j_2/2} \Phi_{L,0} \ket{\Omega_{\rm matt}} = \ket{\Phi_{\rm matt}}.
\end{align}
This allows us to define the gauge-invariant left-boundary operator
\begin{align}
\mathbf{\Phi}_L(t') = e^{\Delta \chi'(t')} e^{i j_1 T'(t')} \Phi_{R,0} e^{-i j_1 T'(t')}.
\end{align}
On the physical Hilbert space \eqref{eq:bosHilbphys}, $\mathbf{\Phi}_L(t')$ acts as
\begin{align}
\Phi_L(t') = e^{i H_L t'} e^{\Delta \chi} \Phi_{L,0} e^{-i H_L t'}.
\end{align}
It is easy to check that 
\begin{align}
[H_R, \Phi_L(t')] = [\Phi^i_R(t), \Phi^j_L(t')] = 0,
\end{align}
so that operators at the left and right boundaries always commute.

In general, local operators in quantum field theory are (despite the name!) not actually densely-defined Hilbert space operators, since correlation functions at coincident points diverge. Instead the local operator needs regulated in some way: two standard approaches are either a) sandwiching the local operator with a small amount of Euclidean time evolution or b) smearing the operator in Lorentzian time. In JT gravity, the same issue exists and the same solutions work. For any state $\ket{\Psi_{JT}}$, we have
\begin{align}
\braket{\Psi_{JT}|\Phi_R(t) \Phi_R(t)|\Psi_{JT}} = \braket{\Psi_{JT}|e^{i H_R t} e^{2 \Delta \chi} \Phi_{R,0}^2 e^{-i H_R t}|\Psi_{JT}}  = \infty.
\end{align}
So $\Phi_R(t)$ is not a densely defined operator. However, operators of the form
\begin{align}\label{eq:lorentzsmear}
\Phi_{R,f} = \int dt f(t) \Phi_R(t)
\end{align}
for some smooth smearing function $f$ or
\begin{align}\label{eq:euclideansandwich}
e^{-\beta_1 H_R} \Phi_R(t) e^{-\beta_2 H_R}
\end{align}
for $\beta_1,\beta_2 > 0$ are densely defined. Bounded operators with a well defined action on the entire Hilbert space $\mathcal{H}_{JT}$ can then be constructed from bounded functions of operators such as  \eqref{eq:lorentzsmear} or \eqref{eq:euclideansandwich}.

\subsection{The structure of the algebras}
The full algebra $\mathcal{A}_R$ of right-boundary observables is generated by (or, more precisely, is the double commutant of) the Hamiltonian $H_R$ and the matter operators $\Phi_R(t)$ (regularised as in \eqref{eq:lorentzsmear} or \eqref{eq:euclideansandwich}). In fact we will see that a single matter operator insertion is always sufficient: operators of the form \eqref{eq:euclideansandwich} (along with $f(H_R)$ for arbitrary bounded functions $f$) span $\mathcal{A}_R$.

The key result, which we prove in Appendix \ref{app:JT2pt}, is that
\begin{align} \nonumber
\braket{\Omega_{\rm matt}, s_1'| {\Phi^i_R} \Pi_{E_R = s^2} \Phi^j_R | \Omega_{\rm matt},s_2'} &=\braket{\Omega_{\rm matt}, s_1'| {\Phi^i_R} U^\dagger \left(\ket{s}\!\bra{s}\otimes \mathds{1}_{\rm matt}\right) U \Phi^j_R | \Omega_{\rm matt}, s_2'} 
\\&= \delta_{ij} \delta(s_1' - s_2') \frac{\rho(s) \Gamma(\Delta_i \pm i s_1' \pm i s)}{\Gamma(2 \Delta_i)}.\label{eq:JT2pt}
\end{align}
There is a lot of notation here that we need to unpack. The delta-function normalised wavefunctions $\ket{s}$ were defined in \eqref{eq:kets}. We are using a shorthand
\begin{align}
\ket{\Omega_{\rm matt}, s_2'} = \ket{\Omega_{\rm matt}} \ket{s_2'} 
\end{align}
for tensor product states where the matter fields are in the vacuum state $\ket{\Omega_{\rm matt}} \in \mathcal{H}_{\rm matt}$. The projection-valued measure $\Pi_{E_R = s^2}$ for the right Hamiltonian $H_R$ is defined so that
\begin{align}\label{eq:HRPVM}
f(H_R) = \int ds f(s^2)\Pi_{E_R = s^2}  
\end{align}
for any function $f$. The unitary $U = U_2 U_1$ was defined in \eqref{eq:conjHR}, where we also showed that conjugation by $U$ turns the Hamiltonian $H_R$ into a tensor product of the pure-JT Hamiltonian acting on $L^2(\R)$ with the identity $\mathds{1}_{\rm matt}$ on $\mathcal{H}_{\rm matt}$ so that
\begin{align}
U \Pi_{E_R = s^2} U^\dagger =  \mathds{1}_{\rm matt} \otimes \ket{s}\!\bra{s}
\end{align}
is a tensor product of the projection-valued measure $\ket{s}\!\bra{s}$ for pure JT gravity with $\mathds{1}_{\rm matt}$. Finally, we are using a standard and convenient convention where there is an implicit product over all choices of sign in gamma functions, so that e.g.
\begin{align}
\Gamma(\Delta_i \pm i s_1' \pm i s) = \Gamma(\Delta_i+ i s_1' + i s)\Gamma(\Delta_i + i s_1' - i s)\Gamma(\Delta_i - i s_1' + i s)\Gamma(\Delta_i - i s_1' - i s).
\end{align}

The basic strategy for proving \eqref{eq:JT2pt} is the following. Recall that $U_1 = \exp(-i \chi j_3)$. For any fixed $\chi$, it follows from \eqref{eq:scaling} and $j_3 \ket{\Omega_{\rm matt}} = 0$ that 
\begin{align}\label{eq:U1Phi}
e^{-i \chi j_3} \Phi^j_R \ket{\Omega_{\rm matt}} = e^{\Delta_j \chi} e^{-i \chi j_3}  \Phi^j_{R,0} \ket{\Omega_{\rm matt}} = e^{2\Delta_j \chi} \Phi^j_{R,0} \ket{\Omega_{\rm matt}},
\end{align}
Meanwhile 
\begin{align}\label{eq:U2s'}
U_2^\dagger \left(\ket{s}\!\bra{s}\otimes \mathds{1}_{\rm matt}\right)  U_2 = \rho(s) K_{2is}(2\sqrt{1+j_+} e^{\chi})K_{2is}(2\sqrt{1+j_+} e^{\chi'})
\end{align} 
only acts on $\mathcal{H}_{\rm matt}$ via $j_+$.  Finally, for any function $f$, we have
\begin{align}\label{eq:intj+}
\braket{\Omega_{\rm matt}| \Phi^i_{R,0} f(j_+) \Phi^j_{R,0}|\Omega_{\rm matt}} = \delta_{ij}\frac{1}{\Gamma(2\Delta_i)}\int_0^\infty dj_+\, j_+^{2 \Delta_i - 1} f(j_+),
\end{align}
which is simply the Fourier transform of the usual AdS$_2$ two-point function in Poincar\'{e} coordinates. Using \eqref{eq:U1Phi} - \eqref{eq:intj+}, we can reduce the left-hand side of \eqref{eq:JT2pt} to an integral over two copies of $\chi$ together with one integral over $j_+$; the full form is given in \eqref{eq:tripleintegral}. These integrals are somewhat nontrivial to carry out. However with sufficient effort one obtains the right-hand side of \eqref{eq:JT2pt}.

Ignoring the delta function, the right-hand side of \eqref{eq:JT2pt} is almost symmetric between $s$ and $s_1'$, except for the existence of a factor $\rho(s)$ with no corresponding factor of $\rho(s_1')$. As a result, the linear functional
\begin{align}\label{eq:trbosonJT}
\Tr[a] = \int ds_1 ds_2 \, \sqrt{\rho(s_1)\rho(s_2)} \braket{\Omega_{\rm matt}, s_1| a|\Omega_{\rm matt}, s_2}
\end{align}
satisfies
\begin{align}\label{eq:tracecyclicspecial}
\Tr\left[{\Phi^i_R} f(H_R) \Phi^j_R g(H_R)\right] &= \delta_{ij}  \int ds ds'\, f(s^2) g(s'^2) \, \rho(s) \rho(s')\frac{ \Gamma(\Delta_i \pm i s \pm i s')}{\Gamma(2 \Delta_i)}
\\&= \Tr\left[\Phi^j_R g(H_R) {\Phi^i_R}^\dagger f(H_R) \right].
\end{align}
Without the factor of $\sqrt{\rho(s_1)\rho(s_2)}$ in \eqref{eq:trbosonJT}, this would not work. 

As one might guess given our choice of notation, the functional \eqref{eq:trbosonJT} will end up being the unique normal semifinite trace on the algebra $\mathcal{A}_R$. In particular, the result \eqref{eq:tracecyclicspecial} is a special case of the trace cyclicity condition
\begin{align}\label{eq:tracialcondition}
\Tr[ab] = \Tr[ba]
\end{align}
for all $a,b \in\mathcal{A}_R$. 

\begin{figure}
  \centering
 \includegraphics[width = 0.85\linewidth]{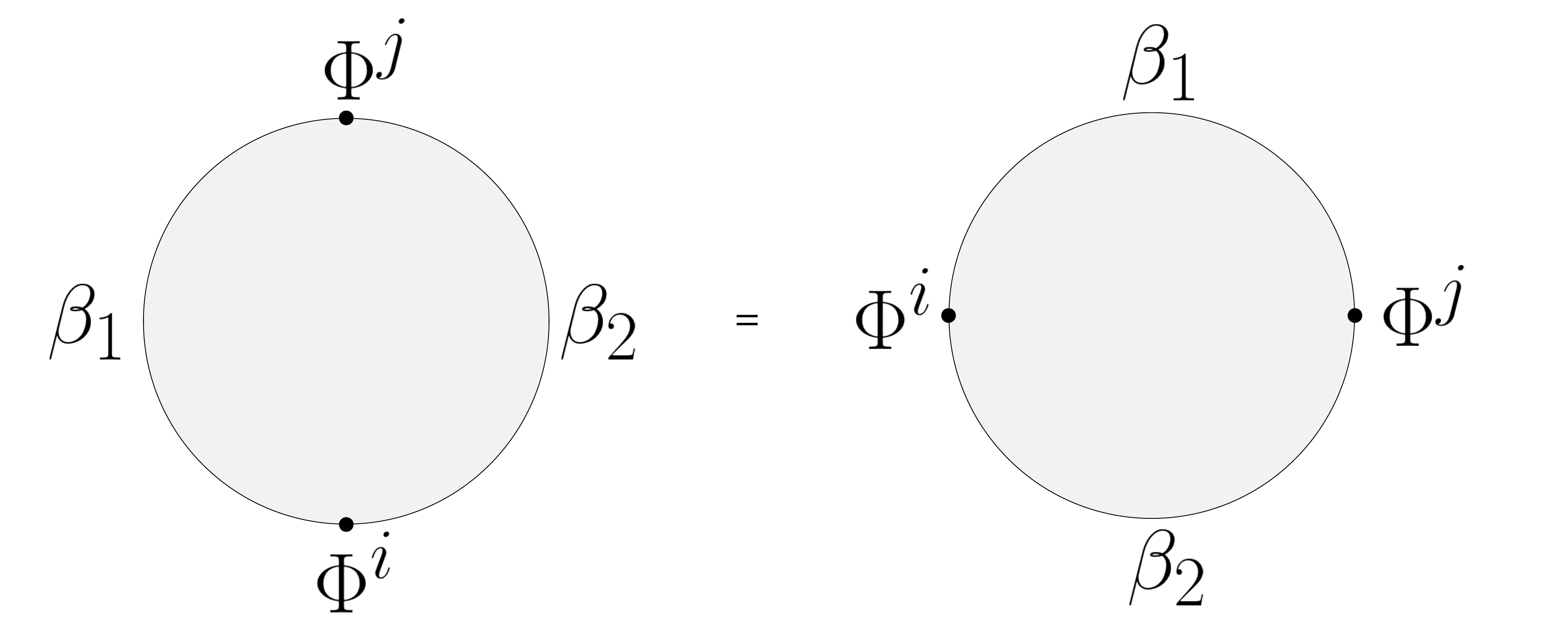}
\caption{\footnotesize The trace $\Tr\left[{\Phi^i_R} e^{-\beta_1 H_R} \Phi^j_R e^{-\beta_2 H_R}\right]$ can be computed (\emph{left}) using a gravitational path integral on the Euclidean disc where the two matter insertions are separated by Euclidean boundary times $\beta_1$ and $\beta_2$. The same Euclidean path integral (\emph{right}) computes a left-right boundary two-point function $\braket{\beta_1| \Phi^i_L \Phi^j |\beta_2}$ where the bra and ket are in the Hartle-Hawking states of the form $\ket{\beta} = \int ds \exp(-\beta s^2)\sqrt{\rho(s)} \ket{s}$. Taking an inverse Laplace transform over $\beta_1$ and $\beta_2$ then leads to the equivalence found in \eqref{eq:leftrightvstrace}, which is not eacy to see directly from the canonically quantised theory.}
\label{fig:leftrighttrace}
\end{figure}

A standard computation, reviewed in \eqref{eq:leftright2ptcalc}, shows that the left-right boundary two-point function is given by
\begin{align}\label{eq:leftrighttwopt}
\braket{\Omega_{\rm matt}, s'| {\Phi^i_L} \Phi^j_R |\Omega_{\rm matt}, s} = \delta_{ij} \sqrt{\rho(s) \rho(s')} \frac{\Gamma(\Delta_i \pm is \pm is')}{\Gamma(2\Delta_i)}.
\end{align}
As a result, we have
\begin{align}\label{eq:leftrightvstrace}
\int ds_1 ds_2 \, \sqrt{\rho(s_1)\rho(s_2)} f(s_1^2) g(s_2^2)\braket{\Omega_{\rm matt}, s_1| {\Phi^i_L} \Phi^j_R |\Omega_{\rm matt}, s_2} &= \Tr\left[{\Phi^i_R} f(H_R) \Phi^j_R g(H_R)\right]. 
\end{align}
From the canonically quantised perspective taken in this paper, there is no obvious reason that the two sides of \eqref{eq:leftrightvstrace} should agree. As shown in Figure \ref{fig:leftrighttrace}, the real reason that they match is because both can be computed using the same gravitational path integral on the Euclidean disc. In both cases, the relevant path integral has two matter insertions and the renormalised distances $\beta_1$ and $\beta_2$ between those insertions are Laplace transforms respectively of $s_1^2$ and $s_2^2$. Crucially, the explicit factors of $\rho(s)$ that needed to be included in \eqref{eq:leftrightvstrace} and \eqref{eq:trbosonJT} to make the agreement work are automatically present in a Euclidean path integral computation. Indeed $\rho(s)$ is simply the inverse Laplace transform of the Euclidean partition function $Z(\beta)$. It is therefore naturally interpreted as the Euclidean density of states of the theory. Our goal is to show that the same function already be understood in the canonically quantised Lorentzian theory as the unique density of states needed for \eqref{eq:trbosonJT} to have the algebraic structure of a trace.

To achieve this goal, by proving \eqref{eq:tracialcondition} in full generality, we will need to prove that the operators on $\mathcal{A}_R$ only ever contain a single matter insertion, or, more precisely that the linear span of operators of the form \eqref{eq:euclideansandwich}, together with functions of $H_R$, are dense in $\mathcal{A}_R$ with respect to the strong operator topology. Recall from Section \ref{sec:spectra} that there exists a unique delta-function normalisable state within any given matter $\hSL(2,\R)$ representation with left-boundary energy $E_L > 0$ and right-boundary energy $E_R > 0$. One way to write these states is as
\begin{align}
\ket{\Phi^j, s', s} = \sqrt{\frac{\Gamma(2 \Delta_j)}{\rho(s) \Gamma(\Delta_j \pm i s' \pm i s)}} \,\,\Pi_{E_R = s^2} \Phi^j_R \ket{\Omega_{\rm matt}, s'},
\end{align}
where $E_L = s'^2$, $E_R = s^2$ and the state is in the same $\hSL(2,\R)$ representation as $\ket{\Phi^j}$. To see that this is true, first note that the Hamiltonian $H_R$ only acts on matter fields via $\hSL(2,\R)$ generators and so functions of $H_R$ preserve the $\hSL(2,\R)$ representations of the matter theory. Because $\Pi_{E_R = s^2}$ projects onto states with $E_R = s^2$, we have
\begin{align}\label{eq:j,s,s'}
H_R \ket{\Phi^j, s', s} = s^2  \ket{\Phi^j, s', s}.
\end{align}
Moreover, since both $\Pi_{E_R = s^2}$ and $\Phi^j_R$ commute with $H_L$, we have
\begin{align}
H_L \ket{\Phi^j, s', s} = s'^2  \ket{\Phi^j, s', s}.
\end{align}
Finally we have
\begin{align}
\nonumber \braket{\Phi^i, s_1', s_1|\Phi^j, s_2', s_2} &=  \frac{\sqrt{\Gamma(2 \Delta_i)\Gamma(2 \Delta_j)}\braket{ \Omega_{\rm matt}, s_1'| {\Phi^i_R} \Pi_{E_R = s_1^2}\Pi_{E_R = s_2^2} \Phi^j_R | \Omega_{\rm matt}, s_2'}}{ \sqrt{\rho(s_1)\rho(s_2)\Gamma(\Delta_j \pm i s_1 \pm i s_1')\Gamma(\Delta_j \pm i s_2 \pm i s_2')}}\\
&= \delta_{ij} \delta(s_1 - s_2) \delta(s_1' - s_2').
\end{align}
We conclude that the states $\ket{\Phi^j,s,s'}$ are indeed the unique delta-functional normalised states in their respective matter representations with $E_L = s'^2$ and $E_R = s^2$. Consequently, together with the states $\ket{\Omega_{\rm matt}, s}$, they form a delta-function normalised basis for $\mathcal{H}_{JT}$.

By an exactly analogous argument, we can also write
\begin{align}\label{eq:js's=PhiL}
\ket{\Phi^j, s', s} = \sqrt{\frac{\Gamma(2 \Delta_j)}{\rho(s') \Gamma(\Delta_j \pm i s' \pm i s)}} \,\,\Pi_{E_L = s'^2} \Phi^j_L \ket{\Omega_{\rm matt}, s}
\end{align}
where
\begin{align}
\Pi_{E_L = s'^2} =   \exp(-i \chi j_3) (1 + j_-)^{i p_\chi/2} \left(\ket{s'}\!\bra{s'}\otimes \mathds{1}_{\rm matt}\right)  (1 + j_-)^{-i p_\chi/2}  \exp(i \chi j_3)
\end{align}
is the projection-valued measure for $H_L$. Indeed it is easy to check using \eqref{eq:leftrighttwopt} that
\begin{align}
\nonumber  \bra{\Omega_{\rm matt}, s_1}{\Phi^i_L} \Pi_{E_L = s_1'^2} \ket{\Phi^j,s_2',s_2 } &= \delta(s_1 - s_2) \delta(s_1' - s_2') \frac{\sqrt{\Gamma(2 \Delta_j)} \braket{\Omega_{\rm matt}, s_2| {\Phi^i_L} \Phi^j_R |\Omega_{\rm matt}, s_2'}}{\sqrt{\rho(s_2) \Gamma(\Delta_j \pm i s_2' \pm i s_2)}}
\\& = \sqrt{\frac{\rho(s_2') \Gamma(\Delta_j \pm i s_2' \pm i s_2)}{\Gamma(2 \Delta_j)}}\,
\delta_{ij} \delta(s_1 - s_2) \delta(s_1' - s_2'),
\end{align}
with $\ket{\Phi^j,s_2',s_2}$ defined as in \eqref{eq:j,s,s'}.

Now suppose that there exist $a, b \in \mathcal{A}_R$ such that
\begin{align}\label{eq:a=b}
a \ket{\Omega_{\rm matt},s} = b \ket{\Omega_{\rm matt}, s}
\end{align}
for all $s > 0$. Since $a$ and $b$ commute with $\mathcal{A}_L$, we then also have
\begin{align}
a \ket{\Phi^j, s', s} &= \sqrt{\frac{\Gamma(2 \Delta_j)}{\rho(s') \Gamma(\Delta_j \pm i s' \pm i s)}} \,\,\Pi_{E_L = s'^2} \Phi^j_L a \ket{\Omega_{\rm matt}, s} 
\\&  = \sqrt{\frac{\Gamma(2 \Delta_j)}{\rho(s') \Gamma(\Delta_j \pm i s' \pm i s)}} \,\,\Pi_{E_L = s'^2} \Phi^j_L b \ket{\Omega_{\rm matt}, s} 
\\& =  b \ket{\Phi^j, s', s}
\end{align}
for all $\ket{\Phi^j, s', s}$. We must therefore have $a=b$. Given any operator $a \in \mathcal{A}_R$, we are going to construct an operator $b \in \mathcal{A}_R$ containing at most a single matter operator insertion and show that \eqref{eq:a=b} is satisfied. We will then be able to conclude that $a = b$ and hence that any operator in $\mathcal{A}_R$ can be written using at most a single matter insertion. Because $a$ commutes with $H_L$, we must have
 \begin{align}\label{eq:aaction}
a \ket{ \Omega_{\rm matt}, s'}  = f(s'^2) \ket{\Omega_{\rm matt}, s'} + \sum_k \int ds \,g_k(s^2,s'^2) \sqrt{\frac{\rho(s) \Gamma(\Delta_k \pm i s' \pm i s)}{\Gamma(2 \Delta_k)}} \ket{\Phi^k,s',s}
\end{align}
for some functions $f(s^2)$ and $g_k(s^2,s'^2)$. (The right-hand side of \eqref{eq:aaction} is the most general state with left-boundary energy $E_L  = s'^2$; the gamma functions and $\rho(s)$ are included explicitly for later convenience.) Let
\begin{align}\label{eq:bexpand}
b = f(H_R) + \sum_{k} \int ds ds' \, g_{k}(s^2,s'^2)  \Pi_{E_R = s^2} \Phi^k_R \Pi_{E_R = s'^2}.
\end{align}
We have
 \begin{align}\label{eq:baction}
b \ket{\Omega_{\rm matt}, s'}  = f(s'^2) \ket{\Omega_{\rm matt}, s'} + \sum_k \int ds\, g_k(s^2,s'^2) \sqrt{\frac{\rho(s) \Gamma(\Delta_k \pm i s' \pm i s)}{\Gamma(2 \Delta_k)}} \ket{\Phi^k,s',s},
\end{align}
which exactly matches \eqref{eq:aaction}. So $a = b$ can indeed be written using only a single matter operator insertion.

As an explicit example, it is helpful to consider the special case where
\begin{align}
a = {\Phi^i_R} e^{-\beta H_R} \Phi^j_R.
\end{align}
so that the expansion \eqref{eq:bexpand} is the operator product expansion for two matter insertions separated by Euclidean time $\beta$. From \eqref{eq:JT2pt}, we have
\begin{align}
\braket{\Omega_{\rm matt}, s_1| a |\Omega_{\rm matt}, s_2} = \delta_{ij} \delta(s_1 - s_2) \int ds\,e^{-\beta s^2} \rho(s) \frac{\Gamma(\Delta_i \pm is_1 \pm is)}{\Gamma(2\Delta_i)}.
\end{align}
So
\begin{align}
f(s_1^2) = \delta_{ij} \int ds\,e^{-\beta s^2} \rho(s) \frac{\Gamma(\Delta_i \pm is_1 \pm is)}{\Gamma(2\Delta_i)}.
\end{align}
To obtain the functions $g_j(s^2,s'^2)$, we need to compute
\begin{align}
\braket{\Phi^k, s_1', s_1| a | \Omega_{\rm matt}, s_2'} = \delta(s_1' - s_2') \sqrt{\frac{\Gamma(2 \Delta_k)}{\rho(s_1') \Gamma(\Delta_k \pm i s_1' \pm i s_1)}} \braket{ \Omega_{\rm matt}, s_1| \Phi_L^k \Phi^i_R e^{-\beta H_R} \Phi^j_R |\Omega_{\rm matt}, s_1'}.
\end{align}
In Appendix \ref{app:JT2pt},  we show that 
\begin{align}\nonumber
\Tr\left[ \Phi^i_R \Pi_{E_R = s_1^2} \Phi^j_R \Pi_{E_R = s_2^2} \Phi^k_R \Pi_{E_R = s_3^2}\right] &= \sqrt{\rho(s_1) \rho(s_3)} \braket{ \Omega_{\rm matt}, s_1| \Phi_L^k \Phi^i_R \Pi_{E_R = s_2^2}  \Phi^j_R | \Omega_{\rm matt}, s_3}
\\&= C_{ijk}\, F_{ijk}(s_1, s_2, s_3),
\end{align}
where  $F_{ijk}$ depends only on the scaling dimensions $\Delta_i$, $\Delta_j$ and $\Delta_k$ and can (in principle) be written in terms of the generalised hypergeometric function ${}_4F_3$. Meanwhile, the constants $C_{ijk}$, which depend on the choice of matter theory, characterise the matter primary three-point function
\begin{align}\label{eq:threepointfunction}
\braket{\Omega_{\rm matt}|{\Phi^k_{L,0}} \Phi^i_{R,0} e^{- j_+ \tau} \Phi^j_{R,0}  |\Omega_{\rm matt}} = \frac{C_{ijk}}{\tau^{\Delta_i + \Delta_j -\Delta_k}}.
\end{align}
The functional form of \eqref{eq:threepointfunction} (up to the coefficient $C_{ijk}$) is completely determined by symmetry since
\begin{align}
\tau \frac{d}{d\tau}\braket{\Omega_{\rm matt}|  {\Phi^k_{L,0}} \Phi^i_{R,0} e^{- j_+ \tau} \Phi^j_{R,0} |\Omega_{\rm matt}}  &= -i \braket{\Omega_{\rm matt}| {\Phi^k_{L,0}} \Phi^i_{R,0} [j_3, e^{- j_+ \tau}] \Phi^j_{R,0}  |\Omega_{\rm matt}}
\\&= \left(\Delta_k -\Delta_i-\Delta_j\right) \braket{\Omega_{\rm matt}|  {\Phi^k_{L,0}} \Phi^i_{R,0}  e^{- j_+ \tau} \Phi^j_{R,0} |\Omega_{\rm matt}}.\nonumber
\end{align}
The analyticity of Euclidean matter correlation functions means that $C_{ijk} = C_{kij} = C_{jki}$, while the fact that the boundary primaries were chosen to be Hermitician means that $C_{ijk} = C_{jik}^*$.\footnote{In general the three-point function coefficients $C_{ijk}$ will not be real and, consequently,  will not be invariant under arbitrary permutations. In particular, if a parity symmetry exists in the matter theory, primary states $\ket{\Phi^i}$ can be labelled by their charge $(-1)^{p_i}$ under parity. It is easy to show that in such theories $C_{ijk} = (-1)^{p_i + p_j + p_k} C_{jik}$ and, as a result, that the three-point function is only permutation invariant if $(p_i + p_j +p_k) = 0 \text{ mod } 2$. For a free scalar field with scaling dimension $\Delta$, the lightest parity-antisymmetric primary is a three-particle state with energy $3 \Delta + 3$. There exist nontrivial three-point functions involving this primary, together with the one- and two-particle parity-symmetric primaries, that are not permutation invariant. \label{foot:CijkCjik}}

It follows that
\begin{align}
{\Phi^i_R} e^{-\beta H_R} \Phi^j_R =  \delta_{ij}& \int ds\,e^{-\beta s^2}  \rho(s) \frac{\Gamma(\Delta_i \pm i\sqrt{H_R} \pm is)}{\Gamma(2\Delta_i)} \\&+ \sum_k \int ds_1 ds_2 d s_3 \,\frac{C_{ijk}e^{-\beta s_1^2} \Gamma(2 \Delta_k) F_{ijk}(s_1, s_2, s_3) }{\rho(s_2) \rho(s_3)\Gamma(\Delta_k \pm i s_2 \pm i s_3)} \Pi_{E_R = s_3^2} \Phi_k  \Pi_{E_R = s_2^2}\nonumber
\end{align}

Now that we have shown that any operator $a, b \in \mathcal{A}_R$ can be written in the form \eqref{eq:bexpand}, the general tracial condition \eqref{eq:tracialcondition} follows immediately from \eqref{eq:tracecyclicspecial}, together with the fact that
\begin{align}\nonumber
\bra{\Omega_{\rm matt}, s_1} a f(H_R) \ket{\Omega_{\rm matt}, s_2} &= \bra{ \Omega_{\rm matt}, s_1}  a f(H_L) \ket{ \Omega_{\rm matt}, s_2} \label{eq:afHRtracial}\\&= \bra{ \Omega_{\rm matt}, s_1} f(H_L) a  \ket{ \Omega_{\rm matt}, s_2} \\&= \bra{ \Omega_{\rm matt}, s_1} f(H_R) a  \ket{ \Omega_{\rm matt}, s_2}\nonumber
\end{align}
for any function $f$, operator $a \in \mathcal{A}_R$ and $s_1, s_2 > 0$. In the first and last steps of \eqref{eq:afHRtracial}, we used the fact that $f(H_L)\ket{ \Omega_{\rm matt}, s} = f(H_R) \ket{ \Omega_{\rm matt}, s} = f(s^2) \ket{ \Omega_{\rm matt}, s}$, while, in the second step, we used the fact that $[H_L, a] =0$. 

We also need to prove that the trace \eqref{eq:trbosonJT} satisfies the appropriate ``nice'' properties of the unique trace on a Type II$_\infty$ factor. The relevant technical conditions are that  \eqref{eq:trbosonJT} should be faithful, normal and semifinite. Since operators $a \in \mathcal{A}_R$ commute with $H_L$, \eqref{eq:trbosonJT} can be written as
\begin{align}
\Tr[a] = \sum_{I} \int_I ds_1 \int_I ds_2\, \sqrt{\rho(s_1)\rho(s_2)} \braket{\Omega_{\rm matt}, s_1| a|\Omega_{\rm matt}, s_2}
\end{align}
for any decomposition of $\R^+$ into finite intervals $I$. Each term in this sum is an expectation value on a normalisable state. As a result, the trace $\Tr$ is ultraweakly lower semicontinuous on the space of positive operators and is said to be \emph{normal}. Since the operators $e^{-\beta H_R}$ form an increasing net of positive operators with finite trace and supremum the identity operator, the trace is said to be \emph{semifinite}. Notably, semifiniteness implies (and in fact is equivalent to) the existence of a (w.ot.) dense set of positive operators with finite trace. Finally, for any nonzero $a \in \mathcal{A}_R$, at least one of the functions $f$ and $g_k(s,s')$ must be nonzero. It follows that
\begin{align}
\Tr[a^\dagger a] = \int ds \rho(s) |f(s)|^2 + \sum_j \int ds ds'\left|g_j(s,s')\right|^2 \frac{\rho(s') \rho(s) \Gamma(\Delta_j \pm i s' \pm i s)}{\Gamma(2 \Delta_j)} > 0
\end{align}
for any $a \neq 0$ and so the trace is \emph{faithful}.

The trace $\Tr$ defines a Hilbert-Schmidt inner product $\braket{a|b} = \Tr(a^\dagger b)$ on the Hilbert space $\mathcal{H}_{\mathcal{A}_R}$ containing a state $\ket{a}$ for each operator $a \in \mathcal{A}_R$ with $\Tr(a^\dagger a) < \infty$. There is an isometric map $V: \mathcal{H}_{\mathcal{A}_R} \to \mathcal{H}_{JT}$ defined by
\begin{align}
V \ket{a} = \int ds \sqrt{\rho(s)} a \ket{s}. 
\end{align}
It follows from \eqref{eq:baction} that any state in $\mathcal{H}_{JT}$ can be written as $V\ket{b}$ for some $b \in \mathcal{A}_R$. The map $V$ therefore defines a unitary isomorphism between $\mathcal{H}_{\mathcal{A}_R}$ and $\mathcal{H}_{JT}$.

We can use this isomorphism to identify the commutant of the algebra $\mathcal{A}_R$. An algebra $\mathcal{A}$ acts on itself in two different ways: by left multiplication $a_L(a') = a a'$ and by right multiplication $a_R(a') = a' a^\dagger$. If we identify a von Neumann algebra $\mathcal{A}$ with a Hilbert space $\mathcal{H}_\mathcal{A}$ via the Hilbert-Schmidt inner product defined using a faithful normal semifinite trace $\Tr$, it is a standard result that the left and right actions of $\mathcal{A}$ act as commutants on $\mathcal{H}_\mathcal{A}$. Under the identification $V$, the left action becomes the usual action of the right-boundary algebra $\mathcal{A}_R$. What about the right action of $\mathcal{A}_R$ on itself? We know from \eqref{eq:js's=PhiL} that for any $a \in \mathcal{A}_R$
\begin{align}
\nonumber\int ds \sqrt{\rho(s)} f(H_L) \Phi^i_L g(H_L) a \ket{s} &= a \int dsds'\, f(s'^2)g(s^2) \sqrt{\frac{\rho(s)\rho(s') \Gamma(\Delta_i \pm i s' \pm i s)}{\Gamma(2 \Delta_i)}} \ket{\Phi^i,s',s}
\\&= \int ds' \sqrt{\rho(s')}\, a\, g(H_R) \Phi^i_L f(H_R)  \ket{s'}.
\end{align}
And of course
\begin{align}
\int ds \sqrt{\rho(s)} f(H_L)  a \ket{s} = \int ds \sqrt{\rho(s)}  a f(H_L)  \ket{s} = \int ds \sqrt{\rho(s)}  a f(H_R)  \ket{s}.
\end{align}
It follows that we can identify the right action of $\mathcal{A}_R$ on itself with the left-boundary algebra $\mathcal{A}_L$ by identifying $H_L$ with the right action of $H_R$ on $\mathcal{A}_R$ and $\Phi^i_L$ with the right action of $\Phi^i_R$ on $\mathcal{A}_R$. An immediate consequence of this identification is that the two boundary algebras $\mathcal{A}_L$ and $\mathcal{A}_R$ are commutants.

Next, we want to show that the algebra $\mathcal{A}_R$ is a factor, which can be done using an elegant argument from \cite{kolchmeyer2023neumann}. Since we have already shown that $\mathcal{A}_L$ and  $\mathcal{A}_R$ are commutants, it suffices to show that any operator in their intersection must be a multiple of the identity. In particular, any operator in the centre $\mathcal{A}_L \cap \mathcal{A}_R$ must commute with both $H_L$ and $H_R$. It must therefore preserve the matter vacuum sector of the Hilbert space, since for a normalisable state $\ket{\Psi_{JT}}$ we have
\begin{align}
(H_L - H_R) \ket{\Psi_{JT}} = 0 ~~~~\Longleftrightarrow~~~~ \ \Pi_{\Omega} \ket{\Psi_{JT}} =\ket{\Psi_{JT}},
\end{align} 
where $\Pi_{\Omega} = \ket{\Omega_{\rm matt}} \bra{\Omega_{\rm matt}}$. But, since any operator $a \in \mathcal{A}_R$ can be written in the form \eqref{eq:bexpand}, the only operators in $\mathcal{A}_R$ that preserve the matter vacuum sector are operators $f(H_R)$. Now $H_R - H_L$ has purely continuous spectrum outside of the matter vacuum sector. So
\begin{align}
[f(H_R), \Phi_R] \Pi_\Omega = (f(H_R) - f(H_L)) \Phi_R \Pi_\Omega
\end{align}
is nonzero unless $f$ is constant. It follows that the algebra $\mathcal{A}_R$ is indeed a factor.

The existence of a trace means that the algebra $\mathcal{A}_R$ is either Type I or II. Because there exist projectors with arbitrarily small trace (e.g. the projector onto energies $E_R \in [E, E + \delta]$ for arbitrarily small $\delta$) the algebra cannot be Type I. Finally because the trace of the identity is infinite, $\mathcal{A}_R$ must be Type II$_\infty$. This completes our derivation of the full algebraic structure for $\mathcal{A}_L$ and $\mathcal{A}_R$ found in \cite{penington2023algebras, kolchmeyer2023neumann}, without any reference to Euclidean gravitational path integrals.

It is worth emphasizing the contrast here with pure JT gravity.  As a canonically quantised theory, pure JT gravity has no concept of a density of states associated to either the right or left boundary. The natural object associated to the boundaries is the commutative algebra $\mathcal{A}_L = \mathcal{A}_R$. From an algebraic perspective, any state on this algebra is a reasonable choice of trace; each choice corresponds to a different density of states. Of course, it is natural, given our knowledge of other theories of gravity, to choose a trace that matches the Euclidean path integral ``density of states''. But the Euclidean partition function does not play any direct role in canonically quantised pure-JT gravity and so this choice would need to be explicitly specified as part of the definition of the theory. 

In contrast, the uniqueness of the faithful, normal, semifinite trace on a von Neumann factor means that JT gravity plus matter has a unique density of states $\rho(s)$ (up to an arbitrary multiplicative factor) and entanglement entropies (up to an arbitrary additive constant). And those densities of states and entropies match ones found using Euclidean path integral calculations, since, as discussed earlier in this section, the trace $\Tr$ can found using a Euclidean path integral on a disc. As a result, algebraic entropies can be computed using a Euclidean gravitational replica trick and, for semiclassical states, are given by the holographic QES prescription \cite{Ryu:2006ef, Hubeny:2007xt, Lewkowycz:2013nqa, Faulkner:2013ana, Engelhardt:2014gca}, which includes a Bekenstein-Hawking contribution in addition to the entropy of matter fields.

Moreover, because the JT gravity boundary algebras become factors when matter is added, while remaining commutants, any operator acting on the JT gravity Hilbert space can be written (in principle) as a limit, in the strong operator topology, of sums of products of left- and right-boundary operators. This includes e.g. the operator $\ell = -2\chi$ that describes the length of a geodesic between the two boundaries and that could not be measured locally at the boundaries in pure JT gravity. By considering sufficiently many matter correlation functions, in JT gravity with matter one can determine the length of the wormhole to arbitrarily high accuracy, without otherwise disturbing the state. 

In fact, the boundary algebras of JT gravity satisfy all the usual axioms of an algebraic quantum field theory, including e.g. the timeslice axiom and Haag duality. The only difference between the boundary algebras in JT gravity plus matter and the CFT boundary algebras in a conventional holographic theory is that the boundary algebras in JT gravity with matter are Type II and hence the entanglement between the two boundaries is infinite. In a ``true'' quantum field theory, each connected component of a Cauchy slice is associated with Hilbert space tensor product subfactors and hence with a Type I algebra of observables. In contrast, the left and right boundaries in JT gravity act like complementary, but touching, causal diamonds that are described by commutant von Neumann factors but not by a Hilbert space tensor product.

The reason that canonically quantised JT gravity has Type II rather than Type I boundary algebras is that the topology of the Lorentzian spacetime is fixed. In path integral constructions of pure JT gravity, there is a duality between a sum over bulk spacetime topologies and an ensemble of boundary Hamiltonians   \cite{Saad:2019lba} with finite densities of states. Topologies in the sum are weighted by  $e^{\chi S_0}$ where $\chi$ is the Euler characteristic of the spacetime and $S_0$ is the coefficient of the topological Gauss-Bonnet term in the JT action. To obtain a limit where the topology is fixed and only the disk contributes one needs to take the limit $S_0 \to \infty$. But $S_0$ gives a universal additive contribution to the Bekenstein-Hawking entropy, so the Euclidean density of states diverges  exactly as expected for a Type II algebra.  

Nevertheless, we have seen that JT gravity with matter is a theory where all the fundamental degrees of freedom are at the boundary, even though semiclassically excitations can be localised anywhere in the bulk. It is also a theory where entanglement entropy is given by a holographic entropy prescription. It therefore seems entirely reasonable to call it a holographic theory, even though it was initially defined via a bulk construction.

\section{Super-JT gravity with matter} \label{sec:superJT}
We now turn to $\mathcal{N} = 2$ super-JT gravity coupled to supersymmetric matter. We closely follow \cite{Lin:2022rzw, Lin:2022zxd} by defining the theory using a generalisation of the group-theoretic method used in Section \ref{sec:group}. For other work on supersymmetric versions of JT gravity and the related supersymmetric SYK models, see e.g. \cite{Fu:2016vas, stanford2017fermionic, Stanford:2019vob, Heydeman:2020hhw, Fan:2021wsb, Turiaci:2023jfa, Boruch:2023bte}. We then turn to study the space of supersymmetric ground states and show that there exists a single normalisable state with exactly zero energy at both boundaries within each representation of the matter theory. 

\subsection{The gauge supergroup and boundary superparticles}

The $\mathcal{N} = 2$ supersymmetric generalization of the Lie group $\hSL(2,\R)$ is the supergroup $\hSU(1,1|1)$. This has four real bosonic generators (the three generators $\mathcal{J}_3$, $\mathcal{J}_\pm = \mathcal{J}_1 \pm \mathcal{J}_2$ of $\hSL(2,\R)$ and a $U(1)$ R-charge $\mathcal{J}$) along with two complex fermionic generators $\mathcal{G}_\pm$ with adjoints $\mathcal{G}_\pm^\dagger$. We have
\begin{align}\label{su111algebra}
\nonumber[\mathcal{J}_3, \mathcal{J}_\pm] &= \pm i \mathcal{J}_\pm, ~~~~ [\mathcal{J}_+, \mathcal{J}_-] = - 2i \mathcal{J}_3, ~~~~[\mathcal{J}, \mathcal{J}_3] = [\mathcal{J}, \mathcal{J}_\pm] = 0, \\
[\mathcal{J}_3, \mathcal{G}_\pm] &= \pm \frac{i}{2} \mathcal{G}_\pm, ~~~~[\mathcal{J}, \mathcal{G}_\pm] = \frac{1}{2} \mathcal{G}_\pm, ~~~~ [\mathcal{J}_\mp, \mathcal{G}_\pm] = \pm i \mathcal{G}_\mp, ~~~~[\mathcal{J}_\pm, \mathcal{G}_\pm] = 0,\\
\nonumber \{\mathcal{G}_\pm, \mathcal{G}_\pm^\dagger\} &= \mathcal{J}_\pm, ~~~~ \{\mathcal{G}_+, \mathcal{G}_-^\dagger\} = \mathcal{J}_3 - i \mathcal{J}, ~~~~ \{\mathcal{G}_\pm,\mathcal{G}_\pm\} = \{\mathcal{G}_+,\mathcal{G}_-\} = 0.
\end{align}
We can parameterise $\hSU(1,1|1)$ by four real $c$-number coordinates $(\tau, \rho, \eta, a)$ and two complex Grassmann coordinates $\theta_+, \theta_-$ as
\begin{align}
g = e^{i \tau \mathcal{J}_+} e^{\theta_+ \mathcal{G}_+ + \bar\theta_+ \mathcal{G}_+^\dagger} e^{i \rho \mathcal{J}_3}  e^{\theta_- \mathcal{G}_- + \bar\theta_- \mathcal{G}_-^\dagger} e^{i \kappa \mathcal{J}_-} e^{i a \mathcal{J}}
\end{align}
As with the parameterisation \eqref{eq:param2}, these coordinates only cover part of $\hSU(1,1|1)$, but they cover the only part that will be relevant after all our gauge choices have been made. The parameter $a$ is normally assumed to be periodic, so that the bosonic part of $\hSU(1,1|1)$ is  $\hSL(2,\R) \times U(1)$ and R-charge is quantised. For consistency with the charges  $\pm 1/2$ of the fermionic generators, $a$ must have period $4\pi N$ for some positive integer $N$. The choice of $N$ is a free parameter of the super-JT gravity theory that must be consistent with the set of R charges appearing in a matter theory to which it is coupled. 

The left action of $\hSU(1,1|1)$ on wavefunctions $\mathbf{\Psi}(g)$ in the parameterisation \eqref{su111algebra} is
\begin{align}
\mathcal{L}(\mathcal{J}_+) =& - p_{\tau}\\
\mathcal{L}(\mathcal{G}_+) =& - \partial_{\theta_+} + \frac{1}{2}\bar{\theta}_+ p_{\tau}\\
\mathcal{L}(\mathcal{G}_+^\dagger) =& - \partial_{\bar\theta_+} + \frac{1}{2}\theta_+ p_{\tau}\\
\mathcal{L}(\mathcal{J}_3) =& + \tau p_{\tau} - \frac{i}{2}\theta_+\partial_{\theta_+} -\frac{i}{2}\bar\theta_+\partial_{\bar\theta_+} - p_{\rho}\\
\mathcal{L}(\mathcal{J}) =& -p_{a} - \frac{1}{2}\theta_+\partial_{\theta_+} + \frac{1}{2} \bar\theta_+ \partial_{\bar\theta_+} - \frac{1}{2}\theta_-\partial_{\theta_-} + \frac{1}{2} \bar\theta_- \partial_{\bar\theta_-}\\
\mathcal{L}(\mathcal{G}_-) =& -e^{-\rho/2} \partial_{\theta_-}+ \frac{1}{2}e^{-\rho/2}\bar\theta_- p_{\kappa} + \tau \partial_{\theta_+} - \frac{1}{2} \tau \bar\theta_+ p_{\tau} + \bar\theta_+ p_{\rho} - i \bar\theta_+ \mathcal{L}(\mathcal{J})\\
\mathcal{L}(\mathcal{G}_-^\dagger) =& -e^{-\rho/2} \partial_{\bar\theta_-}+ \frac{1}{2}e^{-\rho/2}\theta_- p_{\kappa} + \tau \partial_{\bar\theta_+} - \frac{1}{2} \tau \theta_+ p_{\tau} + \theta_+ p_{\rho} + i \theta_+ \mathcal{L}(\mathcal{J})\\
\mathcal{L}(\mathcal{J}_-) = &-e^{-\rho} p_{\kappa} - \tau^2 p_{\tau}  +i \tau \theta_+\partial_{\theta_+} +i \tau \bar\theta_+\partial_{\bar\theta_+} + 2 \tau p_{\rho} + \theta_+ \bar\theta_+ \mathcal{L}(\mathcal{J})\\&~~~~~ -e^{-\rho/2} \left[i\theta_+ \partial_{\theta_-} +i \bar\theta_+ \partial_{\bar\theta_-}  -\frac{i}{2} \theta_+ \bar\theta_- p_{\kappa} - \frac{i}{2} \bar\theta_+ \theta_- p_{\kappa}  \right]\nonumber
\end{align}
The Haar measure is
\begin{align} \label{eq:superHaar}
dg = d\tau d\theta_+ d\bar\theta_+ d\rho d\theta_- d\bar\theta_- d\kappa da.
\end{align}
Note in particular that  the left action $\mathcal{L}(e^{i\rho_0 \mathcal{J}_3})$ maps $\tau \to e^{\rho_0} \tau$, $\theta_+ \to e^{\rho_0/2} \theta_+$ and $\rho \to \rho - \rho_0$. Because Grassmann numbers $\theta$ satisfy $d(k\theta) = k^{-1} d\theta$ for any $c$-number $k>0$, it follows that, unlike the Haar measure \eqref{eq:param2} for the bosonic theory, \eqref{eq:superHaar} has no explicit dependence on $\rho$.

In the bosonic version of JT gravity, we found the dynamics of the right boundary particle by imposing the gauge constraint of $\mathcal{R}(\mathcal{J}_-) + 1 = 0$ . One way to think about that constraint, which admittedly would have been unnecessarily fancy when studying the bosonic theory, is that we gauged a diagonal action $\mathcal{K}$ of  $\R$ on a tensor product of $L^2(\hSL(2,\R))$ with a one-dimensional Hilbert space, where the action on $L^2(\hSL(2,\R))$ was generated by $\mathcal{R}(\mathcal{J}_-)$ and the action on the one-dimensional Hilbert space was multiplication by a phase.

The supersymmetric generalization of the action generated by $\mathcal{R}(\mathcal{J}_-)$ is the right action on $\hSU(1,1|1)$ of the subgroup $G_- \subseteq \hSU(1,1|1)$ generated by $\mathcal{J}_-, \mathcal{G}_-, \mathcal{G}_-^\dagger$. In the parameterisation above we have
\begin{align}
\mathcal{R}(\mathcal{J}_-) &= p_{\kappa}\\
\mathcal{R}(\mathcal{G}_-) &= e^{i a/2}\left[\partial_{\theta_-} + \frac{1}{2}  \bar \theta_- p_{\kappa}\right]\\
\mathcal{R}(\mathcal{G}_-^\dagger) &= e^{-i a/2}\left[\partial_{\theta_-} + \frac{1}{2}  \bar \theta_- p_{\kappa}\right].
\end{align}
Abstractly, $G_-$ is a non-Abelian supergroup that is isomorphic to the universal cover of $SU(1|1)$. It is perhaps most familiar as the symmetry algebra of supersymmetric quantum mechanics with Hamiltonian $H_-$ and a single complex supercharge $\mathcal{Q}_-$ which satisfy
\begin{align}
\{\mathcal{Q}_-,\mathcal{Q}_-\} = \{\mathcal{Q}_-^\dagger, \mathcal{Q}^\dagger_-\} = 0 ~~~~~\{\mathcal{Q}_-,\mathcal{Q}^\dagger_-\} = \mathcal{H}_- ~~~[\mathcal{H}_-,\mathcal{Q}_-] = [\mathcal{H}_-,\mathcal{Q}^\dagger_-]  =0
\end{align}
 The supergroup $G_-$ has no nontrivial one-dimensional representations. Instead the natural supersymmetric analogue of the action of $\R$ by multiplication by a phase is a single fermionic mode, on which $\mathcal{Q}_-$ and $\mathcal{Q}^\dagger_-$ act respectively as the annihilation operator $\psi$ and the creation operator  $\psi^\dagger$, while the Hamiltonian $\mathcal{H}_-$ is equal to the identity (so that $e^{-i\mathcal{H}_- t}$ acts as multiplication by a phase). We therefore gauge the diagonal action $\mathcal{K}$ where
\begin{align}\label{eq:rightsuperparticlegauge}
\mathcal{K} (\mathcal{Q}_-) = \mathcal{R}(\mathcal{G}_-) + \psi &=e^{ia/2} \left[\partial_{\theta_-} + \frac{1}{2} \bar\theta_- p_{\kappa}\right] + \psi\\ 
\mathcal{K}(\mathcal{Q}^\dagger_-) = \mathcal{R}(\mathcal{G}_-^\dagger) + \psi^\dagger  &=e^{-ia/2} \left[\partial_{\bar\theta_-} + \frac{1}{2} \theta_- p_\kappa \right] + \psi^\dagger \\ 
\mathcal{K}(\mathcal{H}_-) = \mathcal{R}(\mathcal{J}_-) + 1 &= p_\kappa + 1.
\end{align}
As usual, we implement these gauge constraints using the method of coinvariants. Let $\ket{\Omega_R} \in \mathcal{H}_{\rm fermion}$ be the fermion vacuum (so that $\psi \ket{\Omega_R} = 0$ and let
\begin{align}
\mathbf{\Psi}_R(g) = \mathbf{\Psi}_{R,0}(\tau,\rho,\kappa,a,\theta_+,\theta_-,\bar\theta_+,\bar\theta_-) \ket{\Omega_R} + \mathbf{\Psi}_{R,1}(\tau,\rho,\kappa,a,\theta_+,\theta_-,\bar\theta_+,\bar\theta_-) \psi^\dagger \ket{\Omega_R}
\end{align}
be a wavefunction on $\hSU(1,1|1)$ valued in the fermion Hilbert space $\mathcal{H}_{\rm fermion}$. Up to gauge equivalence, we can always write
\begin{align}\label{eq:superwave1stgauge}
\mathbf{\Psi}_R(g) \sim \delta(\kappa) \,\theta_- \bar\theta_-\,  \Psi_R(\tau,\rho,a,\theta_+,\bar\theta_+)
\end{align}
where $\Psi_R$ is again valued in the fermion Hilbert space. Note that for a Grassmann number $\theta$, the Dirac delta function satisfies $\delta(\theta) = \theta$; the form of the wavefunction given in \eqref{eq:superwave1stgauge} therefore fixes $\kappa = \theta_- = \bar\theta_- = 0$.

The coinvariant (pseudo-)inner product is
\begin{align}
\braket{\mathbf{\Phi}_R,\mathbf{\Psi}_R} = \int dh (\mathbf{\Phi}_R^*, \mathcal{K}(h) \mathbf{\Psi}_R),
\end{align}
where $(\cdot,\cdot)$ is the pseudo-inner product on $\mathcal{H}_{\rm fermion}$-valued wavefunctions on $\hSU(1,1|1)$ (induced by the Haar measure $dg$) and $dh$ is the Haar measure on $G_-$. Note that $(\cdot,\cdot)$ is not positive semidefinite because of the presence of Grassmann numbers. As a simple example, given a Grassmann number $\theta$ the pseudo-inner product
\begin{align}
\int \, d \theta (1 - \theta)^2 = -2
\end{align}
of $(1-\theta)$ with itself is negative. 

If we parameterise $h \in G_-$ by $h = e^{it_h \mathcal{H}_-} e^{\theta_h \mathcal{Q}_- + \bar\theta_h \mathcal{Q}_-^\dagger}$, the Haar measure on $G_-$ is
\begin{align}
dh = dt_h d\theta_h d\bar\theta_h.
\end{align}
For wavefunctions of the form \eqref{eq:superwave1stgauge}, we therefore have
\begin{align}
\braket{\mathbf{\Phi}_R,\mathbf{\Psi}_R} &= \int dt_h (\mathbf{\Phi}_R, e^{it_h \mathcal{K}(\mathcal{H})} \mathcal{K}(\mathcal{Q}) \mathcal{K}(\mathcal{Q}^\dagger) \mathbf{\Psi}_R)\\
&= \int d\tau d\theta_+ d \bar\theta_+ d\rho da \,\,\Phi_R^*(\tau,\rho,a,\theta_+,\bar\theta_+)\Psi_R(\tau,\rho,a,\theta_+,\bar\theta_+)
\end{align}
where in the second line we are implicitly taking an inner product on the fermion Hilbert space. Importantly, the continued presence of the Grassmann numbers $\theta_+,\bar\theta_+$ means that even this coinvariant pseudo-inner product is not positive semidefinite.

Up to gauge equivalence, the $\hSU(1,1|1)$ generators act on the wavefunctions \eqref{eq:superwave1stgauge} as
\begin{align} \label{eq:superparticleSU111}
J_+^R =& - p_{\tau} \\
G_+^R =& - \partial_{\theta_+} + \frac{1}{2}\bar\theta_+ p_{\tau}\\
G_+^{R\dagger} =& - \partial_{\bar\theta_+} + \frac{1}{2}\theta_+ p_{\tau}\\
J_3^R =& + \tau p_{\tau} - \frac{i}{2}\theta_+\partial_{\theta_+} -\frac{i}{2}\bar\theta_+\partial_{\bar\theta_L+} - p_{\rho}\\
J^R =& -p_{a} - \frac{1}{2}\theta_+\partial_{\theta_+} + \frac{1}{2} \bar\theta_+ \partial_{\bar\theta_+} \\
G_-^R =& e^{(-\rho - ia)/2} \psi +\tau \partial_{\theta_+} - \frac{1}{2} \tau \bar\theta_+ p_{\tau} + \bar\theta_+ p_{\rho} - i \bar\theta_+ J^R\\
{G_-^R}^\dagger =& e^{(-\rho + ia)/2} \psi^\dagger + \tau \partial_{\bar\theta_+} - \frac{1}{2} \tau \theta_+ p_{\tau} + \theta_+ p_{\rho} + i \theta_+ J^R\\
J_-^R = &e^{-\rho} - \tau^2 p_{\tau}  +i \tau \theta_+\partial_{\theta_+} +i \tau \bar\theta_+\partial_{\bar\theta_+} + 2 \tau p_{\rho} - \theta_+ \bar\theta_+ p_a\\&~~~~~ +e^{-\rho/2} \left[ie^{-ia/2}\theta_+ \psi +i e^{ia/2}\bar\theta_+ \psi^\dagger  \right]\nonumber.
\end{align}
Finally the Hamiltonian is the Casimir
\begin{align}\label{eq:superHam}
\mathbf{H}_R &= {J^R}^2 +  {J_3^R}^2 - \frac{1}{2} \left\{J_+^R, J_-^R\right\} - \frac{i}{2} \left[G_+^R,{G_-^R}^\dagger\right] + \frac{i}{2} \left[G_-^R,{G_+^R}^\dagger\right]\\
&= p_a^2 +  p_\rho^2 + e^{-\rho} p_\tau - i e^{(-\rho-ia)/2} \psi\left[\partial_{\bar\theta_+} + \frac{1}{2} \theta_+ p_{\tau}\right] - i e^{(-\rho+ia)/2} \psi^\dagger \left[\partial_{\theta_+} + \frac{1}{2} \bar\theta_+ p_{\tau}\right]
\end{align}
It can be proven that \eqref{eq:superHam} is equivalent to the $\mathcal{N}=2$ super-Schwarzian description of super-JT gravity; see \cite{Lin:2022rzw, Lin:2022zxd} for details. Although it is not obvious from the group-theoretic construction, the Hamiltonian $H_R$ inherits from the super-Schwarzian a worldline supersymmetry. Specifically, we can write 
\begin{align}
H_R = \{Q_R, Q_R^\dagger\}
\end{align}
 with the nilpotent complex supercharge $Q_R$ defined by
\begin{align} \label{eq:QR}
Q_R =& \psi \left(p_\rho - i p_a\right) - e^{(-\rho+ia)/2}\left(\partial_{\theta_+} + \frac{1}{2} \bar\theta_+ p_\tau\right) \\
Q_R^\dagger =& \psi^\dagger \left(p_\rho + i p_a\right) - e^{(-\rho-ia)/2}\left(\partial_{\bar\theta_+} + \frac{1}{2} \theta_+ p_\tau\right).\nonumber
\end{align}
The supercharge $Q_R$ has charge one with respect to the R-charge
\begin{align}
J_R = 2p_a + \frac{1}{2}[\psi ,\psi^\dagger]
\end{align}
It is easy to verify that $Q_R, Q_R^\dagger, J_R$ all commute with the superparticle $\hSU(1,1|1)$ action \eqref{eq:superparticleSU111}. One way to understand this is that they can be built out of a particular combination of products of right $\hSU(1,1|1)$ generators and operators on $\mathcal{H}_{\rm fermion}$ that commute with the gauge constraint \eqref{eq:rightsuperparticlegauge}; again see  \cite{Lin:2022rzw, Lin:2022zxd} for a more detailed discussion. 

The analysis of the left-boundary superparticle is almost identical to the right so we shall describe it as briefly as possible. We parameterise $\hSU(1,1|1)$ as
\begin{align}
g = e^{i \tau' \mathcal{J}_-} e^{\theta_-'\mathcal{G}_- + \bar\theta_-' \mathcal{G}_-^\dagger} e^{-i \rho' \mathcal{J}_3}  e^{\theta_+'\mathcal{G}_+ +  \bar\theta_+' \mathcal{G}_+^\dagger}e^{i\kappa' \mathcal{J}_+} e^{i a' \mathcal{J}}.
\end{align}
The left action $\mathcal{L}$  is
\begin{align}
\mathcal{L}(\mathcal{J}_-) =& - p_{\tau'}\\
\mathcal{L}(\mathcal{G}_-) =& - \partial_{\theta_-'} + \frac{1}{2}\bar\theta_-' p_{\tau'}\\
\mathcal{L}(\mathcal{G}_-^\dagger) =& - \partial_{\bar\theta_-'} + \frac{1}{2}\theta_-' p_{\tau'}\\
\mathcal{L}(\mathcal{J}_3) =& - \tau' p_{\tau'} + \frac{i}{2}\theta_-'\partial_{\theta_-'} +\frac{i}{2}\bar\theta_-'\partial_{\bar\theta_-'} + p_{\rho'}\\
\mathcal{L}(\mathcal{J}) =& -p_{a'} - \frac{1}{2}\theta_-'\partial_{\theta_-'} + \frac{1}{2} \bar\theta_-' \partial_{\bar\theta_-'} - \frac{1}{2}\theta_+'\partial_{\theta_+'} + \frac{1}{2} \bar\theta_+' \partial_{\bar\theta_+'}\\
\mathcal{L}(\mathcal{G}_+) =&  -e^{-\rho'/2} \partial_{\theta_+'}+ e^{-\rho'/2}\frac{1}{2}\bar\theta_+' p_{\kappa'} - \tau' \partial_{\theta_-'} + \frac{1}{2} \tau' \bar\theta_-' p_{\tau'} - \bar\theta_-' p_{\rho'} + i \bar\theta_-' \mathcal{L}(\mathcal{J})\\
\mathcal{L}(\mathcal{G}_+^\dagger) =& -e^{-\rho'/2} \partial_{\bar\theta_+'}+ e^{-\rho'/2}\frac{1}{2}\theta_+' p_{\kappa'} - \tau' \partial_{\bar\theta_-'} + \frac{1}{2} \tau' \theta_-' p_{\tau'} - \theta_-' p_{\rho'} - i \theta_-' \mathcal{L}(\mathcal{J})\\
\mathcal{L}(\mathcal{J}_+) = &-e^{-\rho'} p_{\kappa'} - \tau'^2 p_{\tau'}  +i \tau' \theta_-'\partial_{\theta_-'} +i \tau' \bar\theta_-'\partial_{\bar\theta_-'} + 2 \tau' p_{\rho'} + \theta_-' \bar\theta_-' \mathcal{L}(\mathcal{J})\\&~~~~~ +e^{-\rho'/2} \left[i\theta_-' \partial_{\theta_+'} +i \bar\theta_-' \partial_{\bar\theta_+'}  -\frac{i}{2} \theta_-' \bar\theta_+' p_{\kappa'} - \frac{i}{2} \bar\theta_-' \theta_+' p_{\kappa'}  \right]\nonumber
\end{align}
We again add the Hilbert space $\mathcal{H}_{\rm fermion}'$ of a single fermion with annihilation/creation operators $\psi', \psi'^\dagger$ and gauge the diagonal action $\mathcal{K}$ of the subgroup $G_+ \subseteq \hSU(1,1|1)$ generated by $\mathcal{G}_+, \mathcal{G}_+^\dagger, \mathcal{J}_+$ defined by
\begin{align}
\mathcal{K} (\mathcal{Q}_+) = \mathcal{R}(\mathcal{G}_+) + \psi' &=e^{ia'/2} \left[\partial_{\theta_+'} + \frac{1}{2} \bar\theta_+' p_{\kappa'}\right] + \psi'\\ 
\mathcal{K}(\mathcal{Q}^\dagger_+) = \mathcal{R}(\mathcal{G}_+^\dagger) + \psi'^\dagger  &=e^{-ia/2} \left[\partial_{\bar\theta_+'} + \frac{1}{2} \theta_+' p_{\kappa'} \right] + \psi'^\dagger \\ 
\mathcal{K}(\mathcal{H}_+) = \mathcal{R}(\mathcal{J}_+) + 1 &= p_{\kappa'} + 1.
\end{align}
To do so, we gauge fix to $\kappa' = \theta_+' = \bar\theta_+' = 0$ by working with wavefunctions of the form
\begin{align}\label{eq:superwave1stgaugeL}
\mathbf{\Psi}_L(g) \sim \delta(\kappa') \,\theta_+' \bar\theta_+'\,  \Psi_L(\tau',\rho',a',\theta_-',\bar\theta_-').
\end{align}
where $\mathbf{\Psi}_L$ and $\Psi_L$ are valued in $\mathcal{H}_{\rm fermion}'$. After gauging, the coinvariant pseudo-inner product is
\begin{align}
\braket{\mathbf{\Phi}_L,\mathbf{\Psi}_L} = \int d\tau' d\theta_-' d \bar\theta_-' d\rho' da' \Phi_L^*(\tau',\rho',a',\theta_-',\bar\theta_-')\Psi_L(\tau',\rho',a',\theta_-',\bar\theta_-')
\end{align}
where on the right hand side we have implicitly taken an inner product on $\mathcal{H}_{\rm fermion}'$. Up to gauge equivalences, the action $\mathcal{L}$ acts on $\Psi_L$ as
\begin{align}
J_-^L =& - p_{\tau'} \\
G_-^L =& - \partial_{\theta_-'} + \frac{1}{2}\bar\theta_-' p_{\tau'}\\
{G_-^L}^\dagger =& - \partial_{\bar\theta_-'} + \frac{1}{2}\theta_-' p_{\tau'}\\
J_3^L =& - \tau' p_{\tau'} + \frac{i}{2}\theta_-'\partial_{\theta_-'} +\frac{i}{2}\bar\theta_-'\partial_{\bar\theta_-'} + p_{\rho'}\\
J^L =& -p_{a'} - \frac{1}{2}\theta_-'\partial_{\theta_-'} + \frac{1}{2} \bar\theta_-' \partial_{\bar\theta_-'} \\
G_+^L =& e^{(-\rho' - ia')/2} \psi' - \tau' \partial_{\theta_-'} + \frac{1}{2} \tau' \bar\theta_-' p_{\tau'} - \bar\theta_-' p_{\rho'} + i \bar\theta_-' J^L\\
{G_+^L}^\dagger =& e^{(-\rho' + ia')/2} \psi'^\dagger - \tau' \partial_{\bar\theta_-'} + \frac{1}{2} \tau' \theta_-' p_{\tau'} - \theta_-' p_{\rho'} - i \theta_-' J^L\\
J_+^L = &e^{-\rho'}  - \tau'^2 p_{\tau'}  +i \tau' \theta_-'\partial_{\theta_-'} +i \tau' \bar\theta_-'\partial_{\bar\theta_-'} + 2 \tau' p_{\rho'} + \theta_-' \bar\theta_-' J^L\\&~~~~~ -e^{-\rho'/2} \left[ie^{-ia'/2}\theta_-' \psi' +i e^{ia'/2}\bar\theta_-' \psi'^\dagger  \right]\nonumber
\end{align}
The Hamiltonian is
\begin{align}\label{eq:superHamL}
\mathbf{H}_L &=  {J^L}^2 + {J_3^L}^2 - \frac{1}{2} \left\{J_+^L, J_-^L\right\} - \frac{i}{2} \left[G_+^L,{G_-^L}^\dagger\right] + \frac{i}{2} \left[G_-^L,{G_+^L}^\dagger\right]\\
&= p_{a'}^2 +  p_{\rho'}^2 +  e^{-\rho'} p_{\tau'} + i e^{(-\rho'-ia')/2} \psi'\left[\partial_{\bar\theta_-'} + \frac{1}{2} \theta_-' p_{\tau'}\right] + i e^{(-\rho'+ia')/2} \psi'^\dagger \left[\partial_{\theta_-'} + \frac{1}{2} \bar\theta_-' p_{\tau'}\right]\\
&=  \{Q_L, Q_L^\dagger\}
\end{align}
where
\begin{align}\label{eq:QLunphys}
 Q_L =& \psi' \left(p_{\rho'} - i p_{a'}\right) + e^{(-\rho'+ia')/2}\left(\partial_{\theta_-'} + \frac{1}{2} \bar\theta_-' p_{\tau'}\right) \\
 Q_L^\dagger =& \psi'^\dagger \left(p_{\rho'} + i p_{a'}\right) + e^{(-\rho'+ia')/2}\left(\partial_{\bar\theta_-'} + \frac{1}{2} \theta_-' p_{\tau'}\right)\nonumber.
\end{align}
The supercharge $Q_L$ has charge one with respect to the R-charge
\begin{align}
J_L = 2p_{a'} + \frac{1}{2}[\psi' ,\psi'^\dagger].
\end{align}

\subsection{The physical Hilbert space, boundary supercharges and Hamiltonians}
As in the bosonic theory, wavefunctions in super-JT gravity are built out of a product of the two boundary superparticles together with the bulk matter theory, which is an $\mathcal{N} = 2$ supersymmetric quantum field theory in a fixed $\AdS_2$ background with Hilbert space $\mathcal{H}_{\rm matt}$. The $\hSU(1,1|1)$ matter charges will be denoted using lower-case letters . Super-JT wavefunctions can therefore be written as $\mathbf{\Psi}_{JT}(\tau,\rho,a,\theta_+,\bar\theta_+, \tau',\rho',a',\theta_-',\bar\theta_-')$ with $\mathbf{\Psi}_{JT}$ valued in the super-tensor product $\mathcal{H}_{\rm matt} \otimes \mathcal{H}_{\rm fermion} \otimes \mathcal{H}_{\rm fermion}'$.

All such wavefunctions (with spacelike-separated boundary particles) are gauge equivalent under the diagonal $\hSU(1,1|1)$ action\footnote{Here $\mathcal{J}_i$ represents an arbitrary bosonic or fermionic generator of $SU(1,1,|1)$.} $\mathcal{D}(\mathcal{J}_i) = J_i^L + j_i + J_i^R$ to wavefunctions of the form
\begin{align}\label{eq:superJTgaugefixwavefunction}
\mathbf{\Psi}_{JT} = \delta(\tau) \delta(\tau') \delta(\rho-\rho') \delta(a+a')\theta_+\bar\theta_+\theta_-'\bar\theta_-' \Psi_{JT}(\rho, a)
\end{align}
where $\Psi_{JT}$ is again valued in $\mathcal{H}_{\rm matt} \otimes \mathcal{H}_{\rm fermion} \otimes \mathcal{H}_{\rm fermion}'$. Recall that $a$ and $a'$ were originally defined to be periodic with period $4\pi N$ for some integer $N$. There is a residual gauge invariance $\exp(i 2\pi N \mathcal{D}(\mathcal{J})$ that preserves the space of wavefunctions of the form \eqref{eq:superJTgaugefixwavefunction} while shifting $a \to a + 2\pi N$. This can be used to set $0 \leq a < 2\pi N$, which leads to the physical parameter $a$ appearing in $\Psi_{JT}(\rho, a)$ to have period $2 \pi N$. 

The coinvariant inner product is
\begin{align}
\braket{\mathbf{\Phi}_{JT}|\mathbf{\Psi}_{JT}} = \int d\rho da\, \braket{\Phi_{JT}(\rho,a),\Psi_{JT}(\rho,a)},
\end{align}
 where $\braket{\Phi,\Psi}$ is the inner product on $\mathcal{H}_{\rm matt} \otimes \mathcal{H}_{\rm fermion} \otimes \mathcal{H}_{\rm fermion}'$. Unlike the pseudo-inner products on the boundary superparticles, this inner product is positive, as one would hope for the physical inner product in a well defined gauge theory. In fact, it is the standard inner product on 
\begin{align}
\mathcal{H}_{\rm super-JT} \cong  \mathcal{H}_{\rm matt} \otimes L^2(\R) \otimes L^2([0,a_{\rm max}]) \otimes \mathcal{H}_{\rm fermion} \otimes \mathcal{H}_{\rm fermion}'. 
\end{align}
The gauge-equivalence relation implies
\begin{align}
\left(J_+^L + J_+^R + j_+\right)\mathbf{\Psi}_{JT} &= \left(e^{-\rho} - p_{\tau} + j_+\right)\mathbf{\Psi}_{JT} \sim 0\\
\left(G_+^L + G_+^R + g_+\right)\mathbf{\Psi}_{JT} &= \left(e^{(-\rho+ia)/2} \psi' - \partial_{\theta_+} + g_+\right)\mathbf{\Psi}_{JT} \sim 0\\
\left({G_+^L}^\dagger + {G_+^R}^\dagger + g_+^\dagger\right)\mathbf{\Psi}_{JT} &= \left(e^{(-\rho-ia)/2} \psi'^\dagger - \partial_{\bar\theta_+} + g_+^\dagger\right)\mathbf{\Psi}_{JT} \sim 0\\
\left(J_3^L + J_3^R + j_3 \right)\mathbf{\Psi}_{JT} &= \left(p_{\rho'} - p_{\rho} +j_3\right)\mathbf{\Psi}_{JT}\sim 0\\
\left(J^L + J^R + j \right)\mathbf{\Psi}_{JT} &= \left(-p_{a'} - p_{a} +j\right)\mathbf{\Psi}_{JT}\sim 0\\
\left(G_-^L + G_-^R + g_-\right)\mathbf{\Psi}_{JT} &= \left(- \partial_{\theta_-'} + e^{(-\rho-ia)/2} \psi + g_-\right)\mathbf{\Psi}_{JT} \sim 0\\
\left({G_-^L}^\dagger + {G_-^R}^\dagger + g_-^\dagger\right)\mathbf{\Psi}_{JT} &= \left(- 
\partial_{\bar\theta_-'} + e^{(-\rho+ia)/2} \psi^\dagger  + g_-^\dagger\right)\mathbf{\Psi}_{JT} \sim 0\\
\left(J_-^L + J_-^R + j_-\right)\mathbf{\Psi}_{JT} &= \left(- p_{\tau'} + e^{-\rho}  + j_-\right)\mathbf{\Psi}_{JT} \sim 0.
\end{align}
We also have
\begin{align}
\left(p_\rho + p_{\rho'}\right) \mathbf{\Psi}_{JT} &=  \delta(\tau) \delta(\tau') \delta(\rho-\rho') \delta(a+a')\theta_+\bar\theta_+\theta_-'\bar\theta_-' (p_\rho \Psi_{JT}) \\
\left(p_a - p_{a'}\right) \mathbf{\Psi}_{JT} &=  \delta(\tau) \delta(\tau') \delta(\rho-\rho') \delta(a+a')\theta_+\bar\theta_+\theta_-'\bar\theta_-' (p_a \Psi_{JT}).
\end{align}
Substituting these identities into \eqref{eq:superHam}, \eqref{eq:QR}, we find that the right-boundary Lie superalgebra acts on $\Psi_{JT}$ as
\begin{align}\label{eq:rightsuperalgebra}
 Q_R =& \frac{1}{2} \psi \left(p_\rho + j_3 - i p_a - ij\right) - e^{(-\rho+ia)/2}g_+ - e^{-\rho+ia} \psi'  \\\nonumber
 Q_R^\dagger =& \frac{1}{2} \psi^\dagger \left(p_\rho + j_3 + i p_a + ij\right) - e^{(-\rho-ia)/2}g_+^\dagger - e^{-\rho-ia} \psi'^\dagger \\\nonumber
 H_R =& \frac{1}{4} (p_a + j)^2 + \frac{1}{4}(p_\rho + j_3)^2 +  e^{-\rho} j_+ + e^{-2 \rho} \\&- i e^{-\rho-ia} \psi \psi'^\dagger - i e^{(-\rho-ia)/2} \psi g_+^\dagger - ie^{-\rho+ia} \psi^\dagger \psi' - i e^{(-\rho+ia)/2} \psi^\dagger g_+\nonumber\\
J_R =& \left[p_a+j\right] + \frac{1}{2} [\psi,\psi^\dagger].\nonumber
\end{align}
Similarly, substituting into \eqref{eq:superHamL}, \eqref{eq:QLunphys}, we find that the left-boundary Lie superalgebra acts on $\Psi_{JT}$ as
\begin{align}\label{eq:QL}
 Q_L =& \frac{1}{2} \psi' \left(p_{\rho} - j_3 + i p_{a} - ij\right) +e^{-\rho-ia} \psi + e^{(-\rho-ia)/2}g_- \\\nonumber
Q_L^\dagger =& \frac{1}{2}\psi'^\dagger \left(p_{\rho}-j_3 - i p_{a} + ij\right) + e^{-\rho+ia} \psi^\dagger + e^{(-\rho+ia)/2}g_-^\dagger\\\nonumber
 H_L =& \frac{1}{4}(p_{a}-j)^2 + \frac{1}{4} (p_{\rho}-j_3)^2 +  e^{-\rho} j_- + e^{-2\rho} \\\nonumber&+ i e^{-\rho+ia} \psi' \psi^\dagger + i e^{(-\rho+ia)/2} \psi' g_-^\dagger + i e^{-\rho-ia} \psi'^\dagger \psi+ i e^{(-\rho-ia)/2} \psi'^\dagger g_-\\\nonumber
J_L =& \left[-p_a+j\right] + \frac{1}{2} [\psi',\psi'^\dagger].
\end{align}
It is easy to check that the two algebras commute (or anticommute in the case of the supercharges). We also note that in the pure super-JT gravity theory, where the matter charges in \eqref{eq:rightsuperalgebra} and \eqref{eq:QL} are set to zero, the boundary Hamiltonians $H_L$ and $H_R$ are equal but the supercharges $Q_L$ and $Q_R$ are not.

\subsection{The subspace of ground states} \label{sec:groundstates}

We can simplify the form of the right-boundary Lie superalgebra by conjugating with $U_1 =\exp(i\rho j_3 + iaj)$ to obtain
\begin{align}
U_1Q_RU_1^\dagger =& \frac{1}{2} \psi \left(p_\rho - i p_a \right) - e^{-\rho+ia}\left(g_+ + \psi'\right)  \\
U_1Q_R^\dagger U_1^\dagger =& \frac{1}{2} \psi^\dagger \left(p_\rho  + i p_a \right) - e^{-\rho-ia}\left(g_+^\dagger + \psi'^\dagger\right) \nonumber\\
U_1 H_R U_1^\dagger =& \frac{1}{4} (p_a ^2 + p_\rho^2) +  (1+j_+)e^{-2 \rho} - i e^{-\rho-ia} \psi (g_+^\dagger+ \psi'^\dagger )- i e^{-\rho+ia} \psi^\dagger (g_+ +\psi')\nonumber\\
U_1J_RU_1^\dagger =& p_a + \frac{1}{2} [\psi,\psi^\dagger].\nonumber
\end{align}
Further conjugating by $U_2 = (1+j_+)^{ip_\rho/2}$ leads to
\begin{align}
 U_2 U_1Q_RU_1^\dagger U_2^\dagger=& \frac{1}{2} \psi \left(p_\rho - i p_a \right) - e^{-\rho+ia}(1+j_+)^{-1/2}\left(g_+ + \psi'\right)  \\
 U_2 U_1Q_R^\dagger U_1^\dagger U_2^\dagger =& \frac{1}{2} \psi^\dagger \left(p_\rho  + i p_a \right) - e^{-\rho-ia}(1+j_+)^{-1/2}\left(g_+^\dagger + \psi'^\dagger\right) \nonumber\\
 U_2 U_1 H_R U_1^\dagger U_2^\dagger =& \frac{1}{4} (p_a ^2 + p_\rho^2) + e^{-2 \rho} - i e^{-\rho-ia} \psi (1+j_+)^{-1/2} (g_+^\dagger+ \psi'^\dagger )\nonumber\\&~~~~- i e^{-\rho+ia} \psi^\dagger (1+j_+)^{-1/2} (g_+ +\psi')\nonumber\\
U_2 U_1 J_R U_1^\dagger U_2^\dagger  =& p_a + \frac{1}{2} [\psi,\psi^\dagger].\nonumber
\end{align}
We see that the entire right-boundary Lie superalgebra now only depends on the matter fields through the combination $(1+j_+)^{-1/2}(g_+ + \psi')$, which behaves as a fermionic annihilation operator. We can therefore use a final unitary 
\begin{align}
U_3 = \exp\left(\left[\psi'^\dagger g_+ + \psi' g_+^\dagger\right] j_+^{-1/2} \cos^{-1}\left[\left(1+j_+\right)^{-1/2}\right]\right),
\end{align}
which satisfies
\begin{align}
U_3^\dagger \psi' U_3 &= (1+j_+)^{-\frac{1}{2}}( g_+ + \psi') \\U_3^\dagger g_+ U_3 &= (1+j_+)^{-\frac{1}{2}}( g_+ - j_+ \psi'),
\end{align}
to remove the matter dependence entirely. This leads to
\begin{align}\label{eq:conjtopuresuperJT}
 UQ_RU^\dagger =& \frac{1}{2} \psi \left(p_\rho - i p_a \right) - e^{-\rho+ia} \psi'  \\
 UQ_R^\dagger U^\dagger =& \frac{1}{2} \psi^\dagger \left(p_\rho  + i p_a \right) - e^{-\rho-ia} \psi'^\dagger \nonumber\\
 UH_RU^\dagger  =& \frac{1}{4} (p_a ^2 + p_\rho^2) +   e^{-2 \rho} - i e^{-\rho-ia} \psi \psi'^\dagger - i e^{-\rho+ia} \psi^\dagger \psi'\nonumber\\
UJ_RU^\dagger  =& p_a + \frac{1}{2} [\psi,\psi^\dagger],\nonumber
\end{align}
where $U = U_3 U_2 U_1$. This is the right-boundary Lie superalgebra for pure super-JT gravity. As we saw for bosonic JT gravity in Section \ref{sec:spectra}, the spectra of the right-boundary Hamiltonian (or the left-boundary Hamiltonian) in super-JT gravity with matter is the same as pure JT gravity, except with an additional degeneracy that is isomorphic to the Hilbert space of the matter theory.

The spectrum of canonically quantised pure super-JT gravity was analysed in \cite{Lin:2022rzw, Lin:2022zxd}. For any fixed energies $E_L, E_R > 0$, the supercharges form a four-dimensional Clifford algebra whose irreducible representation has a basis consisting of four orthogonal states that are each annihilated by one of $Q_L$, $Q_L^\dagger$ and one of $Q_R$, $Q_R^\dagger$. The R-charges of these states can be written as 
\begin{align}\label{eq:J*firstdefs}
J_L = J_L^* + \frac{1}{2} H_L^{-1} \left[Q_L, Q_L^\dagger\right] ~~~~\text{and}~~~~ J_R = J_R^* + \frac{1}{2} H_R^{-1} \left[Q_R, Q_R^\dagger\right]
\end{align}
for some constants $J_L^*$, $J_R^*$ that label the irreducible representation. So e.g. the states annihilated by $Q_R$ have $J_R = J_R^* + \frac{1}{2}$, while states annihilated by $Q_R^\dagger$ have $J_R = J_R^* - \frac{1}{2}$.

In pure super-JT gravity, we have 
\begin{align}\label{eq:puresuperJTHequal}
H_L = H_R = \frac{1}{4}\left(p_\rho^2 + p_a^2\right) + e^{-2\rho} - i e^{-\rho-ia} \psi \psi'^\dagger  - ie^{-\rho+ia} \psi^\dagger \psi'
\end{align}
and
\begin{align}
[Q_R, Q_R^\dagger] &= \frac{1}{4} \left[ \psi, \psi^\dagger\right] \left(p_\rho^2 + p_a^2\right) + e^{-2\rho} \left[\psi',\psi'^\dagger\right] \\&\qquad- \frac{1}{2}\left\{e^{-\rho + ia}, p_\rho + i p_a \right\} \psi' \psi^\dagger - \frac{1}{2}\left\{e^{-\rho - ia}, p_\rho - i p_a \right\}  \psi \psi'^\dagger\nonumber\\
[Q_L, Q_L^\dagger] &= \frac{1}{4} \left[ \psi', \psi'^\dagger\right] \left(p_\rho^2 + p_a^2\right) + e^{-2\rho} \left[\psi,\psi^\dagger\right] \\&\qquad+ \frac{1}{2}\left\{e^{-\rho + ia}, p_\rho + i p_a \right\} \psi' \psi^\dagger + \frac{1}{2}\left\{e^{-\rho - ia}, p_\rho - i p_a \right\}  \psi \psi'^\dagger\nonumber.
\end{align}
We therefore have
\begin{align}
H_L J_L + H_R J_R &= \frac{1}{2} H_R \left(\left[\psi,\psi^\dagger\right] + \left[\psi',\psi'^\dagger\right] \right)
\\&= \frac{1}{2}\left(\frac{1}{4}\left(p_\rho^2 + p_a^2\right) + e^{-2\rho}\right) \left(\left[\psi,\psi^\dagger\right] + \left[\psi',\psi'^\dagger\right] \right)
\\&= \frac{1}{2}\left( [Q_R, Q_R^\dagger] +[Q_L, Q_L^\dagger]  \right)
\end{align}
so that
\begin{align}\label{eq:JL*JR*}
J_L^* = -J_R^*.
\end{align}
It turns out that pure super-JT gravity contains a single, delta-function-normalisable irreducible representation of the supercharge Clifford algebra for any $J_R^* \in \mathbb{Z}/N$ and $E_R > {J_R^*}^2/4$  \cite{Lin:2022rzw, Lin:2022zxd}. Thanks to \eqref{eq:puresuperJTHequal} and \eqref{eq:JL*JR*} these representations satisfy $E_L = E_R$ and $J_L^* = - J_R^*$. The representations can be constructed by acting with $Q_L^\dagger$ and $Q_R^\dagger$ on the wavefunction
\be
\ket{s, J_R^*} = \sqrt{\rho(s)} e^{ia J_R^*}  K_{2 is }(2e^{-\rho}) \ket{\Omega_L}\ket{\Omega_R}
\ee
where $s^2 = E_R^2 - {J_R^*}^2/4$.

In addition, for any $J_R  \in \mathbb{Z}/N$ satisfying  $|J_R| < 1/2$, there exists a single, \emph{normalisable} zero-energy state with $J_L = -J_R$. The wavefunction of this state can be written as
\begin{align} \label{eq:puresuperJTgroundstate}
\ket{\tilde\Psi_{J_R}} =  \frac{2}{\pi}\sqrt{\cos(\pi J_R)}e^{iJ_R a -\rho}\left[- e^{ia/2}K_{\frac{1}{2} - J_R}(2e^{-\rho}) \psi^\dagger + ie^{-ia/2}K_{\frac{1}{2} + J_R} (2 e^{-\rho}) \psi'^\dagger \right]\ket{\Omega_L}\ket{\Omega_R}.
\end{align}
These normalisable ground states are supersymmetric and so are annihilated by all the supercharges.

The existence of the states \eqref{eq:puresuperJTgroundstate} in pure super-JT gravity means that, in super-JT gravity with matter,  there exists an infinite set of normalisable states with $E_R = 0$ for any allowed R-charge $J_R$ satisfying $|J_R| < 1/2$. Moreover, the set of normalisable $E_R=0$ states for any fixed $J_R$ is naturally isomorphic to $\mathcal{H}_{\rm matt}$. Explicitly, we can define an isometric map $\tilde V_{J_R}: \mathcal{H}_{\rm matt} \to \mathcal{H}_{\rm super-JT}$ , whose image is the space of zero-energy states with R-charge $J_R$, by
\begin{align}\label{eq:tildeV}
\tilde V_{J_R} \ket{\Psi_{\rm matt}}&= U^\dagger \ket{\tilde \Psi_{J_R}} \ket{\Psi_{\rm matt}}\\
&= \frac{2}{\pi} U_1^\dagger \sqrt{\cos(\pi J_R)}e^{iJ_R a -\rho}\big[- e^{ia/2}K_{\frac{1}{2} - J_R}(2\sqrt{1 + j_+}e^{-\rho}) \left(\sqrt{1 +g_+ g_+^\dagger} \psi^\dagger + \psi'^\dagger \psi^\dagger g_+\right)\nonumber\\&~~~~ + ie^{-ia/2}K_{\frac{1}{2} + J_R} (2 \sqrt{1+j_+}e^{-\rho}) \left(\sqrt{1 + g_+^\dagger g_+} \psi'^\dagger + g_+^\dagger\right) \big]\ket{\Omega_L}\ket{\Omega_R}\ket{\Psi_{\rm matt}}.\nonumber
\end{align}
In deriving the second equality in \eqref{eq:tildeV}, we used the formulas
\begin{align}\label{eq:U3Omegag+dagger}
U_3^\dagger \ket{\Omega_L} g_+^\dagger \ket{\Psi_{\rm matt}} &= \left(1 + j_+\right)^{-\frac{1}{2}} \left(1  - \psi'^\dagger g_+\right) \ket{\Omega_L}  g_+^\dagger \ket{\Psi_{\rm matt}}\\
U_3^\dagger \ket{\Omega_L} g_+ \ket{\Psi_{\rm matt}} &= \ket{\Omega_L}  g_+ \ket{\Psi_{\rm matt}}\\
U_3^\dagger \psi'^\dagger\ket{\Omega_L} g_+^\dagger \ket{\Psi_{\rm matt}} &= \psi'^\dagger\ket{\Omega_L} g_+^\dagger \ket{\Psi_{\rm matt}}\\
U_3^\dagger \psi'^\dagger\ket{\Omega_L} g_+ \ket{\Psi_{\rm matt}} &=\left(1 + j_+\right)^{-\frac{1}{2}} \left(\psi'^\dagger + g_+^\dagger \right)\ket{\Omega_L} g_+ \ket{\Psi_{\rm matt}}.\label{eq:U3daggerfinal}
\end{align}
These can be easily found by noting that e.g. \eqref{eq:U3Omegag+dagger} is the unique state in the four-dimensional subspace generated by acting with $\psi'$, $g_+$ and their adjoints on $\ket{\Omega_L}\ket{\Psi_{\rm matt}}$ that has squared-norm $\braket{\Psi_{\rm matt}| g_+ g_+^\dagger|\Psi_{\rm matt}}$ and that is annihilated by $U_3^\dagger \psi' U_3$ and $U_3^\dagger g_+^\dagger U_3$. Now any state $\ket{\Psi_{\rm matt}}$ that is orthogonal to the vacuum can be written as
\begin{align}
\ket{\Psi_{\rm matt}} =\g_+g_+^\dagger  j_+^{-1} \ket{\Psi_{\rm matt}} + g_+^\dagger g_+ j_+^{-1}  \ket{\Psi_{\rm matt}}.
\end{align}
It therefore follows immediately from \eqref{eq:U3Omegag+dagger} - \eqref{eq:U3daggerfinal} that
\begin{align}\label{eq:U3Psimatt1}
U_3^\dagger \ket{\Omega_L} \ket{\Psi_{\rm matt}} &= \left(1 + j_+\right)^{-\frac{1}{2}} \left(\sqrt{1 + g_+ g_+^\dagger}  - \psi'^\dagger g_+\right) \ket{\Omega_L} \ket{\Psi_{\rm matt}} \\
U_3^\dagger \psi'^\dagger \ket{\Omega_L} \ket{\Psi_{\rm matt}} &= \left(1 + j_+\right)^{-\frac{1}{2}} \left(\sqrt{1 + g_+^\dagger g_+} \psi'^\dagger +  g_+^\dagger\right) \ket{\Omega_L} \ket{\Psi_{\rm matt}}. \label{eq:U3Psimatt2}
\end{align}
Both of these equations are also true when $\ket{\Psi_{\rm matt}} = \ket{\Omega_{\rm matt}}$. The second equality in \eqref{eq:tildeV} can be found fairly easily using \eqref{eq:U3Psimatt1} and \eqref{eq:U3Psimatt2}.

Since the left-boundary Lie superalgebra commutes with $H_R, J_R$, it must preserve the space of zero-energy states with charge $J_R$. It should therefore be possible to write e.g. $Q_L \tilde V_{J_R}  = \tilde V_{J_R}  \tilde Q_L$ where the matter QFT operator $\tilde Q_L$ describes the action of $Q_L$ restricted to zero-energy states with charge $J_R$. A somewhat laborious calculation, described in Appendix \ref{app:gsrestriction}, reveals that
\begin{align}
\tilde Q_L &= - (1 + g_+ g_+^\dagger)^{1/4} (1 + g_+^\dagger g_+ )^{-1/4} g_- (1+g_+^\dagger g_+)^{1/4} (1+g_+g_+^\dagger)^{-1/4} + i\frac{J_R}{2} (1+j_+)^{-1/2} g_+ \nonumber\\
 \tilde Q_L^\dagger &= - (1 + g_+ g_+^\dagger)^{-1/4} (1 + g_+^\dagger g_+ )^{1/4} g_-^\dagger (1+g_+^\dagger g_+)^{-1/4} (1+g_+g_+^\dagger)^{1/4} - i\frac{J_R}{2} (1+j_+)^{-1/2} g_+^\dagger\nonumber\\
\tilde H_L &= \{\tilde Q_L, \tilde Q_L^\dagger\} \label{eq:tildesuperalgebra}\\
\tilde J_L &= -J_R + 2 j.\nonumber
\end{align}

\subsection{Simultaneous ground states}\label{sec:simulground}
It is particularly interesting to understand states where both boundaries are simultaneously in a ground state. Clearly this is true for the image of the matter vacuum $\tilde V_{J_R} \ket{\Omega_{\rm matt}}$. But intuitively it is hard to understand how states can exist where bulk matter is excited and yet both boundaries have exactly zero energy. Nevertheless, it was argued in \cite{Lin:2022rzw, Lin:2022zxd} based on an equivalence of different interpretations of the same Euclidean gravitational path integral that such states must exist. Here, we shall show that this is the case directly in the canonically quantised theory. Moreover, we will directly count the simultaneous ground states and show that there is exactly one within each matter representation of $\hSU(1,1|1)$ with appropriate left-boundary R-charge $\tilde J_L$.

Given a supersymmetric Hamiltonian $\tilde H_L$ and state $\Psi$, we have
\begin{align}\label{eq:susygs}
\braket{\Psi|\tilde H_L |\Psi} = \braket{\tilde Q_L \Psi| \tilde Q_L \Psi} + \braket{\tilde Q_L^\dagger \Psi| \tilde Q_L^\dagger \Psi} \geq 0.
\end{align}
It follows that zero energy states must be annihilated by both $\tilde Q_L$ and $\tilde Q_L^\dagger$. All other states come in equal-energy pairs that are related by, and that are each annihilated by exactly one of, $\tilde Q_L$ and $\tilde Q_L^\dagger$.

We can count supersymmetric ground states by computing the cohomology of the supercharge $\tilde Q_L$, i.e. the kernel of $\tilde Q_L$ quotiented by the image of $\tilde Q_L^\dagger$.  The cohomology of $\tilde Q_L$ is isomorphic to the cohomology of $A \tilde Q_L A^{-1}$ where $A$ is any invertible linear map on $\mathcal{H}_{\rm matt}$. Naively, one would therefore expect that,
when $J_R = 0$, the cohomology of $\tilde Q_L$ is the same  as the cohomology of
\be
-(1+g_+^\dagger g_+)^{1/4} (1+g_+g_+^\dagger)^{-1/4} \tilde Q_L (1 + g_+^\dagger g_+ )^{-1/4} (1 + g_+ g_+^\dagger)^{1/4} = g_-,
\ee
which vanishes in any representation of $\hSU(1,1|1)$ that can appear in the matter theory. However this would be too quick: the operator $ (1+g_+^\dagger g_+)^{1/4} (1+g_+g_+^\dagger)^{-1/4} $ is unbounded and hence is only defined on a dense subspace of Hilbert space. Since there is no reason  that states in the kernel of $\tilde Q_L$ need to lie in the domain of $ (1+g_+^\dagger g_+)^{1/4} (1+g_+g_+^\dagger)^{-1/4}$, there is no reason that the two cohomologies need to be the same.

To make progress, it will again be helpful to realise the matter $\hSU(1,1|1)$ representations as differential operators acting on a single real bosonic variable. Representations of $\hSU(1,1|1)$ that have energy $j_1$ bounded from below are labelled by the smallest eigenvalue $\lambda$ of $j_1$ and the R-charge $|u| \leq \lambda$ of the $j_1 = \lambda$ eigenstate. Representations with $|u| < \lambda$ decompose into four discrete series representations of $\hSL(2,\R)$. We therefore describe them using a single bosonic variable $\sigma$, along with two fermions $\chi_1$ and $\chi_2$. It turns out that the correct generators are
\begin{align}
j_3 = p_\sigma ~~~~~~j_+ &= e^{-\sigma}~~~~~~j_- = \left(p_\sigma + i\hat \lambda\right)  e^{\sigma}\left(p_\sigma - i \hat\lambda\right)\\
g_+ = e^{-\sigma/2} \chi_1 ~~~~~~ g_- &= e^{\sigma/2}\left[p_\sigma \chi_1 + \frac{i}{2} \chi_1\left(-\chi_2^\dagger\chi_2 + 2u \right) - i \sqrt{\lambda^2 -u^2} \chi_2 \right]\\
g_+^\dagger = e^{-\sigma/2} \chi_1^\dagger ~~~~~~ g_-^\dagger &= e^{\sigma/2}\left[p_\sigma \chi_1^\dagger - \frac{i}{2} \chi_1^\dagger\left(\chi\chi_2^\dagger + 2u\right) + i \sqrt{\lambda^2 -u^2} \chi_2^\dagger \right]\\
j &=u + \frac{1}{4} [\chi_1, \chi_1^\dagger] + \frac{1}{4} [\chi_2, \chi_2^\dagger],
\end{align}
where the operator 
\begin{align}
\hat\lambda = \lambda + \frac{1}{2} -\frac{\sqrt{\lambda^2 - u^2}}{2\lambda}\left(\chi_2^\dagger \chi_1 + \chi_1^\dagger \chi_2\right)  +\frac{u}{2 \lambda} \left(\chi_2 \chi_2^\dagger - \chi_1 \chi_1^\dagger\right),
\end{align}
which has eigenvalues $\lambda, \lambda+1/2, \lambda+1$, describes the minimal eigenvalue of $j_1$ in each of the four $\hSL(2,\R)$-representations that make up the $\hSU(1,1|1)$ representation. 

As with the representation of $\hSL(2,\R)$ described in Section \ref{sec:spectra}, when $\lambda < 1$ we have to be careful about boundary conditions as $\sigma \to +\infty$ when defining the domains of $g_-, g_-^\dagger$ and $j_-$. In this case, we need to restrict the domain of $g_-$ to wavefunctions $\Psi(\sigma)$ where $\chi_1\chi_2 \Psi(\sigma)$ vanishes faster than $e^{-\sigma/2}$. The domain of its adjoint $g_-^\dagger$ is then restricted to wavefunctions $\Psi(\sigma)$ where $\chi_1^\dagger\chi_2^\dagger \Psi(\sigma)$ vanishes faster than $e^{-s/2}$. Finally, we see that the domain of $j_- = \{g_-,g_-^\dagger\}$ is restricted to wavefunctions $\Psi(\sigma)$ where $\chi_1 \chi_2 \Psi(\sigma)$ and $\chi_1^\dagger \chi_2^\dagger \Psi(\sigma)$ vanish faster than $e^{-\sigma/2}$ and where $(p_\sigma - i \lambda)\chi_1 \chi_2^\dagger \Psi(\sigma)$ and $(p_\sigma - i\lambda) \chi_1^\dagger \chi_2\Psi(\sigma)$ vanish faster than $e^{-\sigma}$. It is straightforward to check that this picks out the correct $\hSL(2,\R)$ representations with smallest $j_1$ eigenvalues $\lambda$ and $\lambda+1/2$ (rather than the representations with smallest eigenvalues $1-\lambda$ and $1/2 - \lambda$).

There are also BPS representations that have $\lambda = |u|$ and decompose into only two discrete-series representations of $\hSL(2,\R)$. These are described using a single fermion $\chi$.  For $u>0$, the generators can be written as 
\begin{align}
j_3 = p_\sigma ~~~~~~j_+ = e^{-\sigma}~~~~~~j_- &= \left(p_\sigma + iu + \frac{i}{2} \chi^\dagger \chi\right) e^{\sigma}\left(p_\sigma - i u - \frac{i}{2} \chi^\dagger \chi\right)\\
g_+ = e^{-\sigma/2} \chi ~~~~~~ g_- &= \left(p_\sigma + i u \right) e^{\sigma/2} \chi\\
g_+^\dagger = e^{-\sigma/2} \chi^\dagger ~~~~~~ g_-^\dagger &= e^{\sigma/2}\left(p_\sigma - i u \right) \chi^\dagger\\
j =u - &\frac{1}{2}  \chi^\dagger \chi.
\end{align}
For $u< 1/2$, we need to restrict the domain of $g_-$ to wavefunctions $\Psi$ where $\chi \Psi$ vanishes faster than $e^{-\sigma/2}$ as $\sigma\to+ \infty$. Consequently there is no such restriction on the domain of $g_-^\dagger$. The domain of $j_-$ consists of wavefunctions $\Psi$ where $\chi \Psi$ vanishes faster than $e^{-\sigma/2}$ and $(p_\sigma -iu) \chi^\dagger \Psi$ vanishes faster than $e^{-\sigma}$. For $u<0$, we have 
\begin{align}
j_3 = p_\sigma ~~~~~~j_+ = e^{-\sigma}~~~~~~j_- &= \left(p_\sigma - iu + \frac{i}{2}  \chi\chi^\dagger\right) e^{\sigma}\left(p_\sigma + i u - \frac{i}{2} \chi \chi^\dagger\right)\\
g_+ = e^{-\sigma/2} \chi ~~~~~~ g_- &=  e^{\sigma/2} \chi \left(p_\sigma + i u \right)\\
g_+^\dagger = e^{-\sigma/2} \chi^\dagger ~~~~~~ g_-^\dagger &=    \left(p_\sigma - i u \right) e^{\sigma/2}\chi^\dagger\\
j =u + &\frac{1}{2}  \chi \chi^\dagger.
\end{align}
where the domain  of $g_-^\dagger$ is restricted to wavefunctions $\Psi$ where $\chi^\dagger \Psi$ vanishes faster than $e^{-\sigma/2}$, the domain of $g_-$ is unrestricted and the domain of $j_-$ requires $\chi^\dagger \Psi$ and $e^{\sigma/2}(p+iu)\chi \Psi$ to vanish faster than $e^{-\sigma/2}$. 

We first consider a BPS representation with  $u>0$. Restricted to such a representation, \eqref{eq:tildesuperalgebra} becomes
\begin{align} \label{{eq:QLBPSU+}}
\tilde Q_L &= - \left(1 + e^{-\sigma}\right)^{1/4}\chi\left(p_\sigma + i u \right)\left(1 + e^{-\sigma}\right)^{1/4}e^{\sigma/2} + \frac{i}{2} J_R (1+e^{\sigma})^{-1/2} \chi\\
&=  (1+e^\sigma)^{1/2}\left[-p_\sigma - iu +\frac{i}{2} + i\left(\frac{J_R}{2} - \frac{1}{4}\right) (1 + e^{\sigma})^{-1}\right]\chi.
\end{align} 
and
\begin{align} \label{{eq:QLBPSU+dag}}
\tilde Q_L^\dagger &= - e^{\sigma/2} \left(1 + e^{-\sigma}\right)^{1/4}\chi^\dagger\left(p_\sigma - i u \right)\left(1 + e^{-\sigma}\right)^{1/4}- \frac{i}{2} J_R (1+e^{\sigma})^{-1/2} \chi^\dagger \\
&=  (1+e^\sigma)^{1/2}\left[-p_\sigma + iu - i\left(\frac{J_R}{2} + \frac{1}{4}\right) \left(1 + e^{\sigma}\right)^{-1}\right]\chi^\dagger.
\end{align} 
Let $\ket{0}$ be the fermion vacuum (i.e. $\chi \ket{0} = 0$). Any wavefunction of the form $\Psi_0(\sigma)\ket{0}$ is trivially annihilated by $\tilde Q_L$. For $\Psi_0(\sigma)\ket{0}$ to be a supersymmetric ground state, we also require
\begin{align}
 \tilde Q_L^\dagger \Psi_0(\sigma)\ket{0}= 0 ~~~~~\Leftrightarrow ~~~~~ \left[-p_\sigma + iu - i\left(\frac{J_R}{2} + \frac{1}{4}\right) (1 + e^{\sigma})^{-1}\right]\Psi_0(\sigma) = 0.
\end{align}
This is a first-order differential equation and hence has a single smooth solution. All that we need to check is whether that solution is normalisable. We have
\begin{align}
\Psi_0(\sigma) \sim \exp(-u \sigma) ~~~ \sigma \to +\infty~~~~\text{and}~~~~\Psi_0(\sigma) \sim \exp\left(\left[-u+\frac{J_R}{2} + \frac{1}{4}\right]\sigma\right) ~~~~\sigma \to -\infty.
\end{align}
It is therefore normalisable if and only if $-u + J_R/2 + 1/4 >0$ or, equivalently,
\begin{align}\label{eq:JLBPScondition}
\tilde J_L = -J_R + 2j = -J_R + 2 u < \frac{1}{2}.
\end{align}
Since $u>0$, we always have $\tilde J_L >x -J_R > -1/2$. As a result, \eqref{eq:JLBPScondition} is the only nontrivial condition required to have $|J_L| < 1/2$ and hence for a normalisable ground state with left R-charge $J_L$ to exist in pure super-JT gravity. 

We can also consider states of the form $\Psi_1(\sigma) \chi^\dagger \ket{0}$. In this case, however, annihilation by $\tilde Q_L$ requires
\begin{align}
\left[-p_\sigma - iu +\frac{i}{2} + i\left(\frac{J_R}{2} - \frac{1}{4}\right) (1 + e^{-\sigma})^{-1}\right]\Psi_1(\sigma) = 0.
\end{align}
The solution to this equation scales as $\exp((u - 1/2)\sigma)$ as $\sigma \to + \infty$. This is normalisable for $u < 1/2$, but is never contained in the domain of $g_-$ (and hence is not in the domain of $\tilde Q_L$)  for $u>0$.

Before moving on to to BPS representations with $u < 0$, let us briefly analyse the spectrum of states with nonzero energy. The operator $\tilde H_L$ is a second order differential equation and hence has two smooth solutions of the form $\Psi_0(\sigma)\ket{0}$ and two solutions of the form $\Psi_1(\sigma) \chi^\dagger \ket{0}$. As $\sigma \to + \infty$, we have 
\begin{align}
\tilde H_L = \{\tilde Q_L, \tilde Q_L^\dagger\} = e^{\sigma/2}\left(p_\sigma^2 - \left(u +\frac{1}{2} \chi \chi^\dagger \right)^2\right)e^{\sigma/2} + O(e^{\sigma/2}). 
\end{align}
It follows that finite energy states must satisfy
\begin{align}
\Psi_0(\sigma) &\approx A \exp(-u \sigma) + B \exp((u-1) \sigma) \\ \Psi_1(\sigma) &\approx C \exp\left(\left(u - \frac{1}{2}\right) \sigma\right) +D \exp\left(\left(-u - \frac{1}{2}\right) \sigma\right).
\end{align}
For both $\Psi_0(\sigma)$ and $\Psi_1(\sigma)$, there is exactly one solution for any $u > 0$ that is a) normalisable and b) contained within the domain of $j_-$ (and hence $\tilde H_L$. On the other hand, as $\sigma \to -\infty$, we have
\begin{align}\label{eq:tildeHLBPS}
\tilde H_L \approx  p_\sigma^2 + \frac{1}{4}\left(u- \frac{1}{2} - J_R\right)^2.
\end{align}
We therefore have a continuum of delta-function-normalisable states with energies\footnote{For energies below the bound \eqref{eq:BPSu>0continuumbound}, one solution to \eqref{eq:tildeHLBPS} blows up exponentially while the other is normalisable. In principle, there could therefore exist a discrete set of eigenstates with energy $0<E_L<\left(u- 1/4 - J_R/2\right)^2$ that are normalisable at both $\sigma\to \pm \infty$. However, we know that this cannot occur because no such states are contained in the spectrum of $H_L$ in the full theory of super-JT gravity.}
\begin{align}\label{eq:BPSu>0continuumbound}
 \tilde E_L \geq \left(u- \frac{1}{4}- \frac{J_R}{2}\right)^2.
\end{align} 
For each allowed energy $\tilde E_L$ there exists a single delta-function-normalisable state with fermion number zero and a single state with fermion number one. For a state $\Psi_0(\sigma)\ket{0}$ annihilated by $\chi$, we have $j = u$ so that $\tilde J_L = 2u - J_R$. The bound \eqref{eq:BPSu>0continuumbound} can therefore be written as
\begin{align} 
\tilde E_L \geq \frac{1}{4}(\tilde J_L -\frac{1}{2})^2.
\end{align} 
For the state $\Psi_1(\sigma)\chi^\dagger \ket{0}$ annihilated by $\chi^\dagger$, we have $j = u -1/2$ and $\tilde J_L = 2u - J_R - 1$. The same bound \eqref{eq:BPSu>0continuumbound} therefore instead becomes
\begin{align}
 \tilde E_L \geq \frac{1}{4}(\tilde J_L + \frac{1}{2})^2.
\end{align}
 These are exactly the lower bounds on the continuous spectrum of $H_L$ in pure super-JT gravity for states that are annihilated by $Q_L$ and $Q_L^\dagger$ respectively. 

For a BPS representation with $u< 0$, we have
\begin{align} \label{{eq:QLBPSU-}}
 \tilde Q_L &= - e^{\sigma/2} \left(1 + e^{-\sigma}\right)^{1/4}\chi\left(p_\sigma + i u \right)\left(1 + e^{-\sigma}\right)^{1/4}+ \frac{i}{2} J_R (1+e^{\sigma})^{-1/2} \chi\\
&= (1+e^\sigma)^{1/2}\left[-p_\sigma - iu + i\left(\frac{J_R}{2} - \frac{1}{4}\right) (1 + e^{\sigma})^{-1}\right]\chi.
\end{align} 
and
\begin{align} \label{{eq:QLBPSU-dag}}
 \tilde Q_L^\dagger &= - \left(1 + e^{-\sigma}\right)^{1/4}\chi^\dagger\left(p_\sigma - i u \right)\left(1 + e^{-\sigma}\right)^{1/4} e^{\sigma/2} - \frac{i}{2} J_R (1+e^{\sigma})^{-1/2} \chi^\dagger \\
&= (1+e^\sigma)^{1/2}\left[-p_\sigma + iu + \frac{i}{2} - i\left(\frac{J_R}{2} + \frac{1}{4}\right) (1 + e^{\sigma})^{-1}\right]\chi^\dagger.
\end{align} 
The solution to
\begin{align}
\tilde Q_L \Psi_1(\sigma) \chi^\dagger \ket{0} = \left[-p_\sigma - iu + i\left(\frac{J_R}{2} - \frac{1}{4}\right) (1 + e^{\sigma})^{-1}\right] \Psi_1(\sigma) \ket{0} = 0
\end{align}
scales as $\exp(u\sigma)$ as $\sigma \to \infty$ and as $\exp\left(\left[u - J_R/2 + 1/4\right]\sigma\right)$ as $\sigma \to -\infty$. It is therefore normalisable (and hence is a supersymmetric ground state) if and only if $u - J_R/2 + 1/4 >0$ or equivalently if 
$\tilde J_L > -1/2$. Since $\tilde J_L < -J_R < 1/2$ for any $u < 0$, this is the only nontrivial condition required for a ground state with left R-charge $\tilde J_L$ to exist in pure super-JT gravity. On the other hand, the solution to
\begin{align}
\tilde Q_L^\dagger \Psi_0(\sigma) \ket{0} =\left[-p_\sigma + iu + \frac{i}{2} - i\left(\frac{J_R}{2} + \frac{1}{4}\right) (1 + e^{\sigma})^{-1}\right] \Psi_0(\sigma) \chi^\dagger \ket{0} = 0
\end{align}
vanishes as $\exp((-u-1/2)\sigma)$ as $\sigma \to +\infty$ and so is never in the domain of $g_-^\dagger$ or $\tilde Q_L^\dagger$. There is also a continuum of states with energy 
\begin{align}
\tilde E_L \geq \frac{1}{4}(2u+\frac{1}{2} - J_R)^2,
\end{align}
 which again corresponds to 
\begin{align}
 \tilde E_L \geq \frac{1}{4} (\tilde J_L - \frac{1}{4})^2
\end{align} for states annihilated by $\tilde Q_L$ and 
\begin{align}
\tilde E_L \geq \frac{1}{4}(\tilde J_L +\frac{1}{2})^2
\end{align}
 for states annihilated by $\tilde Q_L^\dagger$.

Finally, we consider ordinary representations. We have
\begin{align}
\tilde Q_L = (1 + e^\sigma)^{1/2} \left[-p_\sigma - iu + \frac{i}{2}\chi_2^\dagger \chi_2 + i\left(\frac{J_R}{2} - \frac{1}{4}\right) (1 + e^{\sigma})^{-1}\right]\chi_1 + i e^{\sigma/2}\sqrt{\lambda^2 -u^2} \chi_2
\end{align}
and
\begin{align}
\tilde Q_L^\dagger = (1 + e^\sigma)^{1/2} \left[-p_s + iu + \frac{i}{2} \chi_2\chi_2^\dagger - i\left(\frac{J_R}{2} + \frac{1}{4}\right) (1 + e^{\sigma})^{-1}\right]\chi_1^\dagger - i e^{\sigma/2}\sqrt{\lambda^2 -u^2} \chi_2^\dagger.
\end{align}

We first note that, since the $\chi_2$ term in $\tilde Q_L$ has no kernel, there are no states with $j = u -1/2$ annihilated by $\tilde Q_L$. Similarly, there are no states with $j = u + 1/2$ annihilated by $\tilde Q_L^\dagger$. The only possible supersymmetric states therefore have $j = u$. These can be written as
\begin{align}
\Psi = \left[\Psi_1(\sigma) \chi_1^\dagger + \Psi_2(\sigma) \chi_2^\dagger\right]\ket{0}
\end{align}
where $\chi_1 \ket{0} = \chi_2 \ket{0} = 0$. The conditions $\tilde Q_L \Psi = \tilde Q_L^\dagger \Psi = 0$ give us two coupled 1st order differential equations relating $\Psi_1(\sigma)$ and $\Psi_2(\sigma)$. It is possible to eliminate one of these functions, and thereby to get a second order equation for the other. The space of solutions is therefore two-dimensional. However we still must determine
whether either or both of those solutions is normalisable.

To analyse this, it is helpful to  consider the space of unnormalisable wavefunctions with $j = u-1/2$ that are annihilated by the differential operator $\tilde H_L$.\footnote{In a mild abuse of notation, we will use the same notation for unbounded, densely defined Hilbert space operators and for the corresponding differential operator defined on the space of smooth functions (regardless of normalisability). We will try to make the distinction clear whenever confusion may arise.} Such wavefunctions can exist even though no wavefunction with $j = u-1/2$ (normalisable or otherwise) is annihilated by $\tilde Q_L$ because \eqref{eq:susygs} requires integration by parts and so is only valid for normalisable states. The equation
\begin{align}
\tilde H_L \Psi = 0
\end{align} 
is a second-order differential equation and so the space of solutions is two dimensional. Since we have already established that $\tilde Q_L$ has no kernel with $j = u-1/2$, the image of this space under the differential operator $\tilde Q_L$ must also be two-dimensional. That space is annihilated by $\tilde Q_L$ (since $\tilde Q_L^2 = 0$) and also by $\tilde Q_L^\dagger$ (since $\tilde H_L = \tilde Q_L^\dagger \tilde Q_L$ when restricted to wavefunctions with $j = u -1/2$). It is therefore the space of potential supersymmetric ground states that we want to understand.

Finding a supersymmetric state at $j =u$ that is normalisable is thus equivalent to finding a wavefunction $\Psi = \Psi_0(\sigma)\chi_1^\dagger \chi_2^\dagger \ket{0}$ that (1) satisfies $\tilde H_{L} \Psi = 0$, and (2) has the property that $\tilde Q_{L} \Psi$ is normalisable even though $\Psi$ itself cannot be (or at least $\Psi$ cannot be in the domain of the Hilbert space operator $\tilde H_L$). 

To determine the normalisability of $\tilde Q_{L} \Psi$, we again only need to consider the limits $\sigma \to \pm \infty$. For $\sigma \ll 0$, the condition $\tilde H_L \Psi = 0$ is equivalent to 
\begin{align}\label{eq:HLordinary-inf}
\left[  p_\sigma^2 +  \left(-u+ \frac{J_R}{2} + \frac{1}{4}\right)^2 + O(e^{\sigma}) \right] \Psi_0(\sigma) = 0.
\end{align}
The general solution is therefore of the form
\begin{align}
\Psi_0(\sigma) \approx A\exp\left(\left[u -\frac{J_R}{2} - \frac{1}{4}\right] \sigma\right)   + B \exp\left(\left[-u +\frac{J_R}{2} + \frac{1}{4}\right] \sigma\right).
\end{align}
Up to $O(e^{\sigma/2})$ corrections, the first term is annihilated by 
\begin{align}
\tilde Q_L =  \left[-p_\sigma - iu + \frac{i}{2}\chi_2^\dagger \chi_2 + i\left(\frac{J_R}{2} - \frac{1}{4}\right) \right]\chi_1 +O(e^{\sigma/2}).
\end{align}
 Its image under $\tilde Q_L$ therefore decays exponentially so long as $u - J_R/2 + 1/4 > 0$, or equivalently so long as
\begin{align}
\tilde J_L  = -J_R + 2u > - \frac{1}{2}.
\end{align}
The second term is not annihilated by $\tilde Q_L$ at leading order. Its image under $\tilde Q_L$ therefore decays exponentially so long as $-u + J_R/2+1/4>0$ or equivalently if
\begin{align}
\tilde J_L = -J_R + 2u < \frac{1}{2}.
\end{align}

We also need to consider the limit $\sigma \gg 0$. In this limit we have $\tilde Q_L \approx g_-$, $\tilde Q_L^\dagger \approx g_-^\dagger$ and $\tilde H_L \approx j_-$. It follows that $\tilde H_L \Psi = 0$ requires
\begin{align}
e^{\sigma} \left(p_\sigma^2 -i  p_\sigma + \lambda^2 - \frac{1}{4} + O(e^{-\sigma}) \right) \Psi_0(\sigma) = 0
\end{align}
and hence
\begin{align}
\Psi_0(\sigma) \approx C \exp\left(-\left[\frac{1}{2} + \lambda \right]\sigma \right)+ D \exp\left(-\left[\frac{1}{2} - \lambda \right]\sigma \right).
\end{align}
The action of $\tilde Q_L$ multiplies both solutions by $O(e^{\sigma/2})$. As a result, for any $\lambda > 0$ the wavefunction $\tilde Q_L \Psi$ is normalisable at $\sigma \gg 0$ if and only if $D = 0$.\footnote{The same condition is also sufficient to ensure that it is in the domain of $j_-$ and hence of $\tilde H_L$.} Generically, when $\sigma \ll 0$ this solution will have both $A, B$ nonzero. As a result, there will exist a single normalisable supersymmetric ground state $\tilde Q_L \Psi$ with $j = u$  if and only if $|\tilde J_L| < 1/2$ i.e. if there exist ground states of super-JT gravity with left-boundary R-charge $\tilde J_L$.\footnote{One might worry that solutions with $|\tilde J_L| \geq 1/2$ could exist if we were able to fine tune $u$ and $J_R$ so that either $A = 0$ or $B = 0$ when $\sigma \ll 0$ for the solution with $D = 0$ when $\sigma \gg 0$. However this is impossible because any wavefunction $\Psi$ in the kernel of $\tilde H_L$ such that $\tilde Q_L \Psi$ is normalisable and $|J_R - 2u| \geq 1/2$  would itself be normalisable. We have already shown that no such wavefunction $\Psi$ can exist because it would have to be annihilated by $\tilde Q_L$.}

We also note from the behavior of \eqref{eq:HLordinary-inf} for $\sigma\to -\infty$  that there exists a continuum of delta-function normalisable states with $j = u - 1/2$ and energy 
\begin{align}
\tilde E_L \geq \frac{1}{2} \left(-u+ \frac{J_R}{2} + \frac{1}{4}\right)^2 =\frac{1}{8} (\tilde J_L + 1/2)^2. 
\end{align}
Acting on these states with $\tilde Q_L$ leads to a continuum of states with the same energy but $j = u$ so that $\tilde E_L \geq (\tilde J_L -1/2)^2/8$. Similar arguments show that there exists a continuum of states with $j = u+1/2$ and $\tilde E_L \geq (\tilde J_L - 1/2)^2/8$ and that acting on these states with $\tilde Q_L^\dagger$ gives a continuum of states with $\tilde E_L  \geq (\tilde J_L + 1/2)^2/8$. All these inequalities match the lower bounds on the continuous spectrum of the operator $H_L$ in the full theory of super-JT gravity for states that have left R-charge $\tilde J_L$  and that are annihilated by either $\tilde Q_L$ or $\tilde Q_L^\dagger$.

While the arguments above were somewhat technical, the conclusions are very intuitive and can be stated simply. Within any $\hSU(1,1|1)$ matter representation there exists a single normalisable simultaneous ground state  for each pair of boundary R-charges $J_L$, $J_R$ satisfying
\begin{align}
J_L + J_R= 2u ~~~~\text{and}~~~~|J_L|, |J_R| < \frac{1}{2},
\end{align}
where $u$ is the matter R-charge of the lowest energy state within that representation. In other words, a unique simultaneous ground state exists if and only if the left-boundary R-charge matches that of a normalisable ground state in pure super-JT gravity. 

The spectrum of states with   right-boundary energy $E_R = 0$ but with nonzero left-boundary energy $\tilde E_L$ obeys a similar set of rules: a single delta-function normalisable state annihilated by $\tilde Q_L$ or $\tilde Q_L^\dagger$ exists if and only if pure super-JT gravity contains a state with the same left-boundary energy, the same R-charge and that is annihilated by the same supercharge.

\section{Boundary algebras in super-JT gravity}\label{sec:algebras} 

\subsection{Pure super-JT gravity}
In pure super-JT gravity, the right-boundary algebra of observables $\mathcal{A}_R$ is generated by the Hamiltonian $H_R$,  the supercharges $Q_R$, $Q_R^\dagger$ and the R-charge $J_R$. These charges are given by \eqref{eq:rightsuperalgebra} with all matter $\hSU(1,1|1)$ charges set to zero. The Hamiltonian $H_R$ is central and in addition the R-charge $J_R$ can only be changed by $\pm 1$ by acting with $Q_R$ or $Q_R^\dagger$. This means that the operator
\begin{align}\label{eq:JR*}
J_R^* = J_R - \frac{1}{2} H_R^{-1} [Q_R, Q_R^\dagger],
\end{align}
which was first defined in \eqref{eq:J*firstdefs}, is also central. (The second term in \eqref{eq:JR*} is chosen to have eigenvalue $-1/2$ on states annihilated by $Q_R$ and eigenvalue $+1/2$ on states annihilated by $Q_R^\dagger$, which cancels the R-charge of the supercharges.)
As explained in \cite{Lin:2022rzw, Lin:2022zxd} and briefly reviewed in Section \ref{sec:groundstates}, the pure super-JT gravity Hilbert space $\mathcal{H}^0_{\rm super-JT}$ decomposes with respect to this centre as the direct sum and integral
\begin{align}
\mathcal{H}_{\rm super-JT}^0 \cong \widetilde{\mathcal{H}}_0 \,\oplus\, \left[\underset{J_R^*}{\mathlarger{\mathlarger{\oplus}}} \int^\oplus \!\!dE_R \,\,\mathcal{H}_{E_R, J_R^*}\right].
\end{align}
Here the subspace $\widetilde{\mathcal{H}}_0$ of normalisable ground states with energy $E_R = 0$ is spanned by the single state $\ket{\widetilde{\Psi}_{J_R}}$ for each allowed R-charge with $|J_R| < 1/2$. The states $\ket{\widetilde{\Psi}_{J_R}}$ have $J_R^* = J_R$ because they are annihilated by all the supercharges. Meanwhile $\mathcal{H}_{E_R, J_R^*}$ is a four-dimensional irreducible representation of the left and right supercharge Clifford algebra and exists for all R-charges $J_R^* \in \mathbb{Z}/N$ and energies $E_R > {J_R^*}^2/4$. 

The algebra $\mathcal{A}_R$ is generated by its centre, together with the noncentral operators $Q_R$ and $Q_R^\dagger$. The operators $Q_R$ and $Q_R^\dagger$ annihilate all states in $\widetilde{\mathcal{H}}_0$. On $\mathcal{H}_{E_R, J_R^*}$, the operators $H_R^{-1/2}Q_R^\dagger$ and $H_R^{-1/2}Q_R$ act as fermion creation and annihilation operators. The algebra $\mathcal{A}_R$ is therefore Type I since it is a direct sum and integral over finite-dimensional factors.

The left-boundary algebra is generated by $H_L$, $Q_L$, $Q_L^\dagger$ and $J_L$. Its centre is generated by $H_L = H_R$ and
\begin{align}
J_L^* = J_L -  \frac{1}{2} H_L^{-1} [Q_L, Q_L^\dagger].
\end{align}
We showed in \eqref{eq:JL*JR*} that in pure super-JT gravity $J_L^* = - J_R^*$. So $\mathcal{A}_L$ and $\mathcal{A}_R$ share a common centre. Like $\mathcal{A}_R$, the algebra $\mathcal{A}_L$ is generated by its centre together with $Q_L$ and $Q_L^\dagger$, which annihilate $\widetilde{\mathcal{H}}_0$. And $H_L^{-1/2}Q_L^\dagger$ and $H_L^{-1/2} Q_L$ act as fermionic creation and annihilation operators on $\mathcal{H}_{E_R, J_R^*}$ that anticommute with $H_R^{-1/2}Q_R$ and $H_R^{-1/2}Q_R^\dagger$.

It follows from the discussion above that the algebras $\mathcal{A}_L$ and $\mathcal{A}_R$ are supercommutants, meaning that $\mathcal{A}_L$ consists of all bosonic operators that commute with $\mathcal{A}_R$, and all fermionic operators that a) commute with all bosonic operators in $\mathcal{A}_R$ and b) anticommute with all fermionic operators in $\mathcal{A}_R$. Supercommutants are the natural generalisation of commutants to graded algebras with fermions. However one can also construct the "true" commutant of $\mathcal{A}_R$ (i.e. the set of operators that commute with all operators in $\mathcal{A}_R$ whether bosonic or fermionic). The true commutant consists of bosonic operators in $\mathcal{A}_L$ together with fermionic operators in $\mathcal{A}_L$ multiplied by the fermion parity operator $(-1)^F$. This relationship is very general: the supercommutant of a graded algebra with fermion parity operator $(-1)^F$ always consists of bosonic operators in the commutant together with fermionic operators in the commutant multiplied by $(-1)^F$.

Unlike pure JT gravity,  we have seen that the boundary algebras in pure super-JT gravity are not purely commutative. However, they still have a large centre and, as a result, there exist operators acting on $\mathcal{H}^0_{\rm super-JT}$ that cannot be written using boundary observables. Pure super-JT gravity is not a holographic theory.

\subsection{Adding matter}\label{sec:addingmatter}
As in bosonic JT gravity, adding matter to super-JT gravity both enlarges the Hilbert space and introduces new boundary operators. We construct the matter operators out of supersymmetric matter boundary primaries $\Phi_{R,0}$, acting at $T =0$ on the right boundary. Such operators satisfy
\begin{align}
[j_3, \Phi_{R,0}] = i \Delta \Phi_{R,0}, ~~~ [j, \Phi_{R,0}] = q \Phi_{R,0} ~~\text{and}~~ [j_-, \Phi_{R,0}] = [g_-,\Phi_{R,0}] = [g_-^\dagger,\Phi_{R,0}] = 0.
\end{align}
Here $\Delta$ is the scaling dimension of $\Phi_{R,0}$ and $q$ is its charge. Throughout this section, we assume for simplicity that $\Phi_{R,0}$ is bosonic. However our results can easily be generalised to fermionic matter primaries by replacing all commutators with fermionic charges by anticommutators. Unlike in bosonic JT gravity we cannot take the supersymmetric primaries to all be self-adjoint unless $q=0$. Instead, $\Phi_{R,0}^\dagger$ will be a supersymmetric primary with scaling dimension $\Delta$ and R-charge $-q$.  Also, in contrast to our conventions for bosonic JT gravity, it will turn out to be convenient to count the identity operator as a supersymmetric matter primary with $\Delta = q =0$.

Now, let
\begin{align}
 \Phi_{R,0} (\tau(t),\theta_+(t),\bar\theta_+(t)) = e^{ij_+ \tau}e^{\theta_+ g_+ + \bar\theta_+ g_+^\dagger} \Phi_{R,0} e^{-\theta_+ g_+ - \bar\theta_+ g_+^\dagger}e^{-ij_+ \tau(t)}.
\end{align}
A somewhat laborious but fundamentally trivial calculation, carried out in Appendix \ref{app:gaugeinvariance}, shows that the operators
\begin{align}\label{eq:superJTphiR}
\mathbf{\Phi}_R(t) = e^{-\Delta \rho(t) + i qa(t)} \Phi_{R,0} (\tau(t),\theta_+(t),\bar\theta_+(t))
\end{align}
are gauge invariant under the diagonal $\hSU(1,1|1)$ action $\mathcal{D}(\mathcal{J}_i) = J_i^L + j_i + J_i^R$.

On wavefunctions $\mathbf{\Psi}_{JT}$ of the form \eqref{eq:superJTgaugefixwavefunction}, the operator $\mathbf{\Phi}_R(t)$ acts on $\Psi_{JT}$ as
\begin{align}\label{eq:phiRsuperJT}
\Phi_{R}(t) = e^{iH_R t} e^{-\Delta \rho + i qa} \Phi_{R,0} e^{-i H_R t}.
\end{align}
Let us make a few comments about this operator. Like local operators in quantum field theory, $\Phi_{R}(t)$ is only truly defined in a distributional sense. However smearing $\Phi_{R}(t)$ in Lorentzian boundary time $t$, or sandwiching it with a small amount of Euclidean time evolution leads to densely defined operators acting on the Hilbert space $\mathcal{H}_{\rm super-JT}$. Slightly confusingly, we have
\begin{align}
[J_R,\Phi_{R}] = [p_a + j, \Phi_R] = 2 q \Phi_R.
\end{align}
As a result, the super-JT boundary operator $\Phi_R$ has boundary R-charge $2q$ even though the matter primary $\Phi_{R,0}$ had matter R-charge $q$. The factor of two difference between the two is an unfortunate side effect of the conventions that we are using to normalise the different R-charges. In particular, the boundary R-charge $J_R$ is defined so that the boundary supercharges have charge $\pm 1$, for consistency with the conventions of \cite{Lin:2022rzw, Lin:2022zxd}. Meanwhile,  the matter supercharges have charge $\pm \frac{1}{2}$ with respect to the matter R-charge $j$, so that the normalisation of $j$ matches the normalisation of the $SU(2)$ matter charges $j_-$. 
Finally, it is easy to check that \eqref{eq:phiRsuperJT} commutes with the left-boundary Lie superalgebra \eqref{eq:QL} as expected for a right-boundary operator. 

We can also construct gauge-invariant left-boundary matter operators by an analogous method. Left-boundary matter primaries $\Phi_{L,0}$ satisfy
\begin{align}
[j_3, \Phi_{L,0}] =-i \Delta \Phi_{L,0}, ~~~ [j, \Phi_{L,0}] = q \Phi_{L,0} ~~\text{and}~~ [j_+, \Phi_{L,0}] = [g_+,\Phi_{R,0}] = [g_+^\dagger,\Phi_{R,0}] = 0.
\end{align}
This ensures that the operators
\begin{align}\label{eq:superJTphiL}
\mathbf{\Phi}_L(t') = e^{-\Delta \rho'(t') + i qa'(t')} e^{ij_- \tau'(t')}e^{\theta_-'(t') g_- + \bar\theta_-'(t') g_-^\dagger}\Phi_{L,0} e^{-\theta_-'(t') g_- - \bar\theta_-'(t') g_-^\dagger}e^{-ij_- \tau'(t')}
\end{align}
are gauge-invariant. On the physical Hilbert space $\mathcal{H}_{\rm super-JT}$, $\mathbf{\Phi}_L(t')$ acts as
\begin{align}
\Phi_{L}(t') = e^{iH_L t'} e^{-\Delta \rho - i qa} \Phi_{L,0} e^{-i H_L t'},
\end{align}
which commutes both with $\Phi_R(t)$ and with the right-boundary Lie superalgebra. We therefore have supercommuting boundary algebras $\mathcal{A}_R$, generated by $Q_R, Q_R^\dagger, J_R, \Phi_R$, and $\mathcal{A}_L$, generated by $Q_L, Q_L^\dagger, J_L, \Phi_L$.

Many properties of $\mathcal{A}_L$ and $\mathcal{A}_R$ can be deduced most easily using Euclidean arguments, as was done for bosonic JT gravity in \cite{penington2023algebras, kolchmeyer2023neumann}. However, as we showed explicitly for bosonic JT gravity in Section \ref{sec:JTalgebras}, it should also be possible to obtain the same results directly in the Lorentzian canonically quantised theory (without appeal to Euclidean gravity path integrals). We will not do so in full generality here because the relevant calculations quickly become unnecessarily involved.

The Hartle-Hawking state $\ket{\beta^{(0)}} \in \mathcal{H}^0_{\rm super-JT}$ in canonically quantised pure super-JT gravity was described in detail in \cite{Lin:2022rzw, Lin:2022zxd}. We define the Hartle-Hawking state in super-JT gravity with matter to be
\begin{align}
\ket{\beta} = \ket{\beta^{(0)}} \ket{\Omega_{\rm matt}}.
\end{align}
By construction, the state $\ket{\beta}$ has the property that expectation values of operators in the boundary algebra $\mathcal{A}_R$ can be computed using disc Euclidean path integrals with a Euclidean evolution of time $\beta$ connecting the bra to the ket. Given an operator $a \in \mathcal{A}_R$, we define the trace
\begin{align}\label{eq:susytrace}
\Tr(a) = \lim_{\beta \to 0} \braket{\beta| a |\beta}
\end{align}
For operators of the form
\begin{align}\label{eq:nicestringa}
a = e^{-\beta_0 H_R +\theta_0 Q_R + \bar\theta_0 Q_R^\dagger}e^{-ia_0 J_R} \Phi^{(1)}_R e^{-\beta_1 H_R +\theta_1 Q_R + \bar\theta_1 Q_R^\dagger}e^{-ia_1 J_R} \Phi^{(2)}_R \dots
\end{align}
 this trace can be computed using a Euclidean gravitational path integral on a disc with boundary conditions determined by the operators appearing in $a$, as explained briefly in Section \ref{sec:JTalgebras} and in much more detail in \cite{penington2023algebras}. In particular, the limit $\beta \to 0$ is necessary to remove the additional Euclidean time evolution associated to the bra and the ket Hartle-Hawking states. Since operators of this form span $\mathcal{A}_R$ and the path integral computing $\Tr(a)$ is manifestly cyclic, it follows that \eqref{eq:susytrace} is indeed a trace.

As explained for bosonic JT gravity in \cite{penington2023algebras}, if we assume that any state in $\mathcal{H}_{\rm super-JT}$ can be prepared by a Euclidean path integral, then \eqref{eq:susytrace} allows us to identify the Hilbert space constructed from the algebra $\mathcal{A}_R$ using the Hilbert-Schmidt inner product with $\mathcal{H}_{\rm super-JT}$. This assumption could be established rigorously by showing that states in $\mathcal{H}_{\rm super-JT}$ are uniquely determined by their matter representation, boundary energies, R-charges and the boundary supercharges that annihilate them, as we showed for bosonic JT gravity in Section \ref{sec:spectra} and for states in $\mathcal{H}_{\rm super-JT}$ with $E_R = 0$ in Section \ref{sec:simulground}. We see no reason why the techniques used in those sections would not   generalise to states in $\mathcal{H}_{\rm super-JT}$ with $E_L, E_R > 0$, although working through all the details would be somewhat tedious and probably unenlightening.

As explained in Section \ref{sec:JTalgebras}, it follows from the identification of the algebra $\mathcal{A}_R$ with the Hilbert space $\mathcal{H}_{\rm super-JT}$ that the commutant of $\mathcal{A}_R$ should be identified the right action of $\mathcal{A}_R$ on itself. By arguments analogous to those in Section \ref{sec:JTalgebras}, this right action should be identified with $\mathcal{A}_L$ up to multiplication by a factor of $(-1)^F$ for fermionic operators. The latter factor arises because we need to commute the operator $a' \in \mathcal{A}_L$ past the operator $a \in \mathcal{A}_R$ that created the state in order to have it ``act from the right'', and doing so picks up a sign when both $a$ and $a'$ are fermionic. It follows from this discussion that the boundary algebras in super-JT gravity with matter are supercommutants (just like in pure super-JT gravity).

We would like to show that $\mathcal{A}_R$ is a von Neumann factor. It would then follow immediately that $\mathcal{A}_R$ is a Type II$_\infty$ factor since $\Tr(\mathds{1}) = \infty$ and projectors in $\mathcal{A}_R$ with arbitrarily small trace exist. Recall that the argument we used to show this for JT  gravity in Section \ref{sec:JTalgebras}, which was based on an argument in \cite{kolchmeyer2023neumann}, proceeded by first showing that any operator in $\mathcal{A}_L \cap \mathcal{A}_R$ must preserve the vacuum sector i.e. commute with $\Pi_\Omega$. We then used the fact that operators in $\mathcal{A}_R$ can be written using at most one matter operator insertion to argue that the only such operators are functions $f(H_R)$ of the Hamiltonian $H_R$, which are not central in JT gravity with matter unless $f$ is constant.

In super-JT gravity, we expect that the following claims are true and follow from obvious generalisations of the bosonic arguments in Sections \ref{sec:spectra} and \ref{sec:JTalgebras}: (a) $(H_L - H_R)$ has purely continuous spectrum outside of the matter vacuum sector when $E_R > 0$ and (b) any operator in $\mathcal{A}_R$ can be written using at most a single matter insertion. It follows from (a) that any central operator must preserve the vacuum sector whenever $E_L = E_R > 0$. It then follows from (b) that such operators are contained in the universal enveloping algebra of the right-boundary Lie superalgebra $Q_R, Q_R^\dagger, H_R, J_R$, which should not contain any nontrivial central operators if all matter charges are present.\footnote{If no matter states or operators with the minimal allowed R-charge exist, the super-JT gravity theory will contain a nontrivial centre describing the boundary charge $J_R^*$ modulo the smallest nonzero matter charge present in the theory.} 

The truly novel ingredient that is encountered only in the supersymmetric theory is the presence of the simultaneous ground states from Section  \ref{sec:simulground} that are outside the vacuum sector but that are annihilated by both $H_L$ and $H_R$ (and hence also by $(H_L - H_R)$). It would be consistent with both (a) and (b) for there to exist a nontrivial central operator in $\mathcal{A}_R$ that acts as the identity on all states with $E_R > 0$ but that acts nontrivially on the space of $E_R = 0$ ground states (and in particular that maps the matter vacuum ground state to a simultaneous left- and right-boundary ground state where matter fields are excited). For this reason, along with the other reasons explained in the introduction to this paper,  we shall therefore restrict our attention, for the rest of this section, to the projection of $\mathcal{A}_R$ onto right-boundary ground states.

\subsection{The projection onto ground states}\label{sec:project}
In light of the discussion above, we want to study the algebra $\widetilde{\mathcal{A}}_R = \tilde{\Pi}_{R}\mathcal{A}_R \tilde{\Pi}_R$ constructed by projecting $\mathcal{A}_R$ onto the $E_R = 0$ subspace $\tilde{\Pi}_R \mathcal{H}_{\rm super-JT}$. In fact, for the moment we shall restrict our attention further to the algebra 
\be
\widetilde{\mathcal{A}}_{R,0} = \tilde{\Pi}_{R,0}\mathcal{A}_R \tilde{\Pi}_{R,0} \subseteq \widetilde{\mathcal{A}}_R
\ee
 found by projecting onto states with $E_R, J_R = 0$. (If $J_R$ is quantised to integer/half-integer values then $\widetilde{\mathcal{A}}_{R,0} \cong \widetilde{\mathcal{A}}_{R}$.) Since
\begin{align}\label{eq:traceprojection}
\lim_{\beta\to 0} \braket{\beta|\tilde{\Pi}_{R,0} \,a\, \tilde{\Pi}_{R,0}|\beta} =\bra{\Omega_{\rm matt}} \bra{\tilde\Psi_0} a \ket{\tilde\Psi_0} \ket{\Omega_{\rm matt}},
\end{align}
with $\ket{\tilde\Psi_0}$ given by setting $J_R = 0$ in \eqref{eq:puresuperJTgroundstate}, we expect that $\widetilde{\mathcal{A}}_{R,0}$ should be a Type II$_1$ von Neumann algebra with normalisable tracial state $\ket{\widetilde\Psi_0} \ket{\Omega_{\rm matt}}$.

Recall from Section \ref{sec:groundstates} that we can identify the $E_R = J_R = 0$ subspace of $ \mathcal{H}_{\rm super-JT}$ with the matter Hilbert space $\mathcal{H}_{\rm matt}$ via the isometry $\tilde V_0: \mathcal{H}_{\rm matt} \to \mathcal{H}_{\rm super-JT}$. This identifies $\tilde{\Pi}_{R,0} a \tilde{\Pi}_{R,0}$ with the QFT operator
\begin{align}
\tilde a = \tilde V_0^\dagger a \tilde V_0.
\end{align}
Under this identification, the above discussion suggests that $\widetilde{\mathcal{A}}_{R,0}$ will become a Type II$_1$ factor acting on the QFT Hilbert space $\mathcal{H}_{\rm matt}$ such that the vacuum state $\ket{\Omega_{\rm matt}}$ is tracial (i.e. maximally entangled). The rest of this subsection will be devoted to a) verifying that expectation and b) studying the structure of $\widetilde{\mathcal{A}}_{R,0}$ as a QFT algebra.

Since the right-boundary Lie superalgebra is annihilated by $\tilde{\Pi}_{R,0}$, $\widetilde{\mathcal{A}}_{R,0}$ is generated by the operators
\begin{align}
\tilde\Phi_{R}^i &= \tilde V^\dagger_0 \Phi_{R,0}^i e^{- \Delta_i \rho} \tilde V_0
\\\label{eq:tildephidef}&=\frac{8}{\pi}\int_{-\infty}^\infty d \rho\, \exp(-(2\Delta_i +2)\rho) \,\times\\\nonumber & ~~~~\left[ \sqrt{1 + g_+ g_+^\dagger} K_{1/2}(2 \sqrt{1 + j_+} e^{-\rho}) \Phi_{R,0}^i K_{1/2}(2 \sqrt{1 + j_+} e^{-\rho})  \sqrt{1 + g_+ g_+^\dagger}\right.\\ \nonumber&~~~~~~\left.  + g_+^\dagger K_{1/2}(2 \sqrt{1 + j_+} e^{-\rho}) \Phi_{R,0}^i K_{1/2}(2 \sqrt{1 + j_+} e^{-\rho})  g_+\right.\\\nonumber&~~~~~~\left. + g_+ K_{1/2}(2 \sqrt{1 + j_+} e^{-\rho}) \Phi_{R,0}^i K_{1/2}(2 \sqrt{1 + j_+} e^{-\rho})  g_+^\dagger\right.\\ &~~~~~~\left. + \sqrt{1 + g_+^\dagger g_+ } K_{1/2}(2 \sqrt{1 + j_+} e^{-\rho}) \Phi_{R,0}^i K_{1/2}(2 \sqrt{1 + j_+} e^{-\rho})  \sqrt{1 + g_+^\dagger g_+}\right]\nonumber
\end{align}
for each R-charge neutral supersymmetric primary $\Phi^i_{R,0}$. We are not aware of any significant simplification of the explicit formula \eqref{eq:tildephidef} except in special cases (e.g. if $\Phi_R^i = \mathds{1}$ then $\tilde\Phi_{R}^i = \mathds{1}$). 

We would like to have a general proof that the algebra $\widetilde{\mathcal{A}}_{R,0}$ is a factor. However, this is difficult to find because the product of two operators of the form \eqref{eq:tildephidef} depends on the specific operator product expansion found in the matter theory. While it is in principle possible that there exist quantum field theories for which $\widetilde{\mathcal{A}}_{R,0}$ has a nontrivial centre, it seems likely that any such centre will be highly specific to the matter theory in question, and we expect that in most, or perhaps even all, reasonable matter theories the centre will be trivial. Except where otherwise stated, we shall therefore simply assume for the remainder of this section that we are working with a matter QFT where $\widetilde{\mathcal{A}}_{R,0}$ is a factor and prove that $\widetilde{\mathcal{A}}_{R,0}$ must then satisfy all our other desired properties.  

Without knowledge of super-JT gravity, it would seem quite implausible that, for arbitrary $i,j$, the product
\begin{align} \label{eq:maintextgsalgebraproduct}
\tilde\Phi_R^i \tilde\Phi_R^j = \sum_k A^{ij}_k \tilde\Phi_R^k
\end{align}
with $A^{ij}_k$ $c$-numbers, so that the operators \eqref{eq:tildephidef} actually span (rather than merely generate) the entire algebra $\widetilde{\mathcal{A}}_{R,0}$.\footnote{Here, it is important that the sum over $\tilde\Phi_R^k$ includes the identity operator, which has $\Delta_k = 0$.} Nevertheless we will now show that \eqref{eq:maintextgsalgebraproduct} is indeed true, with the explicit coefficients
\begin{align}\label{eq:Aijk}
A^{ij}_k  = C_{ij\bar{k}} \frac{4^{\Delta_k - \Delta_i -\Delta_j} \Gamma(1 + 2\Delta_i) \Gamma(1+ 2\Delta_j)}{\Gamma(1+ \Delta_i + \Delta_j +\Delta_k)}.
\end{align}
Here, as in \eqref{eq:threepointfunction}, the constants $C_{ij\bar{k}}$ characterise the matter supersymmetric primary three-point function
\begin{align}\label{eq:susythreepointfunction}
\braket{\Omega_{\rm matt}|{\Phi^k_{L,0}}^\dagger \Phi^i_{R,0} e^{- j_+ \tau} \Phi^j_{R,0}  |\Omega_{\rm matt}} = \frac{C_{ij\bar{k}}}{\tau^{\Delta_i + \Delta_j -\Delta_k}}.
\end{align}
The noncommutativity of the product \eqref{eq:maintextgsalgebraproduct} comes from the fact that, in general, $C_{ij\bar{k}}$ is not equal to $C_{ji\bar{k}}$; see Footnote \ref{foot:CijkCjik} for related discussion.

To show \eqref{eq:maintextgsalgebraproduct} it is helpful to project $\widetilde{\mathcal{A}}_{R,0}$ onto the kernel of $\tilde H_L$ and $\tilde J_L$ i.e. the subspace of  simultaneous ground states with $J_L = J_R = j =0$. Since
\begin{align}
[\tilde H_L, \tilde\Phi_R^i] = \tilde V^\dagger_0 [H_L , \Phi_{R}^i]  \tilde V_0 = 0,
\end{align}
the projection $\tilde \Pi_{L,0}$ onto this subspace is contained in the commutant algebra $\widetilde{\mathcal{A}}_{R,0}'$. Since every representation of a von Neumann factor is faithful, it follows that this projection cannot change the structure of $\widetilde{\mathcal{A}}_{R,0}$. This is in fact the only step for which we will need the assumption that $\widetilde{\mathcal{A}}_{R,0}$ is a factor.

Since all $\hSU(1,1|1)$ charges annihilate $\ket{\Omega_{\rm matt}}$, the action of \eqref{eq:tildephidef} on $\ket{\Omega_{\rm matt}}$ simplifies considerably. In fact, the middle two terms of \eqref{eq:tildephidef} vanish and we are left with
\begin{align}\label{eq:tildephiRomega}
\tilde\Phi_{R}^i \ket{\Omega_{\rm matt}} = \frac{8}{\pi}\left[ 1 +\sqrt{1 + j_+}\right]  \int_{-\infty}^\infty d \rho\, e^{-(2\Delta_i +2)\rho} K_{1/2}(2 \sqrt{1 + j_+} e^{-\rho}) K_{1/2}(2 e^{-\rho})\Phi_{R,0}^i  \ket{\Omega_{\rm matt}}.
\end{align}
Here we used the identity
\begin{align}\label{eq:sqrtggdaggeridentity}
\sqrt{1 + g_+^\dagger g_+} + \sqrt{1 + g_+ g_+^\dagger} = 1 + \sqrt{1 + j_+}.
\end{align}
which follows from the splitting of $\mathcal{H}_{\rm matt}$ (outside of the vacuum)  into two orthogonal subspaces annihilated respectively by $g_+$ and $g_+^\dagger$. On the former subspace we have $g_+ g_+^\dagger = j_+$ and on the latter subspace we have $g_+^\dagger g_+ = j_+$.

It follows immediately that the right-boundary two-point function  can be written as
\begin{align}
\braket{\Omega_{\rm matt}|\tilde{\Phi}^{i\dagger}_R \tilde\Phi^j_R |\Omega_{\rm matt}} 
= \braket{\Omega_{\rm matt} |\Phi^{i\dagger} f(j_+) \Phi^j|\Omega_{\rm matt}},
\end{align}
where
\begin{align}
f(j_+) = \frac{64 }{\pi^2}\left(1+ \sqrt{1 + j_+}\right)^2 \left[ \int d \rho\, \exp(-(2\Delta_j +2)\rho)  K_{1/2}(2 \sqrt{1 - j_+} e^{-\rho}) K_{1/2}(2 e^{-\rho})\right]^2.
\end{align}
We can then use \eqref{eq:intj+} to obtain
\begin{align}\label{eq:neutral2pt}
\braket{\tilde{\Phi}^{i\dagger}_R \tilde\Phi^j_R} &=\frac{64 \delta^{ij}}{\pi^2\Gamma(2\Delta_i)} \int dj_+ j_+^{2\Delta_i -1}\left(1+ \sqrt{1 + j_+}\right)^2 \,\\\nonumber&\qquad\qquad~~~~~\times\left[ \int d \rho\, \exp(-(2\Delta_j +2)\rho)  K_{1/2}(2 \sqrt{1 - j_+} e^{-\rho}) K_{1/2}(2 e^{-\rho})\right]^2.
\end{align}
We can also compute the ground left-right boundary two-point function. Projecting left-boundary operators into the subspace of $J_R = 0$ ground states leads to
\begin{align}
\tilde\Phi_L^i = \tilde V_0^\dagger \Phi_L^i V_0^\dagger = \braket{\tilde\Psi_0 |U_3 U_2 \Phi^i_{L,0} U_2^\dagger U_3^\dagger |\tilde\Psi_0} = \Phi_{L,0}^i,
\end{align}
which commutes with $\tilde \Phi_R^j$ as one would hope. It follows that
\begin{align} \label{eq:leftrightneutral2pt}
\braket{\tilde\Phi_{L}^{i\dagger} \tilde\Phi_R^j} &=\braket{\Omega_{\rm matt}| \Phi_{L,0}^i \tilde\Phi_R^j|\Omega_{\rm matt}}
\\&=  \frac{16 \delta^{ij}}{\pi}\int_{-\infty}^\infty d \rho\, \exp(-(2\Delta +2)\rho) K_{1/2}(2 e^{-\rho})  K_{1/2}(2 e^{-\rho})  
\end{align}
The integrals for both \eqref{eq:neutral2pt} and \eqref{eq:leftrightneutral2pt} are done in Appendix \ref{sec:susy2ptcalcs}. We find that
\begin{align}\label{eq:susyneutral2ptfinalformula}
\braket{ \tilde\Phi_{L}^{i\dagger} \tilde\Phi_R^j} = \braket{\tilde \Phi_{R}^{i\dagger} \tilde\Phi_R^j} = 2^{-4\Delta_i} \delta^{ij} \Gamma(2\Delta_i + 1).
\end{align}
From a quantum field theory perspective, there was no obvious reason that these two correlation functions would be related. But, of course, we know from super-JT gravity that they had to match because they are both computed by the same Euclidean path integral, as shown in Figure \ref{fig:betainf}.

Recall from Section \ref{sec:simulground} that there exists a unique state in the kernel of $\tilde H_L$ within each R-charge neutral $\hSU(1,1|1)$ representation. Since the right-boundary two-point function \eqref{eq:susyneutral2ptfinalformula} is nonzero for all $i = j$, this state can be written (up to normalisation) as $\tilde \Phi_R^i \ket{\Omega_{\rm matt}}$. Furthermore, since the left-right and right-boundary two-point functions agree, we can also write
\begin{align}\label{eq:susyleftact=rightact}
\tilde \Pi_{L,0} \Phi_{L,0}^i \ket{\Omega_{\rm matt}} = \tilde \Phi_R^i \ket{\Omega_{\rm matt}} .
\end{align}

We will now show that the operators $\tilde\Phi^i_R$ span $\widetilde{\mathcal{A}}_R$. We already argued that  an operator $\tilde a \in \widetilde{\mathcal{A}}_{R,0}$ is uniquely determined by its projection using $\tilde \Pi_{L,0}$ and hence by the matrix elements
\begin{align}
\braket{\Omega_{\rm matt}|\tilde \Phi^{k\dagger}_R \tilde a \tilde\Phi^{\ell}_R|\Omega_{\rm matt}} = \braket{\Omega_{\rm matt}|\tilde \Phi^{k\dagger}_R \Phi^\ell_{L,0} \tilde a |\Omega_{\rm matt}}.
\end{align}
But, since $\tilde a \ket{\Omega_{\rm matt}}$ is in the image of $\tilde \Pi_{L,0}$,  we must have
\begin{align}
\tilde a \ket{\Omega_{\rm matt}} = \sum_k A_k \tilde\Phi_R^k \ket{\Omega_{\rm matt}}
\end{align}
for some set of $c$-numbers $A_k$. It follows that 
\begin{align}
\tilde a = \sum_k A_k \tilde\Phi_R^k
\end{align}
as was claimed for the product $\tilde a = \tilde\Phi^i_R \tilde\Phi^j_R$ in \eqref{eq:maintextgsalgebraproduct}. To compute the $c$-numbers $A^{ij}_k$ for \eqref{eq:maintextgsalgebraproduct} explicitly, it suffices to consider the three-point function
\begin{align}\label{eq:threepointformula}
\braket{{\Phi^k_{L,0}}^\dagger \tilde\Phi^i_R \tilde\Phi^j_R} &= \braket{\Omega_{\rm matt}|{\Phi^k_{L,0}}^\dagger \Phi^i_R \Phi^j_R|\Omega_{\rm matt}}
\\
&= \frac{64 C_{ijk}}{\pi^2\Gamma(\Delta_i + \Delta_j - \Delta_k)} \int dj_+ j_+^{\Delta_i + \Delta_j - \Delta_k -1}\left(1+ \sqrt{1 + j_+}\right)^2 \,\times\\\nonumber&~~~~~ \left[ \int d \rho\, \exp(-(2\Delta_j +2)\rho)  K_{1/2}(2 \sqrt{1 - j_+} e^{-\rho}) K_{1/2}(2 e^{-\rho})\right]^2.
\end{align}
The integral  in \eqref{eq:threepointformula}, which was found by using \eqref{eq:tildephiRomega} and then taking an inverse Laplace transform of \eqref{eq:susythreepointfunction},\footnote{The inverse Laplace transform is only defined when $\Delta_k < \Delta_i + \Delta_j$. However when this condition is not satisfied we can simply use \eqref{eq:susyleftact=rightact} to cyclically permute $\tilde\Phi^i_R, \tilde\Phi^j_R, \tilde\Phi^k_R$ until it is.} is again computed in Appendix \ref{sec:susy2ptcalcs}. We obtain
\begin{align}\label{eq:threepointanswer}
\braket{{\Phi^k_{L,0}}^\dagger \tilde\Phi^i_R \tilde\Phi^j_R} = C_{ijk} \frac{2^{-2(\Delta_i+ \Delta_j + \Delta_k)}\Gamma(2\Delta_i + 1)\Gamma(2\Delta_j + 1) \Gamma(2\Delta_k + 1)}{\Gamma(1 + \Delta_i + \Delta_j + \Delta_k)}
\end{align}
To our knowledge, this is the first time that the super-JT gravity three-point function has been computed by any method, which shows the technical power of the algebraic approach we are using. It is also worth emphasizing how remarkably simple this formula is compared with the three-point function \eqref{eq:3ptfunctionAppA} in bosonic JT gravity, which was too long to write explicitly but that would have involved three copies of the generalised hypergeometric function ${}_4F_3$. Comparing \eqref{eq:threepointanswer} with \eqref{eq:susyneutral2ptfinalformula}, we immediately obtain the formula for $A^{ij}_k$ given in \eqref{eq:Aijk}.

Abstractly, there would be no reason to expect that the algebra \eqref{eq:maintextgsalgebraproduct} would be associative. However by realising \eqref{eq:maintextgsalgebraproduct} using the explicit QFT operators \eqref{eq:tildephidef}, we have shown that it is. Note that the proof that the product \eqref{eq:maintextgsalgebraproduct} is associative did not require an assumption that $\widetilde{\mathcal{A}}_{R,0}$ is a factor: the product \eqref{eq:maintextgsalgebraproduct} is realised by the associative algebra $\tilde \Pi_{L,0}\widetilde{\mathcal{A}}_{R,0}\tilde \Pi_{L,0}$ found after projecting  \eqref{eq:tildephidef} onto the kernel of $\tilde H_L$, even if that projection isn't faithful and so the algebraic structure of the projection differs from the algebraic structure of $\widetilde{\mathcal{A}}_{R,0}$ itself.

The associativity of \eqref{eq:maintextgsalgebraproduct} implies the crossing relation
\begin{align}\label{eq:crossing}
\tilde\Phi_R^i \tilde\Phi^j_R \tilde\Phi^k_R = \sum_{\ell m} A^{ij}_\ell A^{\ell k}_m \tilde\Phi^m_R = \sum_{\ell m} A^{i \ell}_m A^{jk}_\ell \tilde\Phi^m_R,
\end{align}
which gives an infinite set of constraints that need to be satisfied by any supersymmetric quantum field theory in $\AdS_2$. The relation \eqref{eq:crossing} could presumably be obtained from the crossing relations required by the full matter operator product expansion by integrating against the zero-energy (and zero R-charge) super-JT gravity propagator, which was found explicitly in \cite{Lin:2022rzw, Lin:2022zxd}. However, we have not attempted to demonstrate this directly. Over the last decade, the conformal bootstrap program  has been able to constrain the space of conformal field theories to a remarkable degree using numerical constraints from crossing relations \cite{simmons2017conformal, poland2019conformal}. This is normally achieved by first reducing the full crossing relations to some more manageable set that can be solved numerically. It would be interesting to understand whether $\widetilde{\mathcal{A}}_{R,0}$, or generalisations of $\widetilde{\mathcal{A}}_{R,0}$ to higher-dimensional supersymmetric QFTs in anti-de Sitter space or perhaps to higher-dimensional superconformal field theories, has any value in constraining the space of allowed theories. 

With \eqref{eq:maintextgsalgebraproduct} in hand, it is easy to show,  as hoped, that the vacuum state $\ket{\Omega_{\rm matt}}$ is tracial. Since $\tilde \Phi^i_R$ and $\tilde\Phi^{i\dagger}_R$ have the same scaling dimension $\Delta_i$, it follows immediately that
\begin{align}\label{eq:traceneutraloperators}
\braket{\Omega_{\rm matt}|\tilde\Phi^{i\dagger}_R\tilde\Phi^{j}_R|\Omega_{\rm matt}} = \braket{\Omega_{\rm matt}|\tilde\Phi^{j}_R \tilde\Phi^{i\dagger}_R|\Omega_{\rm matt}},
\end{align}
which is all we need for $\ket{\Omega_{\rm matt}}$ to be tracial. In particular, this means that
\begin{align}
\Tr(\mathds{1}) = 1 ~~~~~\text{and}~~~~~\Tr(\tilde\Phi^i_R) = 0
\end{align}
for all nontrivial matter primaries $\Phi^i_{R,0}$. Since $\widetilde{\mathcal{A}}_{R,0}$ is infinite dimensional but the trace of the identity is finite, $\widetilde{\mathcal{A}}_{R,0}$ must be Type II$_1$. Note that the assumption that $\widetilde{\mathcal{A}}_{R,0}$ is a factor was not actually needed to show that $\ket{\Omega_{\rm matt}}$ is tracial, but it was needed to prove that the trace $\ket{\Omega_{\rm matt}}$ defines is faithful (or equivalently to show that $\ket{\Omega_{\rm matt}}$ is separating on $\widetilde{\mathcal{A}}_{R,0}$) and that it is unique.

The commutant algebra $\widetilde{\mathcal{A}}_{R,0}'$ includes the left-boundary matter operators $\Phi_{L,0}$ and the left-boundary Lie superalgebra $\tilde Q_L, \tilde Q_L^\dagger, \tilde H_L, \tilde J_L$.  Since the left super-JT boundary algebra $\mathcal{A}_L$ commutes with the projection $\tilde \Pi_{R,0}$, the claims made (but not rigorously proved) in Section \ref{sec:addingmatter} are sufficient to show that $\widetilde{\mathcal{A}}_{R,0}'$ is a Type II$_\infty$ factor that is isomorphic to $\mathcal{A}_L$. 

A Hilbert space representation $\mathcal{K}: \mathcal{A} \to \mathcal{B}(\mathcal{H})$ of a von Neumann factor $\mathcal{A}$ can be uniquely characterised, up to isomorphism, by a parameter $d \in [0,\infty]$ called the dimension of $\mathcal{K}$. (This dimension should not be confused with the Hilbert space dimension of the space $\mathcal{H}$, which is infinite for any nontrivial representation of an infinite-dimensional factor.) The dimension $d$ of each representation is fixed, up to an overall normalisation factor, by the requirements that a) a representation $\mathcal{K}'$ with dimension $d'$ is isomorphic to the restriction of $\mathcal{K}$ to a subspace of $\mathcal{H}$ if and only if $d' \leq d$ and, b) for any $\mathcal{K}, \mathcal{K'}$, the representation $\mathcal{K} \oplus \mathcal{K}'$ has dimension $d+ d'$. For a Type II$_1$ factor $\mathcal{A}$,  representations of $\mathcal{A}$ exist for all $d \in [0,\infty]$ and $d$ is conventionally normalised so that the canonical action of $\mathcal{A}$ on itself has dimension $d=1$. For all representations $\mathcal{K}: \mathcal{A} \to \mathcal{B}(\mathcal{H})$ of $\mathcal{A}$ with finite dimension $d > 0$, the commutant algebra $\mathcal{A}' \subseteq \mathcal{B}(\mathcal{H})$ is also a Type II$_1$ factor. In the $d = \infty$ representation, the commutant $\mathcal{A}'$ is a Type II$_\infty$ factor. Finally, the $d=0$ representation is trivial. With these properties in mind, it follows immediately from our discussion above that the matter Hilbert space $\mathcal{H}_{\rm matt}$ is isomorphic to the $d = \infty$ representation of the algebra $\widetilde{\mathcal{A}}_{R,0}$.

There is a natural isomorphism between the algebra $\widetilde{\mathcal{A}}_{R,0}$ (with the Hilbert-Schmidt inner product) to the subspace $\tilde \Pi_{L,0} \mathcal{H}_{\rm matt} \subseteq \mathcal{H}_{\rm matt}$ of states annihilated by $\tilde H_L, \tilde J_L$ where the operator $\tilde\Phi^i_R$ is identified with the state $\tilde\Phi^i_R \ket{\Omega_{\rm matt}}$. The action of $\widetilde{\mathcal{A}}_{R,0}$ on $\tilde \Pi_{L,0} \mathcal{H}_{\rm matt}$ is identified with the left action of $\widetilde{\mathcal{A}}_{R,0}$ on itself, which is the canonical $d = 1$ representation of the algebra $\widetilde{\mathcal{A}}_{R,0}$.  

After projecting using $\tilde \Pi_{L,0}$, the algebra  $\mathcal{A}_L$ becomes the ground state algebra
\begin{align}
\widetilde{\mathcal{A}}_{L,0} = \tilde \Pi_{L,0} \mathcal{A}_L \tilde \Pi_{L,0}.
\end{align}
The algebra $\widetilde{\mathcal{A}}_{L,0}$ is generated by $\tilde \Pi_{L,0} \Phi_{L,0}^i \tilde \Pi_{L,0}$ for  any R-charge neutral supersymmetric primary $\Phi_{L,0}^i$.  It then follows from \eqref{eq:susyleftact=rightact} that the algebra $\widetilde{\mathcal{A}}_{L,0}$ is identified with the right action of $\widetilde{\mathcal{A}}_{R,0}$ on itself, which shows directly that $\widetilde{\mathcal{A}}_{L,0}$ is the supercommutant of $\widetilde{\mathcal{A}}_{R,0}$ on the space of simultaneous ground states $\tilde \Pi_{L,0} \mathcal{H}_{\rm matt}$.\footnote{Note that this analysis of $\widetilde{\mathcal{A}}_{L,0}$ (unlike the analysis of the  left boundary algebra $\mathcal{A}_L$ prior to the projection $\tilde \Pi_{L,0}$) does not require us to use any properties of the original super-JT boundary algebra $\mathcal{A}_{L,0}$.}

\subsection{Charged operators and BPS microstate counts}
It remains to consider ground states and operators with nonzero R-charge. The main difference from the discussion above is that the full Hilbert space of right-boundary ground states is no longer naturally isomorphic to $\mathcal{H}_{\rm matt}$, but instead to the direct sum
\begin{align}
\widetilde{\mathcal{H}}_R = \underset{|J_R| < \frac{1}{2}}{\oplus} \mathcal{H}_{\rm matt}^{(J_R)},
\end{align}
where each subspace $\mathcal{H}_{\rm matt}^{(J_R)}$, which describes the space of right-boundary ground states with R-charge $J_R$, is naturally isomorphic to $\mathcal{H}_{\rm matt}$ but is acted on differently by the right-boundary algebra  $\widetilde{\mathcal{A}}_R$. Indeed, the algebra $\widetilde{\mathcal{A}}_R$ is generated by the R-charge $J_R$ and the matter operators 
\begin{align}\label{eq:chargedtildedef}
\tilde\Phi^i_R &= \sum_{J_R, J_R'} \tilde V_{J_R}^\dagger \Phi^i_R \tilde V_{J_R'},
\end{align}
where $\tilde V_{J_R}: \mathcal{H}_{\rm matt}^{(J_R)} \to \mathcal{H}_{\rm super-JT}$ was defined in \eqref{eq:tildeV} and the sum is over all allowed pairs of R-charges  $J_R, J_R'$. For neutral matter operators, expanding out \eqref{eq:chargedtildedef} explicitly leads to an integral over the Wilson line phase $a$ that fixes $J_R = J_R'$ and so they act diagonal (but differently) within each R-charge sector $\mathcal{H}_{\rm matt}^{(J_R)}$. However, for a matter operator $\Phi^i_R$ with nonzero R-charge $q_i$, the integral over $a$ instead sets $J_R' = J_R + 2 q_i$. As a result, such operators do not commute with $J_R$ and do not preserve $\mathcal{H}_{\rm matt}^{(J_R)}$.\footnote{Note that projecting onto a fixed $\tilde J_L =2 j - J_R$ projects each $\mathcal{H}_{\rm matt}^{(J_R)}$ sector onto a different R-charge sector of $\mathcal{H}_{\rm matt}$. For each allowed $J_L$, this leads to an action of $\widetilde{\mathcal{A}}_R$ on a single copy of $\mathcal{H}_{\rm matt}$ (or, more precisely, on the subspace of $\mathcal{H}_{\rm matt}$ with $|J_L - 2j| < 1/2$) that can be reconstructed out of these pieces).}

The right-boundary two-point function for supersymmetric primaries with scaling dimension $\Delta_i$ and R-charge $q_i$ is computed in Appendix \ref{sec:susy2ptcalcs}. After a somewhat laborious calculation, one finds that
\begin{align}\label{eq:rightcharged2pt}
\braket{\Omega_{\rm matt}^{(J_R)}|\tilde\Phi_R^{i\dagger} \tilde \Phi_R^j|\Omega_{\rm matt}^{(J_R')}} = \delta^{ij} \delta_{J_R, J_R'}  \frac{\cos(\pi (J_R+ 2q_i)) \Gamma(\Delta_i + 1 \pm q_i) \Gamma(\Delta_i  + \frac{1}{2} \pm (J_R + q_i))}{\pi \Gamma(2 \Delta_i + 1)},
\end{align}
whenever $|J_R + 2 q_i| < 1/2$ and is otherwise zero. Here $\ket{\Omega_{\rm matt}^{(J_R)}}$ is the matter vacuum in $\mathcal{H}_{\rm matt}^{(J_R)}$. Crucially, we see that, for any fixed $J_R$, the vacuum state $\ket{\Omega_{\rm matt}^{(J_R)}}$ is not tracial, because the operator $\tilde\Phi_R^{i\dagger}$ has R-charge $-q_i$ and the $\cos(\pi (J_R+ 2q_i))$ term in \eqref{eq:rightcharged2pt} is not invariant under $q_i \to - q_i$.

However, we can easily construct a state
\begin{align}\label{eq:chargedtrace}
\ket{\rm MAX} = \sum_{ |J_R| < \frac{1}{2}} \sqrt{\cos(\pi J_R)} \ket{\Omega_{\rm matt}^{(J_R)}}
\end{align}
such that
\begin{align}
\braket{{\rm MAX}|\tilde\Phi_R^{i\dagger} \tilde \Phi_R^j|{\rm MAX}}  &=  \delta^{ij} \sum_{J_R}  \frac{\cos(\pi (J_R)\cos(\pi (J_R+ 2q_i)) \Gamma(\Delta_i + 1 \pm q_i) \Gamma(\Delta_i  + \frac{1}{2} \pm (J_R + q_i))}{\pi \Gamma(2 \Delta + 1)}
\end{align}
where the sum is over R-charges $J_R$ such that $|J_R|, |J_R + 2 q_i| < 1/2$. Defining $J_R' = J_R + 2 q_i$, it then follows immediately that
\begin{align}\nonumber
\braket{{\rm MAX}|\tilde\Phi_R^{i\dagger} \tilde \Phi_R^j|{\rm MAX}} &=  \delta^{ij} \sum_{J_R'}  \frac{\cos(\pi (J_R')\cos(\pi (J_R' - 2q_i)) \Gamma(\Delta_i + 1 \pm q_i) \Gamma(\Delta_i  + \frac{1}{2} \pm (J_R' - q_i))}{\pi \Gamma(2 \Delta + 1)},
\\&= \braket{{\rm MAX}| \tilde \Phi_R^j \tilde\Phi_R^{i\dagger} |{\rm MAX}}, \label{eq:chargedsinglemattertracial}
\end{align}
where the sum is now over R-charges $J_R'$ such that $|J_R'|, |J_R'- 2 q_i| < 1/2$. This is the natural generalization of \eqref{eq:traceneutraloperators} to charged operators. 

In Appendix \ref{sec:susy2ptcalcs}, we also compute the left-right two-point
\begin{align}\label{eq:leftrightcharged2pt}
\braket{\Omega_{\rm matt}^{(J_R')}|\Phi_{L,0}^{i\dagger} \tilde\Phi_R^j |\Omega_{\rm matt}^{(J_R)}} = \delta_{J_R', J_R + 2q} \delta^{ij} \frac{\sqrt{\cos(\pi J_R)\cos(\pi (J_R + 2q))} \Gamma(\Delta + 1 \pm q) \Gamma(\Delta  + \frac{1}{2} \pm (J_R + q))}{\pi \Gamma(2 \Delta + 1)}.
\end{align}
We therefore have
\begin{align}\label{eq:chargelr=rr}
\braket{{\rm MAX}| \Phi_{L,0}^{i\dagger} \tilde\Phi_R^j|{\rm MAX}} = \braket{{\rm MAX}|\tilde\Phi_R^{i\dagger} \tilde \Phi_R^j|{\rm MAX}}.
\end{align}

With  \eqref{eq:chargelr=rr} in hand, we can repeat many of the arguments from Section \ref{sec:project} to characterise the algebra $\widetilde{\mathcal{A}}_R$. The states $\tilde \Phi_R^i\ket{\Omega_{\rm matt}^{(J_R)}}$ are nonzero for any $i, J_R$ satisfying $\left|J_R + 2 q_i\right| < 1/2$. They are thus in one-to-one correspondence with the set of simultaneous ground states described in Section \ref{sec:simulground}. Since we again have
\begin{align}
[\tilde H_L, \tilde\Phi^i_R] = \sum_{J_R, J_R'} \tilde V_{J_R}^\dagger [H_L, \Phi^i_R] \tilde V_{J_R'} = 0
\end{align}
and hence $\tilde H_L \tilde \Phi_R^i\ket{\Omega_{\rm matt}^{(J_R)}} = \tilde \Phi_R^i \tilde H_L \ket{\Omega_{\rm matt}^{(J_R)}} = 0$, the two sets of states are the same. 

Now that we know the set of states $\tilde\Phi^i_R \ket{\Omega_{\rm matt}^{(J_R)}}$ span the kernel of $\tilde H_L$, we can immediately see from comparing  \eqref{eq:rightcharged2pt} with \eqref{eq:leftrightcharged2pt} that
\begin{align}
\sqrt{\cos(\pi (J_R ))}\tilde\Phi^i_R \ket{\Omega_{\rm matt}^{(J_R)}} =  \sqrt{\cos(\pi (J_R + 2 q_i))} \tilde \Pi_L \Phi^i_{L,0} \ket{\Omega_{\rm matt}^{(J_R + 2 q_i)}},
\end{align}
where $\tilde \Pi_L$ projects onto the kernel of $\tilde H_L$.

As with $\widetilde{\mathcal{A}}_{R,0}$ in Section \ref{sec:project}, we will assume that we are working with a matter QFT where $\widetilde{\mathcal{A}}_R$ is a factor. It then follows from $[\tilde \Pi_L, \widetilde{\mathcal{A}}_R] = 0$ that  $\tilde \Pi_L \widetilde{\mathcal{A}}_R \tilde \Pi_L$ is isomorphic to $\widetilde{\mathcal{A}}_R$ and we can work solely within the subspace of simultaneous ground states. An operator $\tilde a \in \widetilde{\mathcal{A}}_R$ is then uniquely determined by the matrix elements
\begin{align}\label{eq:chargedmatrixelements}
\braket{\Omega_{\rm matt}^{(J_R)}|\tilde \Phi^{k\dagger}_R \tilde a \tilde\Phi^{\ell}_R|\Omega_{\rm matt}^{(J_R')}} = \sqrt{\frac{\cos(\pi (J_R + 2 q_\ell))}{\cos(\pi J_R)}}\braket{\Omega_{\rm matt}|\tilde \Phi^{k\dagger}_R \Phi^\ell_{L,0} \tilde a |\Omega_{\rm matt}^{(J_R' + 2q_\ell)}}.
\end{align}
The state $\tilde a \ket{\Omega_{\rm matt}^{(J_R)}}$ is always a simultaneous ground state. We must therefore have
\begin{align}\label{eq:chargedproduct}
\tilde a  \ket{\Omega_{\rm matt}^{(J_R)}} = \sum_{k, J_R} A_{k}(J_R)  \tilde\Phi^k_R  \ket{\Omega_{\rm matt}^{(J_R)}} 
\end{align}
for some set of coefficients $A_k$ that will in general depend on $J_R$. But then it follows from the uniqueness of the matrix elements \eqref{eq:chargedmatrixelements} that
\begin{align}\label{eq:generalchargedoperator}
\tilde a = \sum_k  \tilde\Phi^k_R A_{k}(J_R)
\end{align}
where, for each $k$, $A_{k}(J_R)$ is now a function of the operator $J_R$. We conclude that $\widetilde{\mathcal{A}}_R$ is spanned by operators of the form $\tilde \Phi^i_R \Pi_{J_R}$ where $\Pi_{J_R}$ is projects onto states with fixed R-charge $J_R$. But it follows immediately from \eqref{eq:rightcharged2pt} that
\begin{align}\nonumber
&\braket{{\rm MAX}|\Pi_{J_R} \tilde \Phi^{i\dagger}_R \Phi^j \Pi_{J_R'}|{\rm MAX}} 
\\&\qquad=\sqrt{\cos(\pi J_R) \cos(\pi J_R')} \braket{\Omega_{\rm matt}^{(J_R)}|\tilde\Phi_R^{i\dagger} \tilde \Phi_R^j|\Omega_{\rm matt}^{(J_R')}} 
\\&\qquad= \delta^{ij} \delta_{J_R, J_R'}  \frac{\cos(\pi J_R)\cos(\pi (J_R+ 2q_i))\Gamma(\Delta_i + 1 \pm q_i) \Gamma(\Delta_i  + \frac{1}{2} \pm (J_R + q_i))}{\pi \Gamma(2 \Delta_i + 1)}
\\&  \qquad=\delta_{J_R, J_R'}\cos(\pi (J_R+ 2q_i))\braket{\Omega_{\rm matt}^{(J_R + 2 q_i)}| \tilde \Phi_R^j \tilde\Phi_R^{i\dagger}|\Omega_{\rm matt}^{(J_R + 2q_i)}} 
\\&\qquad= \braket{{\rm MAX}| \Phi^j \Pi_{J_R'} \Pi_{J_R} \tilde \Phi^{i\dagger}_R|{\rm MAX}}
\end{align}
We conclude that $\ket{\rm MAX}$ is tracial and hence that $\widetilde{\mathcal{A}}_R$ is Type II$_1$. Conventionally the trace on a Type II$_1$ factor is normalised so that $\Tr[\mathds{1}] = 1$. This can be achieved by replacing $\ket{\rm MAX}$ with the normalised state
\begin{align}
\ket{\widehat{\mathrm {MAX}}} = \frac{\ket{\rm MAX}}{\sqrt{\sum_{J_R} \cos(\pi J_R)} }.
\end{align}

The number of zero-energy microstates with R-charge $J_R$ should be proportional to
\begin{align}
\Tr[\Pi_{J_R}] = \braket{\widehat{\rm MAX}| \Pi_{J_R} | \widehat{\rm MAX}} = \frac{\cos(\pi J_R)}{\sum_{J_R'} \cos(\pi J_R')}
\end{align}
Indeed this is exactly the result one finds from Euclidean path integral computations \cite{Lin:2022rzw, Lin:2022zxd}. In other words, the algebra $\widetilde{\mathcal{A}}_R$ knows that the probability of measuring R-charge $J_R$ in a maximally entangled state, and hence also the number of BPS microstates, in super-JT gravity scales as $\cos(\pi J_R)$.

Let us conclude with some very brief comments about the left boundary algebra and various projections thereof. We expect that the commutant algebra $\widetilde{\mathcal{A}}_R'$ on $\widetilde{\mathcal{H}}_R$ is generated by $\tilde Q_L, \tilde Q_L^\dagger, \tilde J_L$ and the matter supersymmetric primaries $\Phi^i_{L,0}$ and is isomorphic to the full Type II$_\infty$ super-JT gravity algebra $\mathcal{A}_L$ described in Section \ref{sec:addingmatter}. After projecting onto simultaneous ground states, the algebra $\widetilde{\mathcal{A}}_R$ can be identified with the Hilbert space on which it acts and we obtain the canonical $d=1$ representation. The commutant algebra 
\begin{align}
 \widetilde{\mathcal{A}}_L \cong \tilde \Pi_L\widetilde{\mathcal{A}}_R' \tilde \Pi_L
\end{align}
 can then be identified with the right action of $\widetilde{\mathcal{A}}_R$ on itself, and is spanned by operators of the form $\tilde \Pi_L \Phi^i_{L,0} \tilde \Pi_L f(\tilde J_L)$.

Since the trace of a projector onto fixed $J_L$ in the full theory of super-JT gravity is infinite, the representation of $\widetilde{\mathcal{A}}_R$ on a subspace of $\mathcal{H}_{\rm matt}$ found by projecting onto a fixed left-boundary R-charge $\tilde J_L = J_L$, with no constraint on $\tilde H_L$, still has dimension $d =\infty$.  This Hilbert space can be naturally identified with the subspace of $\mathcal{H}_{\rm matt}$ with bulk R-charge $j$ satisfying  $|2j-J_L| = |J_R| < 1/2$. The left boundary algebra is generated by $\tilde H_L$ and neutral primaries $\Phi^i_{L,0}$. If we also project onto states that annihilate $\tilde H_L$ (for $|J_L| < 1/2$), we produce a representation of $\widetilde{\mathcal{A}}_R$ with dimension
\begin{align}
d = \frac{\cos(\pi J_L)}{\sum_{|J_R| < \frac{1}{2}} \cos (\pi J_R)}
\end{align}
where the commutant algebra is generated solely by $\tilde \Pi_L \Phi^i_{L,0} \tilde \Pi_L$. If $J_L = 0$, it is isomorphic to the algebra $\widetilde{\mathcal{A}}_{L,0}$.  As a representation of $\widetilde{\mathcal{A}}_{L,0}$ the same Hilbert space has dimension
\begin{align}
d = \sec(\pi J_L) \sum_{|J_R| < \frac{1}{2}} \cos (\pi J_R).
\end{align}

\subsection{Summary}
We end this section with a brief summary of the algebraic structure of super-JT gravity, and of its similarities and differences from bosonic JT gravity.

In pure bosonic JT gravity, the boundary algebras are commutative since they consist only of functions of the Hamiltonian. In pure super-JT gravity this is not true, since the boundary supercharges do not commute (or anticommute). However they still have a large centre generated by the Hamiltonian $H_R$ and the operator $J_R^*$ given in \eqref{eq:JR*}. They are also not commutants because the left and right supercharges anticommute, but they are supercommutants, which is the natural generalisation of commutants to algebras with fermions.

When matter is added, Euclidean arguments suggest that the boundary algebras become Type II$_\infty$ factors while remaining supercommutants, in close analogy with the bosonic story. We expect that, for states with nonzero energy, those arguments could be made rigorous within the confines of the canonically quantised Lorentzian theory by simple generalisations of the arguments made in Section \ref{sec:JTalgebras}. However, we did not attempt to work through all the details.

The most interesting new feature of the supersymmetric theory is the presence of normalisable ground states, and the techniques used in Section \ref{sec:JTalgebras} cannot be used to rule out a central operator that only acts nontrivially within that subspace. Partially because of this, and partially because it provides an interesting example of a Type II$_1$ boundary algebra that exists despite the absence of nontrivial boundary time evolution, we devoted most of our efforts to studying the projection of the boundary algebras onto ground states. A key ingredient in our strategy was our use of the results from Section \ref{sec:simulground} to reinterpret this algebra as a somewhat peculiar algebra acting on the matter quantum field theory.

We were not able to give a general argument to exclude the possibility that a nontrivial central operator exists within the projected algebra. However, under the assumption that we were working with a matter theory where that was not the case, we were able to show that a) the algebra of operators contains only a single operator for each supersymmetric primary in the matter theory, b) the product of neutral operators in this algebra has a remarkably simple form, given in \eqref{eq:maintextgsalgebraproduct},  with products of primaries given simply by a sum over primaries multiplied by the matter three-point coefficient and some gamma functions, and c) the matter vacuum state is tracial for neutral operators.

To construct a state that is also tracial for charged operators required us to take a superposition over copies of the matter vacuum within different R-charge sectors, with each sector weighted in proportion to the number of BPS microstates found within it using a Euclidean path integral. Consequently, the quantum field theory algebra describing the supersymmetric ground states knew about the ratio of the number of BPS microstates within each R-charge sector in the theory of quantum gravity that we used to construct it. 

\vskip1cm
 \noindent {\it {Acknowledgements}}  We would like to thank Henry Lin, Juan Maldacena and Steve Shenker for valuable discussions. GP was supported by the Department of Energy through QuantISED Award DE-SC0019380 and an Early Career Award DE-FOA-0002563, by AFOSR award FA9550-22-1-0098 and by a Sloan Fellowship. Work of EW partly supported by NSF
 Grant  PHY-2207584.

\appendix

\section{Two- and three-point functions in JT gravity} \label{app:JT2pt}

We first verify the normalisation of the states $\ket{s}$ defined in \eqref{eq:kets}. The modified Bessel function $K_\nu(z)$ can be defined by the integral representation
\begin{align}\label{eq:besselkintegral}
K_{\nu}(z) = \frac{1}{2}\int_{-\infty}^\infty dt e^{\nu t} e^{-z \cosh(t)}.
\end{align}
A useful identity will be the Mellin transform
\begin{align}\label{eq:besselkmellin}
\mathcal{M}(K_{\nu})(s)&= \int_0^\infty \frac{dz}{z} z^{s} K_{\nu}(z)
\\&= \frac{1}{2}\int_0^\infty \frac{dz}{z} \int_{-\infty}^{\infty} dt z^{s} e^{\nu t -z \cosh(t)}
\\&= 2^{s-2}\int_0^\infty \frac{dv}{v} v^{(s+\nu)/2} e^{-v}  \int_0^\infty \frac{dw}{w} w^{(s-\nu)/2} e^{-w}
\\&= 2^{s-2} \Gamma\left(\frac{s \pm \nu }{2}\right).
\end{align}
where in the third step we defined $v = ze^t/2$ and $w = ze^{-t}/2$ under which the integration measures transforms as 
\begin{align}
\frac{dv}{v}\frac{dw}{w} = 2 \frac{dz}{z} dt.
\end{align}
Consequently the momentum basis wavefunction for the state $\ket{s}$ is
\begin{align}\label{eq:ftbesselk}
\frac{\sqrt{\rho(s)}}{\sqrt{2\pi}} \int_{-\infty}^\infty d \chi e^{ip_\chi \chi} K_{2 is}(2e^{\chi}) &= \frac{\sqrt{\rho(s)}
}{\sqrt{2\pi}}\int_0^\infty \frac{dz}{z} \left(\frac{z}{2}\right)^{i p_\chi} K_{2 is}(z)
\\&= \frac{\sqrt{\rho(s)}}{4\sqrt{2\pi}} \Gamma\left(\frac{i p_\chi \pm 2 i s }{2}\right),\nonumber
\end{align}
where $ \rho(s) = 8 s \sinh(2 \pi s)/\pi^2$. We therefore have
\begin{align}
\braket{s'|s} &= \sqrt{\rho(s)\rho(s')} \int d\chi K_{2 is'}(2e^{\chi})K_{2 is}(2e^{\chi})\nonumber
\\&=\frac{\sqrt{\rho(s)\rho(s')}}{32\pi} \int dp_{\chi} \,\Gamma\left(\frac{ i p_\chi \pm 2 i s'}{2}\right)   \Gamma\left(\frac{- i p_\chi \pm 2 i s }{2}\right)\nonumber
\\&=\frac{1}{8}\sqrt{\rho(s)\rho(s')}\sum_{n=0}^\infty \frac{(-1)^n}{n!} \Gamma( \pm is + is' + n) \Gamma(-2is' - n) + \text{c.c.}\nonumber
\\&=\frac{1}{8}\sqrt{\rho(s)\rho(s')}\frac{\Gamma( is' \pm is)\Gamma(-2is')\Gamma(1+2is')}{\Gamma(1 + is' \pm is)} + \text{c.c.}\nonumber
\\&=\frac{\pi i \sqrt{\rho(s)\rho(s') }}{8 (s^2 - s'^2) \sinh(2 \pi s')} + \text{c.c.}\nonumber
\\&= \delta(s -s'),\label{eq:ketsnorm}
\end{align}
In the second step, we Fourier transformed the inner product to the momentum basis, using \eqref{eq:ftbesselk}.
 In the third step, we used the fact that $\Gamma(z)$ has simple poles at $z = -n$ for integer $n \geq 0$ with residue $(-1)^n/n!$ to complete the contour and sum over poles at $p_\chi  = \pm is' + in$ for $n\geq 0$. In the fourth step, we used Gauss' summation theorem. In the fifth step, we used Euler's reflection formula
\begin{align}
\Gamma(z)\Gamma(1-z) = \frac{\pi}{\sin(\pi z)}. 
\end{align}
and
\begin{align}
\Gamma(1+z) = z \Gamma(z)
\end{align}
Finally, in the last step we used
\begin{align}\label{eq:im1/x}
\mathrm{Im}((s-s')^{-1}) =- \pi \delta(s -s')
\end{align}
where the sign is determined by the requirement that we need $\mathrm{Im}(s') > \mathrm{Im}(s)$ for the infinite sum to converge.

An intuitive explanation of \eqref{eq:ketsnorm}, originally explained in \cite{Maldacena:2016vs}, is the following. We know that  \eqref{eq:ketsnorm} must vanish when $s \neq s'$ because the states $\ket{s}$ are eigenfunctions of a self-adjoint Hamiltonian with different eigenvalues. We therefore expect to get an answer proportional to $\delta(s-s')$. There is no divergence in \eqref{eq:ketsnorm} when $s = s'$ as $\chi \to + \infty$ because the Bessel function $K_\nu(z)$ decays exponentially at $z \to \infty$ (as can be seen immediately from \eqref{eq:besselkintegral}). On the other hand, as $z \to 0$, we have
\begin{align}\label{eq:besselksmallz}
K_\nu(z)\approx \frac{1}{2}\left(\left(\frac{2}{z}\right)^\nu \Gamma(\nu)+\left(\frac{2}{z}\right)^{-\nu} \Gamma(-\nu)\right).
\end{align}
This can again be found from \eqref{eq:besselkintegral} by approximating $z\cosh t \approx z \exp(|t|)$. The first term comes from the integral over $t >0$ while the second term comes from the integral over $t <0$. As a result, the part of the integral \eqref{eq:ketsnorm} with $\chi < -\Lambda \ll 0$ can be approximated as
\begin{align}
&\sqrt{\rho(s)\rho(s')} \frac{1}{4}\int_{-\infty}^{-\Lambda}\d \chi \biggl( \Gamma(2\i s') e^{-2\i s'\chi}+\Gamma(-2\i s')e^{2\i s'\chi}\biggr)  
 \biggl( \Gamma(2\i s) e^{-2\i s\chi}+\Gamma(-2\i s)e^{2\i s\chi}\biggr)\nonumber
\\&=\sqrt{\rho(s)\rho(s')} \frac{1}{4}\int_{-\infty}^{-\Lambda}\d \chi \left(\Gamma(2\i s')\Gamma(-2\i s) e^{2\i (s'-s)\chi} +\Gamma(-2\i s')\Gamma(2\i s) e^{-2\i(s'-s)\chi}\right) + \mathrm{finite}\nonumber
\\&= \frac{1}{4} \rho(s)\Gamma(2\i s)\Gamma(-2\i s) \pi \delta(s-s') + \mathrm{finite}\nonumber
\\&= \delta(s-s') + \mathrm{finite}\nonumber.
\end{align}
In the first step we dropped terms that remain finite as $s \to s'$ since we know such terms vanish in the full integral (as indeed we verified explicitly above). In the last step we again used Euler's reflection formula.

It will also be helpful to be able to compute expectation values of the form
\begin{align}\label{eq:fj+}
\braket{\Omega_{\rm matt}| \Phi_{R,0}^\dagger f(j_+) \Phi_{R,0} |\Omega_{\rm matt} }= \int dj_+ p(j_+) f(j_+)
\end{align}
where $f$ is an arbitrary function and $p(j_+)$ is the spectral distribution for $j_+$ in the state $\Phi_{R,0} \ket{\Omega_{\rm matt}}$. We have
\begin{align}
2 \Delta \braket{\Omega_{\rm matt}| \Phi_{R,0} f(j_+) \Phi_{R,0} |\Omega_{\rm matt} } &= \braket{\Omega_{\rm matt}| \Phi_{R,0}^\dagger i [j_3, f(j_+)] \Phi_{R,0} |\Omega_{\rm matt}} 
\\&= -\braket{\Omega_{\rm matt}| \Phi_{R,0}^\dagger j_+ f'(j_+) \Phi_{R,0} |\Omega_{\rm matt}} 
\\& = - \int dj_+ p(j_+) j_+ f'(j_+)
\\& = \int dj_+ \left[p(j_+) + j_+ p'(j_+)\right] f(j_+).
\end{align}
Here, the first equality follows from \eqref{eq:scaling} and $j_3 \ket{\Omega_{\rm matt}} = 0$, while the second equality uses $[j_3,j_+] = i j_+$. Finally the last equality requires integrating by parts. Since this must be true for arbitrary $f(j_+)$, it follows that
\begin{align}
p(j_+) = C j_+^{2\Delta - 1}
\end{align}
for some constant $C$. Setting $f(j_+) = e^{-\tau j_+}$ leads to
\begin{align}
\braket{\Omega_{\rm matt}| \Phi_{R,0}^\dagger e^{-\tau j_+} \Phi_{R,0} |\Omega_{\rm matt} } = C \int_0^\infty dj_+ j_+^{2\Delta - 1} e^{-\tau j_+} = \frac{C \Gamma(2\Delta)}{\tau^{2\Delta}}.
\end{align}
So to recover the usual Poincar\'{e} two-point function with no prefactor we need to set 
\begin{align}\label{eq:C}
C = \Gamma(2\Delta)^{-1}.
\end{align}
To check that this is indeed the correct normalisation with our conventions, recall that we defined $\Phi_{R,0}$ so that $\ket{\Phi} = e^{-\pi j_2/2} \Phi_{R,0}\ket{\Omega_{\rm matt}}$ is a normalised state. Now
\begin{align}
\frac{\partial}{\partial \theta} \left(e^{-\theta j_2} \Phi_{R,0} \ket{\Omega_{\rm matt}}\right) &= -e^{-\theta j_2} j_2 \Phi_{R,0} \ket{\Omega_{\rm matt}}
\\&=\frac{1}{\cos \theta + 1} e^{-\theta j_2}\left[- \cos \theta j_1 - j_2 +  \sin \theta (i j_3 + \Delta)\right] \Phi_{R,0} \ket{\Omega_{\rm matt}}
\\&=\frac{1}{\cos \theta + 1} \left[ - j_+ + \Delta  \sin \theta\right] e^{-\theta j_2} \Phi_{R,0} \ket{\Omega_{\rm matt}},
\end{align}
where in the second step we used \eqref{eq:scaling} and \eqref{eq:primary}. But we also have
\begin{align}
\frac{\partial}{\partial \theta} &\left(\frac{1}{\cos^{2\Delta}\left(\frac{\theta}{2}\right)}e^{-\tan\left(\frac{\theta}{2}\right) j_+} \Phi_{R,0} \ket{\Omega_{\rm matt}} \right) \\&= \frac{1}{\cos \theta + 1} \left[ - j_+ + \Delta  \sin \theta\right] \left(\frac{1}{\cos^{2\Delta}\left(\frac{\theta}{2}\right)}e^{-\tan\left(\frac{\theta}{2}\right) j_+} \Phi_{R,0} \ket{\Omega_{\rm matt}} \right).\nonumber
\end{align}
So
\begin{align}
e^{-\theta j_2} \Phi_{R,0} \ket{\Omega_{\rm matt}} = \frac{1}{\cos^{2\Delta}\left(\frac{\theta}{2}\right)}e^{-\tan\left(\frac{\theta}{2}\right) j_+} \Phi_{R,0} \ket{\Omega_{\rm matt}}
\end{align}
and
\begin{align}
1 = \braket{\Phi|\Phi} = C\, 2^{2\Delta}  \int_0^\infty dj_+ j_+^{2\Delta - 1} e^{-2 j_+} = C\, \Gamma(2\Delta),
\end{align}
which confirms \eqref{eq:C}.

We now wish to compute
\begin{align} \label{eq:tripleintegral}
\bra{s_1',\Omega_{\rm matt}}\Phi_R^\dagger \Pi_{E_R = s^2} \Phi_R\ket{s_2', \Omega_{\rm matt}} = &\int d j_+ \frac{j_+^{2\Delta -1}}{\Gamma(2\Delta)} \rho(s) \sqrt{\rho(s_1')  \rho(s_2')} \,\,\times \\\nonumber&\int d\chi'  K_{2is_1'}(2e^{\chi'}) e^{2 \Delta \chi'} K_{2is}(2\sqrt{1+j_+}e^{\chi'}) \,\times \\\nonumber& \int d\chi K_{2is}(2\sqrt{1+j_+} e^{\chi}) e^{2 \Delta \chi} K_{2is_2'}(2e^{\chi}),
\end{align}
where on the right-hand side we have used \eqref{eq:U1Phi}, \eqref{eq:U2s'} and \eqref{eq:intj+}.

Fourier transforming to the $p_\chi$ basis leads to
\begin{align}\label{eq:besselkinnerproduct}
\int d\chi K_{2is}&(2\sqrt{1+j_+} e^{\chi}) e^{2 \Delta \chi} K_{2is'}(2e^{\chi}) \\\nonumber&= \frac{1}{32\pi} \int dp_\chi (1+j_+)^{-i p_\chi/2}  \Gamma\left(\frac{ i p_\chi \pm 2 i s}{2}\right)   \Gamma\left(\frac{2\Delta \pm 2 i s' - i p_\chi}{2}\right)
\\&= \frac{1}{8} \sum_{n=0}^\infty \frac{(-1)^n}{n!}  (1+ j_+)^{-\Delta -is' - n}\Gamma( \Delta + is' \pm is + n)\Gamma(-2is' - n) \, + \text{c.c.}\label{eq:hypergeomsum}
\\&= \frac{\Gamma(\Delta + is' \pm is)\Gamma(-2is'){}_2F_1(\Delta +is'+is, \Delta +is' - is; 1 + 2is'; (1+j_+)^{-1})}{8(1+ j_+)^{\Delta +is'} }  + \text{c.c.}
\\&= \frac{\Gamma(\Delta  \pm is \pm is')}{8 \Gamma(2\Delta)}(1+ j_+)^{is}{}_2F_1(\Delta +is+is',\Delta +is - is';2\Delta;-j_+)\label{eq:laststepchiintegral}
\end{align}
where ${}_2F_1(a,b;c;z)$ is the Gaussian hypergeometric function. In the first step, we used \eqref{eq:ftbesselk}. In the second step, we completed the contour and summed over poles at $p_\chi =2( \pm s - i \Delta - in)$ for $n \geq 0$.  In the last step, we used a fractional linear transformation of ${}_2F_1$ to combine the two terms.  

This leaves one integral left to compute. A crucial tool will be the Mellin transform
\begin{align}
\mathcal{M}({}_2F_1)(s)  =\int_0^\infty dz z^{s -1} {}_2F_1(a,b;c;-z) = \frac{\Gamma(s)\Gamma(a-s)\Gamma(b-s) \Gamma(c)}{\Gamma(a)\Gamma(b)\Gamma(c-s)},
\end{align}
which can be derived from an integral representation of ${}_2F_1$. We therefore have
\begin{align}\label{eq:2F1horrorstart}
\int_0^{\infty} &d j_+ j_+^{2\Delta -1}  {}_2F_1(\Delta +is_1'+is,\Delta -is_1' + is;2\Delta;-j_+){}_2F_1(\Delta +is_2'-is,\Delta -is_2' - is;2\Delta;-j_+)\nonumber
\\&=\frac{1}{2\pi} \int_{-\infty}^\infty d \kappa \frac{\Gamma(\Delta + i \kappa)\Gamma(is \pm is_1'- i\kappa)\Gamma(2\Delta)}{\Gamma(\Delta + is \pm is_1') \Gamma(\Delta - i\kappa)} \frac{\Gamma(\Delta - i \kappa)\Gamma(-is \pm is_2' + i\kappa)\Gamma(2\Delta)}{\Gamma(\Delta-is \pm is_2')\Gamma(\Delta + i \kappa)}
\\&=\frac{1}{2\pi} \int_{-\infty}^\infty d \kappa \frac{\Gamma(is \pm is_1'- i\kappa) \Gamma(-is \pm is_2' + i\kappa)\Gamma(2\Delta)^2}{\Gamma(\Delta + is \pm is_1') \Gamma(\Delta -is \pm is_2')}\label{eq:gammacancellation}
\\& =  \sum_{n=0}^\infty \frac{(-1)^n}{n!} \frac{\Gamma(2 is_1' - n) \Gamma(n \pm is_2' - is_1' )\Gamma(2\Delta)^2}{\Gamma(\Delta + is
 \pm is_1')\Gamma(\Delta-is \pm is_2')} ~+~  (s_1' \leftrightarrow - s_1')
\\&= \frac{ \Gamma( \pm is_2' - is_1' ) \Gamma(2 i s_1') \Gamma(1 - 2is_1')\Gamma(2\Delta)^2}{\Gamma(\Delta\pm is \pm is_2')\Gamma(1 - is_1' \pm i s_2')} ~+~  (s_1' \leftrightarrow - s_1')
\\&= \frac{-i \pi}{ (s_2'^2 - s_1'^2) \sinh(2 \pi s_1') \Gamma(\Delta + is \pm is_1')\Gamma(\Delta-is \pm is_2')} ~+~  (s_1' \leftrightarrow - s_1')
\\&= \frac{8\, \Gamma(2\Delta)^2 \delta(s_1' - s_2')}{\rho(s_1') \Gamma(\Delta \pm is \pm is_1')}.\label{eq:2F1horrorend}
\end{align}
In the first step we used Parseval's theorem to rewrite the integral in terms of the momentum $\kappa$ conjugate to $\log j_+$). In the third step we evaluated this integral by summing over poles at $\kappa = s \pm s_1' - i n$. In the fourth step, we evaluated the hypergeometric series using Gauss' summation theorem. In the fifth step, we use Euler's reflection formula. Finally in the last step we used \eqref{eq:im1/x}. Putting everything together, we conclude that
\begin{align}
\bra{s_1',\Omega_{\rm matt}}\Phi_R^\dagger U^\dagger \ket{s}\!\bra{s} U \Phi_R\ket{s_2', \Omega_{\rm matt}} = \delta(s_1' - s_2') \frac{\rho(s) \Gamma(\Delta \pm is_1' \pm is)}{8 \,\Gamma(2 \Delta)}.
\end{align}

As with \eqref{eq:ketsnorm}, the integral \eqref{eq:tripleintegral} can be understood more intuitively by understanding the origin of the divergence when $s_1'= s_2'$. We know the integral must vanish when $s_1' \neq s_2'$ because the right-boundary operators commute with $H_L$. The only possible source of a divergence when $s_1'= s_2'$ is from large $j_+$. (The integrand scales as $j_+^{2\Delta -1}$ as $j_+ \to 0$ and the integrals over $\chi$ and $\chi'$ depend only on one of $s_1'$ and $s_2'$. So large $j_+$ is the only possible source of a divergence that exists only when $s_1' = s_2'$. ) At large $j_+$, \eqref{eq:besselkinnerproduct} is dominated by large negative $\chi$ (otherwise $K_{2is'}(2\sqrt{1+j_+} e^{\chi})$ vanishes superexponentially) and so we can use the approximation \eqref{eq:besselksmallz} for $K_{2is'}(2 e^{\chi})$. This reduces \eqref{eq:besselkinnerproduct} to the standard integral \eqref{eq:ftbesselk}. One finds that
\begin{align}\nonumber
\int d\chi K_{2is}(2\sqrt{1+j_+} e^{\chi}) e^{2 \Delta \chi} K_{2is'}(2e^{\chi}) &\approx \frac{1}{2}\int d\chi K_{2is}(2j_+^{\frac{1}{2}} e^{\chi}) e^{(2 \Delta +2is') \chi} \Gamma(-2is') \,+ \text{c.c.}
\\&= \frac{j_+^{-\Delta -is'}}{8}  \Gamma( \Delta + is' \pm is)\Gamma(-2is') \, + \text{c.c.},
\end{align}
which is just the leading term of the sum \eqref{eq:hypergeomsum} after setting $(1+ j_+) \approx j_+$. Finally the integral over $j_+ > \Lambda \gg 0$ becomes
\begin{align}\nonumber
\frac{\rho(s') \sqrt{\rho(s_1')  \rho(s_2')}} {64 \Gamma(2\Delta)} \int_\Lambda^{\infty}&\frac{d j_+}{ j_+}\left[j_+^{i(s_2' -s_1')}  \Gamma( \Delta + is_1' \pm is)\Gamma(-2is_1') \Gamma( \Delta - is_2' \pm is)\Gamma(2is_2') + \text{c.c.}\right] + \text{finite}
\\&=\pi \delta(s_1' -s_2')\frac{\rho(s) \rho(s_1') \Gamma( \Delta \pm  is_1' \pm is)\Gamma(\pm 2is_1')}{32\Gamma(2\Delta)} + \text{finite}
\\&=\delta(s_1' - s_2') \frac{\rho(s) \Gamma(\Delta \pm is_1' \pm is)}{8 \,\Gamma(2 \Delta)} + \text{finite}.
\end{align}
In the last step, we again used Euler's reflection formula.

We can also compute the left-right two-point function. We first note that
\begin{align}
\braket{\Omega_{\rm matt}|\Phi_{L,0}^\dagger \Phi_{R,0}|\Omega_{\rm matt}} = \braket{\Phi_{\rm matt}|e^{-\pi j_2/2} e^{\pi j_2/2}|\Phi_{\rm matt}} = 1
\end{align}
We therefore have
\begin{align}\label{eq:leftright2ptcalc}
\braket{s', \Omega_{\rm matt}|\Phi_L \Phi_R|s, \Omega_{\rm matt}} &=\sqrt{\rho(s')\rho(s)} \int d\chi K_{2 is'}(2 e^\chi) e^{2\Delta \chi} K_{2is}(2e^\chi)
\\&= \sqrt{\rho(s')\rho(s)} \frac{\Gamma(\Delta  \pm is \pm is')}{8\Gamma(2\Delta)},
\end{align}
which is simply \eqref{eq:laststepchiintegral} with $j_+ = 0$ since ${}_2F_1(a,b;c;0) = 1$.

Finally, we want to compute the JT gravity three-point function 
\begin{align}
\Tr\left[ \Phi^i_R \Pi_{E_R = s_1^2} \Phi^j_R \Pi_{E_R = s_2^2} \Phi^k_R \Pi_{E_R = s_3^2}\right] &= \sqrt{\rho(s_2) \rho(s_3)} \braket{s_3, \Omega_{\rm matt}| \Phi_L^k \Phi^i_R \Pi_{E_R = s_1^2}  \Phi^j_R |s_2, \Omega_{\rm matt}}. \label{eq:trace3ptapp}
\end{align}
 Taking an inverse Laplace transform of \eqref{eq:threepointfunction}, we see that, for $\Delta_k < \Delta_i + \Delta_j$ and any function $f(j_+)$, we have
\begin{align}
\bra{\Omega_{\rm matt}}\Phi^k_L \Phi^i_R f(j_+) \Phi^j_R\ket{ \Omega_{\rm matt}} = \frac{C_{ijk}}{\Gamma(\Delta_{i+j-k}) } \int_0^\infty dj_+ j_+^{\Delta_{i+j-k}-1} f(j_+).
\end{align}
Here we have introduce notation where e.g. $\Delta_{i+j-k} = \Delta_i +\Delta_j - \Delta_k$. The required condition $\Delta_k < \Delta_i + \Delta_j$ for the inverse Laplace transform to converge can always be achieved by a cyclic permutation of the trace \eqref{eq:trace3ptapp}.  The calculation then proceeds using \eqref{eq:besselkinnerproduct} and manipulations similar to, but somewhat messier than, \eqref{eq:2F1horrorstart} to \eqref{eq:2F1horrorend}. Explicitly, we have
\begin{align}\label{eq:3ptfunctionAppA}
\Tr\left[ \Phi^i_R \Pi_{E_R = s_1^2} \Phi^j_R \Pi_{E_R = s_2^2} \Phi^k_R \Pi_{E_R = s_3^2}\right] = C_{ijk} F_{ijk}(s_1, s_2, s_3)
\end{align}
where
\begin{align}
F_{ijk}(s_1, s_2, s_3)&= \sqrt{\rho(s_2)\rho(s_3)}\frac{\Gamma(\Delta_i  \pm is_1 \pm is_3) \Gamma(\Delta_j  \pm is_1 \pm is_2)}{64  \Gamma(\Delta_{i+j -k}) \Gamma(2\Delta_i)\Gamma(2 \Delta_j)}  \\\nonumber& \qquad\qquad\times\int_0^\infty \frac{dj_+}{j_+} j_+^{\Delta_{i+j-k}} {}_2F_1(\Delta_i +is_1'+is_1,\Delta_i -is_3 + is_1;2\Delta_i;-j_+)\\\nonumber&\qquad\qquad\qquad\times{}_2F_1(\Delta_j +is_2-is_1,\Delta_j -is_2 - is_1;2\Delta_j;-j_+)  .\nonumber
\\\label{eq:3ptgammas}&=\sqrt{\rho(s_2)\rho(s_3)}\frac{\Gamma(\Delta_i  - is_1 \pm is_3) \Gamma(\Delta_j  + is_1 \pm is_2)}{\Gamma(\Delta_{i+j -k})}\\\nonumber&\qquad\qquad \times \int_{-\infty}^\infty \frac{d \kappa}{128\pi} \frac{\Gamma(i \kappa)\Gamma(\Delta_i + is_1 \pm is_3- i\kappa)}{ \Gamma(2\Delta_i - i\kappa) }\\\nonumber&\qquad\qquad\qquad\times \frac{\Gamma(\Delta_{i+j-k} - i \kappa)\Gamma(-is_1 \pm is_2 +\Delta_{i+j-k} + i\kappa)}{ \Gamma(\Delta_{j + k - i} + i \kappa)}\nonumber
\end{align}
The last integral can be evaluated by summing poles at (for integer $n\geq 0$) $\kappa = -i \Delta_i + s_1 + s_3 - i n$,  $\kappa = -i \Delta_i + s_1 - s_3 - i n$ and $\kappa = \Delta_{i+j - k} - i n$. Each of these three infinite sums gives a generalized hypergeometric function $_4F_3(a_1 \dots a_4; b_1\dots b_3;z)$, evaluated at $z=1$, multiplied by a number of gamma functions.\footnote{In comparison with \eqref{eq:gammacancellation}, there is no cancellation of two gamma functions between the numerator and denominator of \eqref{eq:3ptgammas} and so the hypergeometric functions are ${}_4F_3$ rather than ${}_2F_1$.} The full formula is unfortunately difficult to fit onto a single page and so we have not written it explicitly. In the special case where e.g. $\Delta_j = \Delta_i + \Delta_k +m$ for integer $m \geq 0$, the poles at $\kappa = \Delta_{i+j - k} - i n$ vanish and the answer simplifies simplifies slightly to a product of gamma functions times a Wilson polynomial \cite{Jafferis:2022wez, kolchmeyer2023neumann}.

\section{The restriction of the left-boundary Lie superalgebra to right-boundary ground states}\label{app:gsrestriction}
In this appendix, we prove some basic properties of the ground state wavefunctions $\ket{\tilde\Psi_{J_R}}$ in pure super-JT gravity and then derive \eqref{eq:tildesuperalgebra} as the action of the left-boundary Lie superalgebra restricted to states with $H_R = 0$ and to some fixed $J_R$ with $|J_R| < 1/2$. We first verify that $\ket{\tilde\Psi_{J_R}}$ defined in \eqref{eq:puresuperJTgroundstate} is indeed annihilated by the pure super-JT charges $Q_R^0 = UQ_RU^\dagger$ and ${Q_R^0}^\dagger = UQ_R^\dagger U^\dagger$. We have
\begin{align}
\frac{1}{2}\psi(p_\rho - ip_a) \ket{\tilde\Psi_{J_R}} = -\frac{2i}{\sqrt{\pi}}\sqrt{\cos(\pi J_R)}e^{i\left(J_R+\frac{1}{2}\right) a -2\rho}\left[\frac{1}{2}\left(\frac{1}{2}- J_R \right) e^{\rho}K_{\frac{1}{2} - J_R}(2e^{-\rho}) + K'_{\frac{1}{2} - J_R}(2e^{-\rho}) \right]\ket{\Omega}
\end{align}
where $\ket{\Omega} = \ket{\Omega_L}\ket{\Omega_R}$ and $K'_s(z) = dK_s/dz$. Meanwhile
\begin{align}
-e^{-\rho +ia} \psi' \ket{\tilde\Psi_{J_R}} =-\frac{2 i}{\pi}\sqrt{\cos(\pi J_R)}e^{i\left(J_R+\frac{1}{2}\right) a -2\rho} K_{\frac{1}{2} + J_R}(2e^{-\rho})\ket{\Omega}.
\end{align}
The result $Q_R^0 \ket{\tilde\Psi_{J_R}} =0$ then follows from the standard identity 
\begin{align}\label{eq:besselkid}
K'_\nu(z) + K_{1-\nu}(z) + (\nu/z)K_\nu(z) = 0,
\end{align}
which itself follows immediately from \eqref{eq:besselkintegral} by integrating by parts.

Similarly, we have
\begin{align}
\frac{1}{2}\psi^\dagger(p_\rho + ip_a) \ket{\tilde\Psi_{J_R}} = \frac{2}{\pi}\sqrt{\cos(\pi J_R)}e^{i\left(J_R-\frac{1}{2}\right) a -2\rho}\left[ \frac{1}{2}\left(J_R +\frac{1}{2}\right)e^{\rho} K_{\frac{1}{2} +J_R}(2e^{-\rho}) + K'_{\frac{1}{2}+J_R}(2e^{-\rho}) \right]\psi'^\dagger \psi^\dagger \ket{\Omega}
\end{align}
while
\begin{align}
-e^{-\rho -ia} \psi'^\dagger \ket{\tilde\Psi_{J_R}} =\frac{2}{\pi}\sqrt{\cos(\pi J_R)}e^{i\left(J_R-\frac{1}{2}\right) a -2\rho} K_{\frac{1}{2} - J_R}(2e^{-\rho})\psi'^\dagger \psi^\dagger \ket{\Omega}.
\end{align}
It follows that ${Q_R^0}^\dagger \ket{\tilde\Psi_{J_R}}$ vanishes by \eqref{eq:besselkid}. Since the boundary Hamiltonians $H_L$ and $H_R$ are equal in the absence of matter, $\ket{\tilde\Psi_{J_R}}$ is also annihilated by $H_L$ and hence by the pure-super-JT left-boundary supercharges
\begin{align}
 Q_L^0 &=\frac{1}{2} \psi' (p_\rho + ip_a) + e^{-\rho -ia} \psi\\
 {Q_L^0}^\dagger &=\frac{1}{2} \psi'^\dagger (p_\rho - ip_a) + e^{-\rho -ia} \psi^\dagger.
\end{align}
This is easy to verify by calculations that are essentially identical to the ones above. We can also verify that
\begin{align}
\braket{\tilde\Psi_{J_R}| \tilde\Psi_{J_R}} &=\frac{8}{\pi} \cos(\pi J_R) \int_{-\infty}^\infty d\rho e^{-2\rho} \left(K_{\frac{1}{2} -J_R}(2e^{-\rho})^2 + K_{\frac{1}{2} +J_R}(2e^{-\rho})^2\right)\\
&= \frac{2}{\pi} \cos(\pi J_R) \int_0^\infty dz z \left(K_{\frac{1}{2} -J_R}(z)^2 + K_{\frac{1}{2} +J_R}(z)^2\right)\\
&= 1,
\end{align}
where in the first line we assumed the measure $da$ was normalised to have integral $2\pi$,\footnote{Recall that $a $ has period $ 2\pi N$ for some  integer $N \geq 1$. So, for $N > 1$, we are implicitly rescaling the measure $da$ here by a factor of $1/N$ relative to the standard measure.} and in the second step we used the standard integral 
\begin{align}
\int_0^\infty dz z K_\nu(z)^2 &= \frac{1}{8\pi} \int_{-\infty}^\infty d\kappa \Gamma\left(\frac{1 \pm i \kappa \pm \nu}{2}\right)
\\&=  \frac{\pi}{8}   \int_{-\infty}^\infty d\kappa \frac{1}{\cos\left(\frac{\pi(i \kappa+\nu)}{2}\right)\cos\left(\frac{\pi(i \kappa-\nu)}{2}\right)}
\\&=  \frac{i}{4 \sin(\pi \nu)}   \left[\log\left(\cos\left(\frac{\pi}{2}(i\kappa + \nu\right)\right)-  \log\left(\cos\left(\frac{\pi}{2}(i\kappa - \nu\right)\right)\right]^\infty_{-\infty}
\\&=\frac{\pi \nu}{2 \sin(\pi \nu)}
\end{align}
which holds for all $0\leq \nu<1$. In the first step, we applied Parseval's theorem using the Mellin transform \eqref{eq:besselkmellin}. In the second step, we used Euler's reflection formula.

We now want to show that
\begin{align}\label{eq:QLTildeprop}
Q_L \tilde V_{J_R} \stackrel{?}{=} \tilde V_{J_R} \tilde Q_L
\end{align}
where $Q_L$ is defined in \eqref{eq:QL}, $\tilde V_{J_R}$ is defined in \eqref{eq:tildeV} and $\tilde Q_L$ is given in \eqref{eq:tildesuperalgebra}. 
After conjugation by $U_1 = \exp(i \rho _3 + i a j)$, we obtain
\begin{align}
 U_1 Q_L U_1^\dagger =  \psi'\left(\frac{1}{2} p_{\rho} + \frac{i}{2} p_a -  j_3 -  i j\right) + g_- + \psi e^{-\rho -ia},
\end{align}
After conjugation by $U_2 = (1+j_+)^{i p_\rho/2}$ we obtain
\begin{align} \label{eq:QLmostlyconj}
U_2 U_1 Q_L U_1^\dagger U_2^\dagger &=   \psi' \left(\frac{i}{2}p_a-  j_3  -  i j\right) +  (1+j_+)^{-1}\frac{p_\rho}{2}\left(\psi' + g_+\right)  + g_- + \psi \frac{e^{-\rho -ia}}{(1 + j_+)^{1/2}}
\end{align}
where we used the fact that for any function $f$, we have $[j_3, f(j_+)] = i j_+ df/dj_+$ and $[g_-, f(j_+)] = i g_+ df/dj_+$. Now note that we have already shown that
\begin{align}
(1 + j_+)^{-1/2}  U_3^\dagger Q_L^0 U_3 = \frac{1}{2} (\psi'+ g_+) \frac{p_\rho + ip_a}{1+j_+} +  \frac{e^{-\rho -ia}}{(1+j_+)^{1/2}} \psi
\end{align}
annihilates $U_3^\dagger \ket{\tilde \Psi_{J_R}} \ket{\Psi_{\rm matt}}$ for any state $\ket{\Psi_{\rm matt}}$. We can therefore subtract it from \eqref{eq:QLmostlyconj} when acting on $U_3^\dagger \ket{\tilde \Psi_{J_R}} \ket{\Psi_{\rm matt}}$ to obtain
\begin{align}\label{eq:QLtildeVaction}
 Q_L \tilde V_{J_R} \ket{\Psi_{\rm matt}} = U_1^\dagger U_2^\dagger\left[ \left(j_+ \psi' - g_+\right)\frac{ip_a }{2(1 + j_+)}-  \psi'\left(j_3  +  i j\right)  + g_- \right] U_3^\dagger \ket{\tilde \Psi_{J_R}} \ket{\Psi_{\rm matt}}.
\end{align}

Finally, it is convenient to define the operator $\Lambda: \mathcal{H}_{\rm matt} \to \mathcal{H}_{\rm matt}$ by
\begin{align}
\Lambda \ket{\Psi_{\rm matt}} = (1+g_+^\dagger g_+)^{\frac{1}{4}}(1+g_+ g_+^\dagger)^{-\frac{1}{4}} \ket{\Psi_{\rm matt}}
\end{align}
and $T: \mathcal{H}_{\rm matt} \to \mathcal{H}_{\rm super-JT}$ by
\begin{align}
T \ket{\Psi_{\rm matt}} &= e^{iJ_R a} e^{-\rho} \big[-F_- \left((1+g_+ g_+^\dagger)\psi^\dagger + g_+ \psi'^\dagger \psi^\dagger  \right) + iF_+ (\psi'^\dagger + g_+^\dagger) \big]\ket{\Omega} \ket{\Psi_{\rm matt}}.
\end{align}
where
\begin{align}
F_\pm = (1+j_+)^{-\frac{1}{2} \pm \frac{1}{4}} K_{\frac{1}{2} \pm J_R}(2e^{-\rho})e^{\mp ia/2}.
\end{align}
A short calculation shows that
\begin{align}\nonumber
U_2 U_1\tilde V_{J_R} \ket{\Psi_{\rm matt}}   &= U_3^\dagger  \ket{\tilde\Psi_{J_R}} \ket{\Psi_{\rm matt}}
\\&= \frac{2}{\pi}  (1 +j_+)^{-\frac{1}{2}} \sqrt{\cos(\pi J_R)}e^{iJ_R a -\rho}\left[- e^{ia/2}K_{\frac{1}{2} - J_R}(2e^{-\rho}) \left(\sqrt{1 +g_+ g_+^\dagger} \psi^\dagger + \psi'^\dagger \psi^\dagger g_+\right)\right.\nonumber\\&~~~~ \left.+ ie^{-ia/2}K_{\frac{1}{2} + J_R} (2 e^{-\rho}) \left(\sqrt{1 + g_+^\dagger g_+} \psi'^\dagger + g_+^\dagger\right) \right]\ket{\Omega_L}\ket{\Omega_R}\ket{\Psi_{\rm matt}}\nonumber
\\&= \frac{2}{\pi}\sqrt{\cos(\pi J_R)}T \Lambda \ket{\Psi_{\rm matt}}. \label{eq:TLambdaaction}
\end{align}
The second equality follows from \eqref{eq:U3Psimatt1} and \eqref{eq:U3Psimatt2} and in the last equality we used the fact for any function $f$ we have $f(g_+ g_+^\dagger) f(g_+^\dagger g_+) = f(j_+)f(0)$. 

Since \eqref{eq:QLtildeVaction} and \eqref{eq:TLambdaaction} are true for all $\ket{\Psi_{\rm matt}}$, we therefore have
\begin{align}
U_2 U_1 Q_L \tilde V_{J_R} = \frac{2}{\pi}\sqrt{\cos(\pi J_R)} \left[ \left(j_+ \psi' - g_+\right)\frac{ip_a }{2(1 + j_+)}-  \psi'\left(j_3  +  i j\right)  + g_- \right] T \Lambda
\end{align}
and
\begin{align}
U_2 U_1 \tilde V_{J_R} = \frac{2}{\pi}\sqrt{\cos(\pi J_R)} T \Lambda.
\end{align}
It follows that the result \eqref{eq:QLTildeprop} that we want to prove can be rewritten as
\begin{align}
U_2 U_1 Q_L \tilde V_{J_R} \Lambda^{-1} &\stackrel{?}{=} U_2 U_1 \tilde V_{J_R} \Lambda^{-1} \left(\Lambda \tilde Q_L  \Lambda^{-1}\right)\\
\left[ \left(j_+ \psi' - g_+\right)\frac{ip_a }{2(1 + j_+)}-  \psi'\left(j_3  +  i j\right)  + g_- \right]T &\stackrel{?}{=} T \left[-g_- + \frac{i J_R g_+}{2(1+ j_+)}\right].\label{eq:target}
\end{align}
Let us now prove \eqref{eq:target} by a brute-force evaluation of all the terms involved. For any state $\ket{\Psi_{\rm matt}}$, we have
\begin{align}\label{eq:mess1}
\frac{i j_+ \psi' p_a}{2(1+j_+)} T\ket{\Psi_{\rm matt}} &= \frac{ e^{i J_R a-\rho}}{4(1+j_+)}\left[ i j_+\left(2J_R+1\right) F_- g_+ \psi^\dagger  - j_+ \left(2 J_R-1\right)F_+\right]\ket{\Omega}\ket{\Psi_{\rm matt}},
\\ \frac{i g_+ (J_R - p_a)}{2(1+j_+)}  T\ket{\Psi_{\rm matt}}&=\frac{e^{iJ_R a - \rho}}{4(1+j_+)}\left(i F_- g_+ \psi^\dagger - F_+g_+ (\psi'^\dagger + g_+^\dagger)\right)\ket{\Omega}\ket{\Psi_{\rm matt}}\\
\frac{-i J_R\{g_+, T\}}{2(j_++1)} \ket{\Psi_{\rm matt}} &= \frac{ e^{iJ_R a - \rho}}{4(1+j_+)}\left(-2i J_RF_- j_+ g_+ \psi^\dagger + 2 J_R j_+ F_+ \right)\ket{\Omega}\ket{\Psi_{\rm matt}}\\
-  \psi'\left([j_3  +  i j,T]\right) \ket{\Psi_{\rm matt}} &= \frac{e^{i J_Ra - \rho}}{4(1+j_+)}\left[ - i (j_++4) F_- g_+\psi^\dagger -j_+ F_+\right]\ket{\Omega}\ket{\Psi_{\rm matt}}\\
\left( \{g_-,T\}  - \psi' T(j_3+ij)\right)\ket{\Psi_{\rm matt}} &= \frac{e^{i J_Ra - \rho}}{4(1+j_+)}\left[3i F_-g_+\psi^\dagger + F_+ g_+(\psi'^\dagger + g_+^\dagger)\right]\ket{\Omega}\ket{\Psi_{\rm matt}}.\label{eq:mess5}
 \end{align}
In deriving these results, we used the identities
\begin{align}
[j_3, f(j_+)] = i j_+ f'(j_+) ~~~~\text{and}~~~~ [g_-, f(j_+)] = + i g_+ f'(j_+),
\end{align}
which hold for any function $f$ and follow directly from the algebra structure of $\hSU(1,1|1)$ given in \eqref{su111algebra}.

Adding the left hand sides of \eqref{eq:mess1} -- \eqref{eq:mess5} gives the left hand side minus the right hand side of \eqref{eq:target}. Meanwhile, all terms on the right hand side of \eqref{eq:mess1} -- \eqref{eq:mess5} cancel. This completes the derivation.

With \eqref{eq:QLTildeprop} in hand, the remaining results are easy to show. We have
\begin{align}
Q_L^\dagger \tilde V_{J_R} = Q_L^\dagger \tilde V_{J_R} \tilde V^\dagger_{J_R} \tilde V_{J_R} = \tilde V_{J_R} \tilde V_{J_R}^\dagger Q_L^\dagger \tilde V_{J_R} = \tilde V_{J_R} \tilde Q_L^\dagger \tilde V_{J_R}^\dagger \tilde V_{J_R} = \tilde V_{J_R} \tilde Q_L^\dagger, 
\end{align}
where in the first step we inserted the identity, in the second we used the fact that $Q_L^\dagger$ commutes with $H_R, J_R$ and hence with $\tilde V_{J_R} \tilde V_{J_R}^\dagger$ and in the third step we used the adjoint of \eqref{eq:QLTildeprop}. It then follows that
\begin{align}
H_L  \tilde V_{J_R} = \{Q_L,Q_L^\dagger\}  \tilde V_{J_R} =  \tilde V_{J_R} \{\tilde Q_L, \tilde Q_L^\dagger\}.
\end{align}
Finally, we have 
\begin{align}
U_1 J_L U_1^\dagger  &= -p_a + 2j + \frac{1}{2} [\psi' , \psi'^\dagger ]\\
\end{align}
It follows immediately from \eqref{eq:tildeV} that
\begin{align}
J_L \tilde V_{J_R} = \tilde V_{J_R} (-J_R + 2j).
\end{align}

\section{The gauge invariance of matter operators in super-JT gravity}\label{app:gaugeinvariance}
In this appendix, we verify that the super-JT matter operators defined in \eqref{eq:superJTphiR} are indeed invariant under the diagonal action $\mathcal{D}(\mathcal{J}_-) = J_i^L + j_i + J_i^R$. To do so, we first note that $\Phi_R(t)$ commutes with all left-boundary charges. Meanwhile 
\begin{align}
(e^{-\theta_+ g_+ - \bar\theta_+ g_+^\dagger}e^{-ij_+ \tau}) \,J_+^R\, (e^{ij_+ \tau}e^{\theta_+ g_+ + \bar\theta_+ g_+^\dagger} ) &= J_+^R - j_+
\\(e^{-\theta_+ g_+ - \bar\theta_+ g_+^\dagger}e^{-ij_+ \tau}) \,G_+^R\, (e^{ij_+ \tau}e^{\theta_+ g_+ + \bar\theta_+ g_+^\dagger})  &= G_+^R - g_+ + \bar\theta j_+\nonumber
\\(e^{-\theta_+ g_+ - \bar\theta_+ g_+^\dagger}e^{-ij_+ \tau}) \,{G_+^R}^\dagger \,(e^{ij_+ \tau}e^{\theta_+ g_+ + \bar\theta_+ g_+^\dagger} ) &= {G_+^R}^\dagger - g_+^\dagger + \theta j_+\nonumber
\\(e^{-\theta_+ g_+ - \bar\theta_+ g_+^\dagger}e^{-ij_+ \tau(t)})\, J_3^R\, (e^{ij_+ \tau}e^{\theta_+ g_+ + \bar\theta_+ g_+^\dagger} ) &= J_3^R + \tau  j_+  - \frac{i}{2} \theta_+ g_+ - \frac{i}{2} \bar\theta_+ g_+^\dagger \nonumber
\\(e^{-\theta_+ g_+ - \bar\theta_+ g_+^\dagger}e^{-ij_+ \tau}) \,J^R\, (e^{ij_+ \tau}e^{\theta_+ g_+ + \bar\theta_+ g_+^\dagger})  &= J^R  - \frac{1}{2} \theta_+ g_+ + \frac{1}{2} \bar\theta_+  g_+^\dagger +\frac{1}{2} \theta_+ \bar\theta_+ j_+\nonumber
 \\(e^{-\theta_+ g_+ - \bar\theta_+ g_+^\dagger}e^{-ij_+ \tau}) \,G_-^R\, (e^{ij_+ \tau}e^{\theta_+ g_+ + \bar\theta_+ g_+^\dagger})  &= G_-^R -  \tau \bar\theta_+ j_+ + \tau g_+ -\frac{i}{2} \theta_+ \bar\theta_+ g_+\nonumber
 \\(e^{-\theta_+ g_+ - \bar\theta_+ g_+^\dagger}e^{-ij_+ \tau}) \,{G_-^R}^\dagger \,(e^{ij_+ \tau}e^{\theta_+ g_+ + \bar\theta_+ g_+^\dagger})  &= {G_-^R}^\dagger -\tau \theta_+ j_+ + \tau g_+^\dagger - \frac{i}{2} \theta_+\bar\theta_+ g_+^\dagger\nonumber
 \\(e^{-\theta_+ g_+ - \bar\theta_+ g_+^\dagger}e^{-ij_+ \tau})\, J_-^R (e^{ij_+ \tau}e^{\theta_+ g_+ + \bar\theta_+ g_+^\dagger})  &= J_-^R  - \tau^2 j_+ + i \tau \left(\theta_+ g_+ +\bar\theta_+ g_+^\dagger\right) \nonumber
\end{align}
and
\begin{align}
(e^{-\theta_+ g_+ - \bar\theta_+ g_+^\dagger}e^{-ij_+ \tau})\, j_+\, (e^{ij_+ \tau}e^{\theta_+ g_+ + \bar\theta_+ g_+^\dagger}  )&=  j_+
\\(e^{-\theta_+ g_+ - \bar\theta_+ g_+^\dagger}e^{-ij_+ \tau} )\,g_+\, (e^{ij_+ \tau}e^{\theta_+ g_+ + \bar\theta_+ g_+^\dagger} ) &= g_+ - \bar\theta j_+ \nonumber
\\(e^{-\theta_+ g_+ - \bar\theta_+ g_+^\dagger}e^{-ij_+ \tau})\, g_+^\dagger\, (e^{ij_+ \tau}e^{\theta_+ g_+ + \bar\theta_+ g_+^\dagger} ) &=  g_+^\dagger - \theta j_+\nonumber
\\(e^{-\theta_+ g_+ - \bar\theta_+ g_+^\dagger}e^{-ij_+ \tau}) \,j_3\,(e^{ij_+ \tau}e^{\theta_+ g_+ + \bar\theta_+ g_+^\dagger})  &= j_3  -\tau j_+ + \frac{i}{2} \theta_+ g_+ +  \frac{i}{2} \bar\theta_+ g_+^\dagger \nonumber
\\(e^{-\theta_+ g_+ - \bar\theta_+ g_+^\dagger}e^{-ij_+ \tau})\, j \,(e^{ij_+ \tau}e^{\theta_+ g_+ + \bar\theta_+ g_+^\dagger})  &= j + \frac{1}{2} \theta_+ g_+ - \frac{1}{2} \bar\theta_+ g_+^\dagger - \frac{1}{2} \theta_+ \bar \theta_+ j_+\nonumber
\\(e^{-\theta_+ g_+ - \bar\theta_+ g_+^\dagger}e^{-ij_+ \tau}) \,g_-\, (e^{ij_+ \tau}e^{\theta_+ g_+ + \bar\theta_+ g_+^\dagger})  &= g_- -\tau g_+ + \tau \bar\theta_+ j_+ - \bar\theta_+ (j_3 + ij) + \frac{i}{2}\theta_+\bar\theta_+ g_+\nonumber
\\(e^{-\theta_+ g_+ - \bar\theta_+ g_+^\dagger}e^{-ij_+ \tau}) \,g_-^\dagger\, (e^{ij_+ \tau}e^{\theta_+ g_+ + \bar\theta_+ g_+^\dagger})  &= g_-^\dagger -\tau g_+^\dagger + \tau \theta_+ j_+ - \theta_+ (j_3 - ij) + \frac{i}{2}\theta_+\bar\theta_+ g_+^\dagger\nonumber
\\(e^{-\theta_+ g_+ - \bar\theta_+ g_+^\dagger}e^{-ij_+ \tau})\, j_-\, (e^{ij_+ \tau}e^{\theta_+ g_+ + \bar\theta_+ g_+^\dagger})  &= j_- - 2  \tau j_3 +\tau^2 j_+ - i \tau \left(\theta_+ g_+ + \bar\theta_+ g_-^\dagger \right)\nonumber\\&\nonumber ~~~~~~~~~~~~~+i \theta_+ g_- + i \bar\theta_+ g_+^\dagger +\theta_+\bar\theta_+ j.
 \end{align}
In the last equality, we made use of the identity
\begin{align}
[j_-, f(j_+)] = 2i f'(j_+) j_3 -  f''(j_+) j_+.
\end{align}
It follows that
\begin{align}
[j_+ + J_+^R,  \Phi_{R,0} (\tau,\theta_+,\bar\theta_+)] &= 0 \label{eq:matter+right}
\\ [g_+ + G_+^R,  \Phi_{R,0} (\tau,\theta_+,\bar\theta_+)] &=  0\nonumber
\\ [g_+^\dagger + {G_+^R}^\dagger,  \Phi_{R,0} (\tau,\theta_+,\bar\theta_+)] &= 0\nonumber
\\ [j_3 + J_3^R,  \Phi_{R,0} (\tau,\theta_+,\bar\theta_+)] &=  i \Delta \Phi_{R,0} (\tau,\theta_+,\bar\theta_+)\nonumber
\\ [j + J^R,  \Phi_{R,0} (\tau,\theta_+,\bar\theta_+)] &=  q \Phi_{R,0} (\tau,\theta_+,\bar\theta_+)\nonumber
\\ [g_- + G_-^R,  \Phi_{R,0} (\tau,\theta_+,\bar\theta_+)] &= - i\bar\theta_+ (q + \Delta )\Phi_{R,0} (\tau,\theta_+,\bar\theta_+)\nonumber
\\ [g_-^\dagger + {G_-^R}^\dagger,  \Phi_{R,0} (\tau,\theta_+,\bar\theta_+)] &=  i\theta_+ (q - \Delta )\Phi_{R,0} (\tau,\theta_+,\bar\theta_+)\nonumber
 \\ [j_- + J_-^R,  \Phi_{R,0} (\tau,\theta_+,\bar\theta_+)] &= (-2i \tau \Delta + \theta_+ \bar\theta_+ q) \Phi_{R,0} (\tau,\theta_+,\bar\theta_+).\nonumber
\end{align}
Finally we have
\begin{align}\label{eq:rho+a}
[J_+^R, e^{-\Delta \rho(t) + i qa(t)}] &= [G_+^R, e^{-\Delta \rho(t) + i qa(t)}] = [{G_+^R}^\dagger, e^{-\Delta \rho(t) + i qa(t)}] = 0
\\\nonumber [J_3^R, e^{-\Delta \rho(t) + i qa(t)}] &= - i \Delta e^{-\Delta \rho(t) + i qa(t)} ~~~~~~~~~~~~ [J^R, e^{-\Delta \rho(t) + i qa(t)} ] = -q e^{-\Delta \rho(t) + i qa(t)}
\\\nonumber [G_-^R, e^{-\Delta \rho(t) + i qa(t)}] &= i \bar\theta_+(\Delta + q) e^{-\Delta \rho(t) + i qa(t)} ~~~~~~~ [{G_-^R}^\dagger,e^{-\Delta \rho(t) + i qa(t)}] = i \theta_+(\Delta - q) e^{-\Delta \rho(t) + i qa(t)}
\\\nonumber [J_-^R, e^{-\Delta \rho(t) + i qa(t)}] &= 2i \tau \Delta - \theta_+ \bar\theta_+ q.
\end{align}
Combining \eqref{eq:matter+right} and \eqref{eq:rho+a}, the gauge invariance of \eqref{eq:superJTphiR} follows immediately.

\section{Ground state two- and three-point functions in super-JT gravity}\label{sec:susy2ptcalcs}
We first compute the right-boundary neutral-operator two-point function with $J_R = 0$ given in \eqref{eq:neutral2pt}. Taking a Fourier transform, we have
\begin{align}
\int d \rho\, &\exp(-(2\Delta +2)\rho)  K_{1/2}(2 \sqrt{1 + j_+} e^{-\rho}) K_{1/2}(2 e^{-\rho}) \\\nonumber&=\frac{1}{32\pi} \int dp_\rho (1 + j_+)^{-i p_\rho/2}\Gamma\left(\frac{i p_\rho}{2} \pm \frac{1}{4}\right)\Gamma\left(\Delta + 1 \pm \frac{1}{4} - \frac{i p_\rho}{2}\right)
\\&= \frac{2^{-2\Delta}}{16} \int dp_\rho (1 + j_+)^{-i p_\rho/2}\Gamma\left(i p_\rho - \frac{1}{2}\right)\Gamma\left(2\Delta -i p_\rho + \frac{3}{2}\right)
\\&= \sum_{n = 0}^\infty\frac{(-1)^n 2^{-2\Delta}\pi}{8 n!} (1 + j_+)^{-\Delta -3/4 - n/2} \Gamma(2 \Delta +1 + n)
\\&= \frac{\pi}{8} 2^{-2\Delta} \Gamma(2\Delta +1) (1 + j_+)^{-1/4}\left(1 + \sqrt{1 + j_+}\right)^{-2\Delta -1}.\label{eq:K1/2integral}
\end{align}
In the second step we used the Legendre duplication formula
\begin{align}
\Gamma(z) \Gamma(z+\frac{1}{2}) = 2^{1-2z}\sqrt{\pi} \Gamma(2z).
\end{align}
It follows that
\begin{align}
\braket{{\tilde\Phi_R}^\dagger \tilde\Phi_R} &= 2^{-4\Delta} \frac{\Gamma(2\Delta + 1)^2}{ \Gamma(2\Delta)}\int_0^\infty dj_+ j_+^{2\Delta -1}(1 + j_+)^{-1/2}\left(1 + \sqrt{1 + j_+}\right)^{-4\Delta }
\\&= 2^{-4\Delta}\Gamma(2\Delta + 1)\left[ j_+^{2\Delta} \left(1 + \sqrt{1 + j_+}\right)^{-4\Delta }\right]^\infty_0
\\&=2^{-4\Delta} \Gamma(2\Delta + 1).
\end{align}
The left-right boundary two-point function \eqref{eq:leftrightneutral2pt} can be computed from \eqref{eq:K1/2integral} by setting $j_+ = 0$. Using  \eqref{eq:K1/2integral}, the three-point function
\begin{align}\label{eq:3ptsuperjtintegral}
\braket{{\Phi^k_{L,0}}^\dagger \tilde\Phi^i_R \tilde\Phi^j_R  } &= 2^{-2(\Delta_i+ \Delta_j)} \frac{\Gamma(2\Delta_i + 1)\Gamma(2\Delta_j + 1)}{ \Gamma(\Delta_i + \Delta_j - \Delta_k)} \, \times\\\nonumber&~~~~\int_0^\infty dj_+ j_+^{\Delta_i + \Delta_j - \Delta_k -1}(1 + j_+)^{-1/2}\left(1 + \sqrt{1 + j_+}\right)^{-2(\Delta_i + \Delta_j) }
\\& = 2^{-2(\Delta_i+ \Delta_j + \Delta_k)} \frac{\Gamma(2\Delta_i + 1)\Gamma(2\Delta_j + 1)}{ \Gamma(\Delta_i + \Delta_j - \Delta_k)} \int_0^\infty dz \frac{z^{\Delta_i + \Delta_j - \Delta_k -1}}{\left(z+1\right)^{\Delta_i + \Delta_j + \Delta_k + 1 }}\nonumber
\\&= \frac{2^{-2(\Delta_i+ \Delta_j + \Delta_k)}\Gamma(2\Delta_i + 1)\Gamma(2\Delta_j + 1) \Gamma(2\Delta_k + 1)}{\Gamma(1 + \Delta_i + \Delta_j + \Delta_k)}.\nonumber
\end{align}
where the subsitution $z = (\sqrt{1+j_+} -1)/2$ (so that $j_+= 4z(z+1)$, $ 1+ \sqrt{1 + j_+} = 2 (z+1)$ and $dj_+ = 4 \sqrt{j_+ + 1} \,dz$) put the integral into a standard form for the Euler beta function 
\begin{align}
B(\Delta_i + \Delta_j - \Delta_k, 2\Delta_k + 1) = \frac{\Gamma(\Delta_i + \Delta_j - \Delta_k)\Gamma(2\Delta_k + 1)}{ \Gamma(\Delta_i + \Delta_j + \Delta_k + 1 )}.
\end{align}

We now turn to computing the charged two-point functions \eqref{eq:rightcharged2pt} and \eqref{eq:leftrightcharged2pt}. It follows from \eqref{eq:chargedtildedef} that
\begin{align}
\tilde\Phi_R \ket{\Omega^{(J_R)}_{\rm matt}} &= \frac{8}{\pi} \sqrt{\cos(\pi J_R)\cos(\pi (J_R+ 2q))} \int d \rho e^{ (- 2 \Delta -2)\rho}\,\times\\\nonumber&~~~\left[\sqrt{1 +g_+ g_+^\dagger} K_{\frac{1}{2} - J_R- q}(2\sqrt{1 + j_+}e^{-\rho})K_{\frac{1}{2} - J_R}(2e^{-\rho}) +\right.\\\nonumber&~~~~\left.\sqrt{1 + g_+^\dagger g_+} K_{\frac{1}{2} + J_R + q} (2 \sqrt{1+j_+}e^{-\rho})K_{\frac{1}{2} + J_R} (2 e^{-\rho})  \right]  \Phi^i_{R,0} \ket{\Omega_{\rm matt}^{(J_R + 2q)}}
\end{align}
if $|J_R + 2 q| < 1/2$ and vanishes otherwise. For  $0 < \nu_1, \nu_2 < 1$, we have
\begin{align}
\int d \rho\,& e^{ (- 2 \Delta -2)\rho} K_{\nu_1}(2\sqrt{1 + j_+}e^{-\rho})K_{\nu_2}(2e^{-\rho}) \\\nonumber&= \frac{1}{32\pi}\int dp_\rho (1 + j_+)^{-i p_\rho/2}\Gamma\left(\frac{i p_\rho\pm \nu_1}{2} \right)\Gamma\left(\Delta + 1  - \frac{i p_\rho \pm \nu_2}{2}\right)
\\\nonumber&= \frac{1}{8}\sum_{n = 0}^\infty \frac{(-1)^n}{n!} (1+j_+)^{-\Delta -1- \nu_2/2 -n} \Gamma(\Delta+1 + \frac{\nu_2 }{2}\pm \frac{\nu_1}{2} + n)\Gamma(- \nu_2 - n) + (\nu_2 \leftrightarrow -\nu_2)
\\\nonumber&=\frac{\Gamma(\Delta +1+ \frac{\nu_2 \pm \nu_1}{2}) \Gamma(- \nu_2) }{8(1+j_+)^{\Delta+1 +\nu_2/2}} {}_2F_1(\Delta+1 +\frac{\nu_1+\nu_2}{2},\Delta + 1  +\frac{\nu_2 - \nu_1}{2}; 1 +  \nu_2; (1+j_+)^{-1})\nonumber \\&~~~~~~~~~+ (\nu_2 \leftrightarrow -\nu_2)
\\\nonumber& = \frac{\Gamma(\Delta +1\pm \frac{\nu_2 \pm \nu_1}{2}) }{8\Gamma(2\Delta + 2)} (1+j_+)^{\nu_1/2} {}_2F_1(\Delta+1 +\frac{\nu_1+\nu_2}{2},\Delta + 1  +\frac{\nu_1-\nu_2}{2}; 2 \Delta + 2; -j_+).
\end{align}
Hence
\begin{align}\label{eq:chargeoperatorvacaction1}
\tilde\Phi_R \ket{\Omega^{(J_R)}_{\rm matt}}= \nonumber&\frac{ \sqrt{\cos(\pi J_R)\cos(\pi (J_R+ 2q))}}{\pi \Gamma(2 \Delta + 2)}\sqrt{1 +g_+ g_+^\dagger} \Gamma\left(\frac{1}{2}\left(2\Delta + 2 \pm \left[\frac{1}{2} - J_R- 2q\right]\pm \left[\frac{1}{2} - J_R\right]  \right)\right)  \,\times\\ \nonumber&~~(1 + j_+)^{1/4 - J_R/2 - q/2} {}_2F_1(\Delta + \frac{3}{2} - J_R -q, \Delta + 1 - q;2\Delta + 2; -j_+) \Phi^i_{R,0} \ket{\Omega_{\rm matt}^{(J_R + 2q)}}\, \\&~~~~+ \left( J_R \leftrightarrow -J_R ~~~~q \leftrightarrow -q ~~~~g_+ \leftrightarrow g_+^\dagger \right).
\end{align}
Since
$
{}_2F_1(a,b;c;0) = 1,
$
this leads immediately to the left-right two-point function
\begin{align}
\braket{\Omega_{\rm matt}^{(J_R + 2q)}|\Phi_{L,0}^\dagger \tilde\Phi_R |\Omega_{\rm matt}^{(J_R)}} = \frac{\sqrt{\cos(\pi J_R)\cos(\pi (J_R+ 2q))} \Gamma(\Delta + 1 \pm q) \Gamma(\Delta  + \frac{1}{2} \pm (J_R + q))}{\pi \Gamma(2 \Delta + 1)}.
\end{align}

To compute the right-boundary two-point function, it will be helpful to decompose $\tilde\Phi_R \ket{\Omega^{(J_R)}_{\rm matt}}$ into a piece annihilated by $g_+$ and a piece annihilated by $g_+^\dagger$. Let $\Pi_{+}$ and $\Pi_{-}$ be projectors onto those respective subspaces.  We have
\begin{align}
f(g_+ g_+^\dagger )\Pi_{+} = f(j_+) ~~~ f(g_+^\dagger g_+  )\Pi_{+} = f(0) ~~~ f(g_+ g_+^\dagger )\Pi_{-} = f(0) ~~~ f(g_+^\dagger g_+  )\Pi_{-} = f(j_+)
\end{align}
along with
\begin{align}
\braket{\Omega_{\rm matt}| \Phi_{R,0}^\dagger f(j_+) \Pi_{\pm} \Phi_{R,0}|\Omega_{\rm matt}}= \frac{\Delta \pm q}{\Gamma(2\Delta+1)} \int dj_+  j_+^{2\Delta-1} f(j_+),
\end{align}
which follows from \eqref{eq:intj+} together with
\begin{align}
\braket{\Omega_{\rm matt}| \Phi_{R,0}^\dagger f(j_+)' g_+ g_+^\dagger  \Phi_{R,0}|\Omega_{\rm matt}} &= 2 i \braket{\Omega_{\rm matt}| \Phi_{R,0}^\dagger [f(j_+), g_-] g_+^\dagger \Phi_{R,0}|\Omega_{\rm matt}}
\\&=-i \braket{\Omega_{\rm matt}| \Phi_{R,0}^\dagger f(j_+) \{g_-, g_+^\dagger\} \Phi_{R,0}|\Omega_{\rm matt}}
\\&=-i \braket{\Omega_{\rm matt}| \Phi_{R,0}^\dagger f(j_+) (j_3 + i j)\Phi_{R,0}|\Omega_{\rm matt}}
\\&= (\Delta + q ) \braket{\Omega_{\rm matt}| \Phi_{R,0}^\dagger f(j_+) \Phi_{R,0}|\Omega_{\rm matt}}.
\end{align}
Since the remaining integral required to compute the right-boundary two-point function is somewhat complex, we introduce the convenient shorthand notation
\begin{align}
\tilde F_{\pm} = {}_2F_1(\Delta + \frac{1}{2} \pm J_R \pm q, \Delta  + 1	\pm q ;2\Delta + 2; -j_+).
\end{align}
Using Euler's transformation, we can then write
\begin{align}
\frac{2 \Delta + 1}{\braket{\Phi_{L,0}^\dagger \tilde\Phi_R} }\Pi_{+} \tilde\Phi_R \ket{\Omega^{(J_R)}_{\rm matt}}  &=\left[(1 + j_+)^{\frac{1}{4} + \frac{J_R}{2} +q}\left(\Delta +\frac{1}{2} - J_R -q\right) \tilde F_{+}\, +\right.\\\nonumber&\left.~~~~~~(1 + j_+)^{-\frac{1}{4} - \frac{J_R}{2} - q} \left(\Delta +\frac{1}{2} + J_R + q\right)\tilde F_{-}\right] \Pi_+\Phi^i_{R,0} \ket{\Omega_{\rm matt}^{(J_R + 2q)}}
\end{align}
and similarly
\begin{align}
\frac{2 \Delta + 1}{\braket{\Phi_{L,0}^\dagger \tilde\Phi_R} }\Pi_{-} \tilde\Phi_R \ket{\Omega^{(J_R)}_{\rm matt}}  &=\left[(1 + j_+)^{-\frac{1}{4} + \frac{J_R}{2} + q}\left(\Delta +\frac{1}{2} - J_R - q\right) \tilde F_{+}\, +\right.\\\nonumber&\left.~~~~~~(1 + j_+)^{\frac{1}{4} - \frac{J_R}{2} - q} \left(\Delta +\frac{1}{2} + J_R + q\right)\tilde F_{-}\right] \Pi_-\Phi^i_{R,0} \ket{\Omega_{\rm matt}^{(J_R + 2q)}}
\end{align}
Finally, using one of Gauss' contiguous relations to write the derivative in a convenient form, we have
\begin{align}
j_+ \frac{\partial}{\partial j_+} \tilde F_{\pm} &= \left(\Delta + \frac{1}{2} \pm J_R \pm q\right)\left[{}_2F_1(\Delta + \frac{3}{2} \pm J_R \pm q, \Delta  + 1	\pm  q;2\Delta + 2; -j_+) - \tilde F_{\pm}\right]\nonumber
\\&= \left(\Delta + \frac{1}{2} \pm J_R \pm q\right)\left[ (1 + j_+)^{-\frac{1}{2} \mp J_R \mp 2q}\tilde F_{\mp} - \tilde F_{\pm} \right],
\end{align}
where in the second step we again used Euler's transformation.

Putting everything together we have
\begin{align}
\frac{\Gamma(2\Delta + 2)\braket{{\tilde\Phi_R}^\dagger \tilde\Phi_R}}{\braket{\Phi_{L,0}^\dagger \tilde\Phi_R} ^2}&= \int dj_+\frac{ j_+^{2\Delta - 1}}{2 \Delta + 1}\left[(1 + j_+)^{-\frac{1}{2} + J_R + 2q}\left(2\Delta + [\Delta + q]j_+\right) \left(\Delta +\frac{1}{2} - J_R - q\right)^2 \tilde F_{+}^2 \right.\nonumber\\&~~~~~~~~~~~~\left. +\,4 \Delta \left(\Delta +\frac{1}{2} - J_R - q\right)\left(\Delta +\frac{1}{2} + J_R + q\right) \tilde F_- \tilde F_+ \right.
\\ &\left.  ~~~~~~~~~~~~  +(1 + j_+)^{-\frac{1}{2} - J_R - 2q}\left(2\Delta + [\Delta - q]j_+\right) \left(\Delta +\frac{1}{2} + J_R + q\right)^2 \tilde F_{-}^2\right]\nonumber
\\&=\left[ \frac{j_+^{2\Delta}}{2\Delta+1} \left(\left(\Delta+ \frac{1}{2} - J_R -q\right)^2 (1+j_+)^{\frac{1}{2} + J_R + 2q} \tilde F_+^2\right.\right.\\\nonumber&\left.\left.~~~~~~+\left(\Delta+ \frac{1}{2} - J_R -q\right)\left(\Delta+ \frac{1}{2} + J_R +q\right)(2 +  j_+) \tilde F_+ \tilde F_- \right.\right.
\\\nonumber& \left.\left.~~~~~~~~~~+\left(\Delta+ \frac{1}{2} + J_R +q\right)^2 (1+j_+)^{\frac{1}{2} - J_R - 2q} \tilde F_-^2\right) \right]^\infty_0
\\&= \frac{\pi \Gamma(\frac{1}{2}-J_R) \Gamma(2\Delta + 1) }{\cos(\pi J_R)\Gamma(\Delta + 1 \pm q)\Gamma(\Delta + \frac{1}{2} \pm (J_R + q))}
\end{align}
In the last step, we used Euler's reflection formula and the standard formula
\begin{align}
\lim_{z \to \infty} z^a {}_2F_1(a,b;c;-z) = \frac{\Gamma(b-a)\Gamma(c)}{\Gamma(b)\Gamma(c-a)} \qquad\text{when}\qquad a<b.
\end{align}
We conclude that
\begin{align}
\braket{\Omega_{\rm matt}^{(J_R)}|\tilde\Phi_R^\dagger \tilde\Phi_R| \Omega_{\rm matt}^{(J_R)}} = \frac{\cos(\pi (J_R+ 2q)) \Gamma(\Delta + 1 \pm q) \Gamma(\Delta  + \frac{1}{2} \pm (J_R + q))}{\pi \Gamma(2 \Delta + 1)}.
\end{align}

\section{JT gravity with matter in the highest weight gauge} \label{app:fixhighestweight}
The physical JT gravity Hilbert space is the Hilbert space of coinvariants for the diagonal $\hSL(2,\R)$ action on the tensor product of the left and right boundary particles with the matter Hilbert space. However for practical purposes it is helpful to analyse this Hilbert space by picking a "gauge" where one uses the equivalence of states in the Hilbert space of coinvariants related by gauge transformations to restrict to wavefunctions of a certain form. In Section \ref{sec:JT}, and in \cite{penington2023algebras}, we restricted to wavefunctions of the form
\begin{align}\label{eq:ordinarywavefunctions}
\mathbf{\Psi}_{JT} =  \delta\left(T\right) \delta\left(T'\right) \delta\left(\chi-\chi'\right)\Psi_{JT}(\chi)
\end{align}
with $\Psi_{JT}(\chi)$ valued in $\mathcal{H}_{\rm matt}$. In other words we chose a gauge where $T = T' = 0$ and $\chi = \chi'$.

There is an alternative approach that connects nicely to the analysis of the chord Hilbert space of the double-scaled SYK model studied in \cite{Lin:2022rbf, Lin:2023trc}.  The matter Hilbert space $\mathcal{H}_{\text {matt }}$ decomposes into a single $\hSL(2,\R)$ invariant vacuum state $\ket{\Omega_{\rm matt}}$ together with irreducible discrete series representations of  $\hSL(2,\R)$.  The diagonal $\hSL(2,\R)$ action  respects this decomposition and so can be studied in each representation separately. If we put the matter in the vacuum state, we will just get the Hilbert space of JT gravity without matter.

Let us instead restrict our attention to a single discrete series representation within the matter theory, which we will label $\mathcal{H}_{\text {matt }}^{\lambda}$.  The Hilbert space $\mathcal{H}_{\text {matt }}^{\lambda}$ contains a ``highest-weight state'' $\ket{\psi_0}$ satisfying
\begin{align}\label{eq:highestweightapp}
j_{1} \ket{\psi_0}=\lambda\ket{\psi_0}~~~~~~~~~~~(j_{2} - i j_3) \ket{\psi_0}=0 .
\end{align}
We claim that the space of coinvariants with the matter state restricted to $\mathcal{H}_{\text {matt }}^{\lambda}$ is spanned by states of the form
\begin{align}\label{eq:highestweightwavefunctions}
\mathbf{\Psi}_{JT} =  \Psi_{JT}\left(\chi, \chi'\right) \delta\left(T\right) \delta\left(T'\right) \ket{\psi_0} \text {. }
\end{align}
where $\Psi_{JT}$ is valued in $\mathcal{H}_{\text {matt }}^{\lambda}$. This is because fixing $T = T' = 0$ unique fixing the gauge up to the diagonal action $\mathcal{D}(\mathcal{J}_3)$. But states of the form $\exp(i \alpha j_3)\ket{\psi_0}$ span $\mathcal{H}_{\text {matt }}^{\lambda}$. So any state is gauge equivalent to a wavefunction of the form \eqref{eq:highestweightwavefunctions}.

Let us first compute the action of the left and right Hamiltonians $\mathbf{H}_{L}, \mathbf{H}_{R}$ on states of the form \eqref{eq:highestweightwavefunctions}. Without changing how they act on the Hilbert space of coinvariants, we can simplify $\mathbf{H}_{L}, \mathbf{H}_{R}$ by adding to them $\sum_{a} \mathcal{D}(\mathcal{J}_{a}) K_{a}$ (for any operators $K_{a}$ ). We have
\begin{align}
&\left[\mathcal{D}( \mathcal{J}_{1})+\mathcal{D}(\mathcal{J}_{2})-i \mathcal{D}(\mathcal{J}_{3})\right] \mathbf{\Psi}_{JT}=\left[-2 p_{T}-e^{\chi} + e^{\chi'}-i\left(p_{\chi}-p_{\chi'}\right)+\lambda\right]\mathbf{\Psi}_{JT}\\
& \left[\mathcal{D}(\mathcal{J}_{1})-\mathcal{D}(\mathcal{J}_{2})+i\mathcal{D}( \mathcal{J}_{3})\right]\mathbf{\Psi}_{JT}=\left[-2 p_{T'} + e^{\chi}-e^{\chi'}+i\left(p_{\chi}-p_{\chi'}\right)+\lambda\right]\mathbf{\Psi}_{JT} .
\end{align}
Using these formulas to "solve" for $p_{T}, p_{T'}$ modulo the constraints, we find that the action of $\mathbf{H}_{L}, \mathbf{H}_{R}$ on states of the form \eqref{eq:highestweightwavefunctions} becomes
\begin{align}\label{eq:highestweighthams}
H_{R} & = p_{\chi}^{2}+ e^{\chi+\chi'}+ \lambda e^{\chi}-i\left(p_{\chi}-p_{\chi'}\right) e^{\chi} \\
H_{L} & =p_{\chi'}^{2}+ e^{\chi+\chi'}+\lambda e^{\chi'}+i\left(p_\chi -p_\chi'\right) e^{\chi'} .\nonumber
\end{align}
Note that in this derivation, it is necessary to be careful with operator ordering: operators $\sum_{a} \mathcal{J}_{a} K_{a}$, but not $\sum_{a} K_{a} \mathcal{J}_{a}$, act trivially on the space of coinvariants. The formulas in \eqref{eq:highestweighthams} are symmetrical under $\chi, p_\chi \leftrightarrow \chi', p_\chi'$, as they should be.

WIth $p_{\chi}=-i \partial_{\chi}, p_{\chi'}=-i \partial_{\chi'}$, the operators $H_{R}, H_{L}$ in \eqref{eq:highestweighthams} are manifestly real, but they are not self-adjoint with respect to the obvious inner product. However, there is a not so obvious inner product with respect to which $H_{R}, H_{L}$ are self-adjoint. Let $\chi_{ \pm}=\frac{1}{2}\left(\chi \pm \chi'\right)$. Then the inner product with respect to which $H_{L}, H_{R}$ are self-adjoint is
\begin{align}\label{eq:altinnerproduct}
\left\langle\Phi_{JT}, \Psi_{JT}\right\rangle=\int d \chi_{+} d \chi_{-} d \tilde\chi_{-}^{\prime} \bar{\Phi}_{JT}\left(\chi_{+}, \chi_{-}\right) \frac{1}{\left(\cosh \left(\frac{\chi_- -\tilde\chi_-}{2}\right)\right)^{2 \lambda}} \Psi_{JT}\left(\chi_{+}, \tilde\chi_{-}\right) .
\end{align}
To verify that this inner product is positive definite, we need to investigate the integral kernel
\begin{align}
F\left(\chi_- - \tilde\chi_-\right)=\frac{1}{\left(\cosh \left(\frac{\chi_- - \tilde\chi_-}{2}\right)\right)^{2 \varepsilon}} .
\end{align}
By translation invariance, this integral kernel is diagonalized by exponential functions
\begin{align}
\int_{-\infty}^{\infty} d \tilde\chi_- F\left(\chi_- - \tilde\chi_-\right) e^{-i s \tilde\chi_-}=A(s) e^{-i \chi_- s}
\end{align}
with
\begin{align}
A(s)=\int_{-\infty}^{\infty} d \chi F(\chi) e^{-i s \chi} .
\end{align}

Thus positivity of the inner product is equivalent to the statement that the Fourier transform $A(s)$ of the function $F(\chi)$ is positive-definite. Indeed, \footnote{The substitution $x=e^{\chi}$ turns this integral into a standard representation of the Euler beta function.}
\begin{align}
\int_{-\infty}^{\infty} d \chi e^{-i \chi s}\left(\cosh \left(\frac{\chi}{2}\right)\right)^{-2 \lambda}=2^{2 \lambda} \frac{\Gamma(\lambda-i s) \Gamma(\lambda+i s)}{\Gamma(2 \lambda)}
\end{align}
and the right hand side is manifestly positive for real $s$ and $\lambda>0$.

Positivity of the inner product \eqref{eq:altinnerproduct} means that we can define a Hilbert space $\widetilde{\mathcal{H}}$ consisting of functions of $\chi_{+}, \chi_{-}$that are square integrable with respect to that inner product. We expect this to be the coinvariant Hilbert space for wavefunctions of the form \eqref{eq:highestweightwavefunctions}.

The fact that wavefunctions of the form \eqref{eq:ordinarywavefunctions} generate the space of coinvariants is fairly obvious from the definitions, but this is less obvious for \eqref{eq:highestweightwavefunctions}. We can show that the two quantizations are equivalent by finding an explicit unitary transformation between them. In the process, we will also understand the origin of the formula \eqref{eq:altinnerproduct} for the inner product.

For real $\alpha$, let $\ket{\psi_{\alpha}}=e^{i\alpha \mathrm{j}_{3}}\ket{\psi_0}$. A computation in $\widetilde{S L}(2, \mathbb{R})$ group theory shows that 
\begin{align}\label{eq:keygrouptheoryaltquantisation}
\left\langle\psi_{\alpha^{\prime}}| \psi_{\alpha}\right\rangle=\left\langle\psi_0\left|e^{i\left(\alpha-\alpha^{\prime}\right) j_{3}}\right| \psi_0\right\rangle=\frac{1}{\left(\cosh \left(\frac{\alpha-\alpha^{\prime}}{2}\right)\right)^{2 \lambda}}.
\end{align}
This formula follows from the identity in $S L(2, \mathbb{C})$
\begin{align}
\exp \left(i \alpha j_{3}\right)=\exp \left(\tanh (\alpha / 2)\left(j_{2}+i j_{3}\right)\right)\left(\cosh \frac{\alpha}{2}\right)^{-2 j_{3}} \exp \left(-\tanh (\alpha / 2)\left(j_{2}-i j_{3}\right)\right),
\end{align}
together with the properties \eqref{eq:highestweightapp} of the highest weight state $\ket{\psi_0}$. This identity is easily proved by a computation in a two-dimensional matrix representation of $S L(2, \mathbb{C})$.

It follows from eqn. (13) that if, for some complex-valued function $f(\alpha)$, we define
\begin{align}
\ket{\psi_{f}}=\int_{-\infty}^{\infty} d \alpha f(\alpha) \ket{\psi_{\alpha}}
\end{align}
then
\begin{align}
\braket{\psi_f|\psi_f} =\int_{-\infty}^{\infty} d \alpha d \alpha^{\prime} \bar{f}(\alpha) f\left(\alpha^{\prime}\right) \frac{1}{\left(\cosh \left(\frac{\alpha-\alpha^{\prime}}{2}\right)\right)^{2 \lambda}}.
\end{align}
This is positive because of the positivity of the integral kernel as we have just shown.

Thus the map $f \rightarrow \Psi_{f}$ from functions of $f$ to $\mathcal{H}_{\text {matt }}^{\lambda}$ is injective. It is also true that this map has a dense image. This can be proved as follows. If a collection of vectors $\ket{\psi_{s}} \in\mathcal{H}_{
\text {matt }}^\lambda$, with $s$ running over some set $S$, does not generate $\mathcal{H}_{\text {matt }}^\lambda$, then it generates a proper subspace $\widetilde{\mathcal{H}}_{\text {matt }}^\lambda$ of $\mathcal{H}_{\text {matt }}^\lambda$. Then there is a nonzero vector $\ket{\chi} \in \mathcal{H}_{\text {matt }}^\lambda$ orthogonal to this subspace and thus to all $\ket{\psi_s}$. Thus to prove that the states $\ket{\psi_{\alpha}}$ generate $\mathcal{H}_{\text {matt }}^\lambda$, it suffices to show that a vector $\ket{\chi}$ that is orthogonal to all of these states vanishes. Any vector $\ket{\chi}$ has an expansion $\ket{\chi}=\sum_{n=0}^{\infty} c_{n} \ket{\psi_{n}}$ with complex coefficients $c_{n}$ and vectors 
\begin{align}
\ket{\psi_n} = (j_2+ij_3)^n \ket{\psi_0}. 
\end{align}
For a given nonzero $\ket{\chi}$, there is some $k \geq 0$ such that $c_{k} \neq 0$ but $c_{k^{\prime}}=0$ for $k^{\prime}<k$. Then $g(\alpha)=\left\langle\chi| \psi_{\alpha}\right\rangle$ is not identically zero, since in fact
\begin{align}
\left.\frac{d^{k}}{\mathrm{~d} \alpha^{k}} g(\alpha)\right|_{\alpha=0} \neq 0 \text {. }
\end{align}

Now let us go back to the usual quantization used in this paper in which the Hilbert space of JT gravity with matter is $\mathcal{H}_{JT} \cong L^2(\R)  \otimes \mathcal{H}_{\text {matt }}$. From what was explained in the last paragraph, it follows that vectors in $L^2(\R) \otimes \mathcal{H}_{\text {matt }}^\lambda$ of the form
\begin{align}\label{eq:vectorinweirdbasis}
\Psi = \int_{-\infty}^{\infty} d \alpha f(\chi, \alpha)\ket{ \psi_{\alpha}}
\end{align}
are dense in $L^2(\R) \otimes \mathcal{H}_{\text {matt }}^\lambda$. Consequently, we can define a map $U: L^2(\R) \otimes \mathcal{H}_{\text {matt }}^\lambda  \rightarrow \widetilde{\mathcal{H}}$ between the two quantizations by saying that $U$ maps the state $\Psi$ given in \eqref{eq:vectorinweirdbasis} to the function $f\left(\chi_{+}, \chi_{-}\right) \in \widetilde{\mathcal{H}}$. To begin with $U$ is only densely defined with dense image, but in view of what has been explained, $U$ is an isometry and therefore extends to a unitary mapping between Hilbert spaces. Hence the two quantizations are equivalent.

We can also use \eqref{eq:keygrouptheoryaltquantisation} to show more directly that $\widetilde{\mathcal{H}}$ describes the coinvariant Hilbert space for wavefunctions of the form \eqref{eq:highestweightwavefunctions}. The delta functions $\delta(T)$ and $\delta(T')$ from each of the ket wavefunction and the bra wavefunction kill the integrals over $T$ and $T'$ and reduce the integral over the $\hSL(2,\R)$ gauge group to an integral over $g = e^{i\alpha \mathcal{D}(\mathcal{J}_3)}$. We are left with the inner product
\begin{align}
\braket{\mathbf{\Phi}_{JT}, \mathbf{\Psi}_{JT}} &= \int d\chi d\chi' d\alpha \, \bar\Phi(\chi, \chi')e^{i \alpha(p_\chi - p_{\chi'})} \Psi(\chi , \chi' ) \left\langle\psi_0\left|e^{i\alpha j_{3}}\right| \psi_0\right\rangle
\\&=  \int d\chi d\chi' d\alpha \, \bar\Phi(\chi - \frac{\alpha}{2}, \chi' + \frac{\alpha}{2}) \frac{1}{\left(\cosh \left(\frac{\alpha}{2}\right)\right)^{2 \lambda}} \Psi(\chi +\frac{\alpha}{2} , \chi' - \frac{\alpha}{2}).
\end{align}
This is the same as \eqref{eq:altinnerproduct} with $\chi_+ = (\chi + \chi')/2$, $\chi_- = (\chi - \chi' -\alpha)/2$ and $\tilde\chi_- = (\chi - \chi' +\alpha)/2$.

The Hamiltonians \eqref{eq:highestweighthams} were previously found for a single free matter particle in Eqn. 30 of \cite{Lin:2022rbf} by taking the JT gravity limit of the chord Hilbert space of the double-scaled SYK model. The inner product \eqref{eq:altinnerproduct} is the JT gravity limit of the  inner product found for single-particle chord Hilbert space states in \cite{Lin:2023trc}.

\bibliographystyle{unsrt}
\bibliography{all.bib}

\end{document}